\documentclass[aps,10pt,singlecolumn,superscriptaddress,amsmath,amssymb]{revtex4-2}

\usepackage[utf8]{inputenc}
\usepackage{subfiles}
\usepackage{graphicx}
\usepackage[colorlinks,allcolors=cyan!70!black]{hyperref}
\usepackage{float}
\usepackage[normalem]{ulem}
\usepackage{comment}
\usepackage{xcolor}
\DeclareMathOperator\erf{erf}

\usepackage{lipsum}

\begin{document}

\preprint{APS/123-QED}

\title{On distinguishability among cell-division models based on population and single-cell-level distributions}

\author{Vikas}
 \email{vikaskaushikkarora@gmail.com}
 \affiliation{Department of Physics, Indian Institute of Technology Delhi, Hauz Khas, 110016, New Delhi, India}
\author{Rahul Marathe}
 \email{maratherahul@physics.iitd.ac.in}
 \affiliation{Department of Physics, Indian Institute of Technology Delhi, Hauz Khas, 110016, New Delhi, India}
\author{Anjan Roy}
 \email{anjanroy@dbeb.iitd.ac.in}
 \affiliation{Department of Biochemical Engineering and Biotechnology, Indian Institute of Technology Delhi, Hauz Khas, 110016, New Delhi, India}

\date{\today}

\begin{abstract}
It is well known that the different cell-division models, such as Timer, Sizer, and Adder, can be distinguished based on the correlations between different single-cell-level quantities such as birth-size, division-time, division-size, and division-added-size. Here, we show that other statistical properties of these quantities can also be used to distinguish between them. Additionally, the statistical relationships and different correlation patterns can also differentiate between the different types of single-cell growth, such as linear and exponential. Further, we demonstrate that various population-level distributions, such as age, size, and added-size distributions, are indistinguishable across different models of cell division despite them having different division rules and correlation patterns. Moreover, this indistinguishability is robust to stochasticity in growth rate and holds for both exponential and linear growth. Finally, we show that our theoretical predictions are corroborated by simulations and supported by existing single-cell experimental data.


\end{abstract}

\maketitle

\section{Introduction}

Understanding bacterial cell growth and division is crucial in microbiology, biotechnology, and biochemical engineering. Despite the underlying complexity of these processes, extensive research conducted over several decades has revealed that bacterial cells often follow relatively robust and simple principles, such as the so-called bacterial growth laws and size laws \cite{schacMaloe1958, schacMaloe1958B, scottMatthew2010, scottMatthew2014, donachie1968, helmstCooper1968}. In particular, it was found that cell division is governed by regulatory mechanisms that can be effectively captured using coarse-grained models. Earlier models of bacterial cell division suggested that cells follow either the Timer or the Sizer mechanism \cite{powell1956, kochSchac1962, powell1964, wheals1982, diekmann1983, tysonDiekmann1986}. The Timer model posits that the cells in the culture divide after a fixed time from birth, implying that division is governed purely by age \cite{powell1956}. In contrast, the Sizer model suggests that division occurs once a cell reaches a critical size, ensuring a threshold for division \cite{kochSchac1962, powell1964, wheals1982, diekmann1983, tysonDiekmann1986}. While the Sizer model gained early traction, Koppes et.al. criticized the model based on various inferred correlations and the skewness in division-time, which did not agree with the model's predictions \cite{koppes1980, voornKoppes1997}.

Although some microorganisms may indeed follow the Sizer division strategy \cite{facchettiChang2017, miottoGosti2024}, recent experimental studies on various bacterial species \cite{iyerSrividya2014, santiDhar2013, taheriJun2015, campos2014, deforetDitmarsch2015, fievetDucret2015, messelink2021, chungKarAmir2024, logsdonAmir2017}, including \textit{Escherichia coli}, \textit{Bacillus subtilis}, \textit{Caulobacter crescentus}, \textit{Mycobacterium smegmatis}, \textit{Mycobacterium tuberculosis}, \textit{Desulfovibrio vulgaris}, and \textit{Corynebacterium glutamicum}, have shown that most of the bacterial cells instead follow a third strategy, known as the Adder principle, or the Adder model \cite{voornKoppes1993, amir2014, junReview2015, junReview2016}. According to the Adder principle, a cell divides after adding a constant amount of biomass, regardless of its initial size or age. Besides bacterial cells, yeasts and mammalian cells also follow division strategies close to the Adder model \cite{chandlerBrown2017, soiferAmir2016, nobsMaerkl2014, cadartGrilli2018}.

\begin{figure}[h]
    \centering
    \begin{tabular} {cc}
        \includegraphics[width=0.47\textwidth]{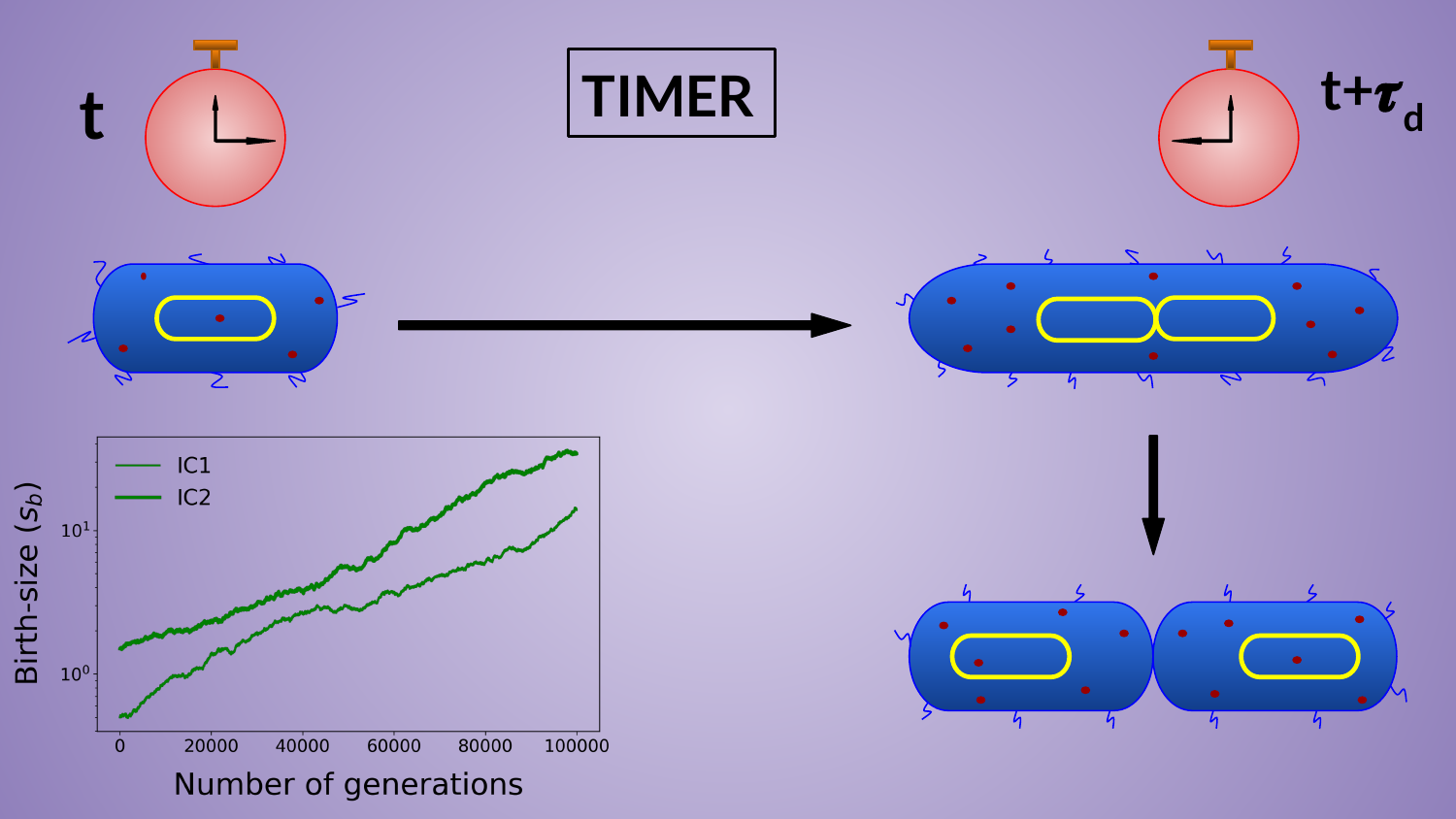}&
        \includegraphics[width=0.47\textwidth]{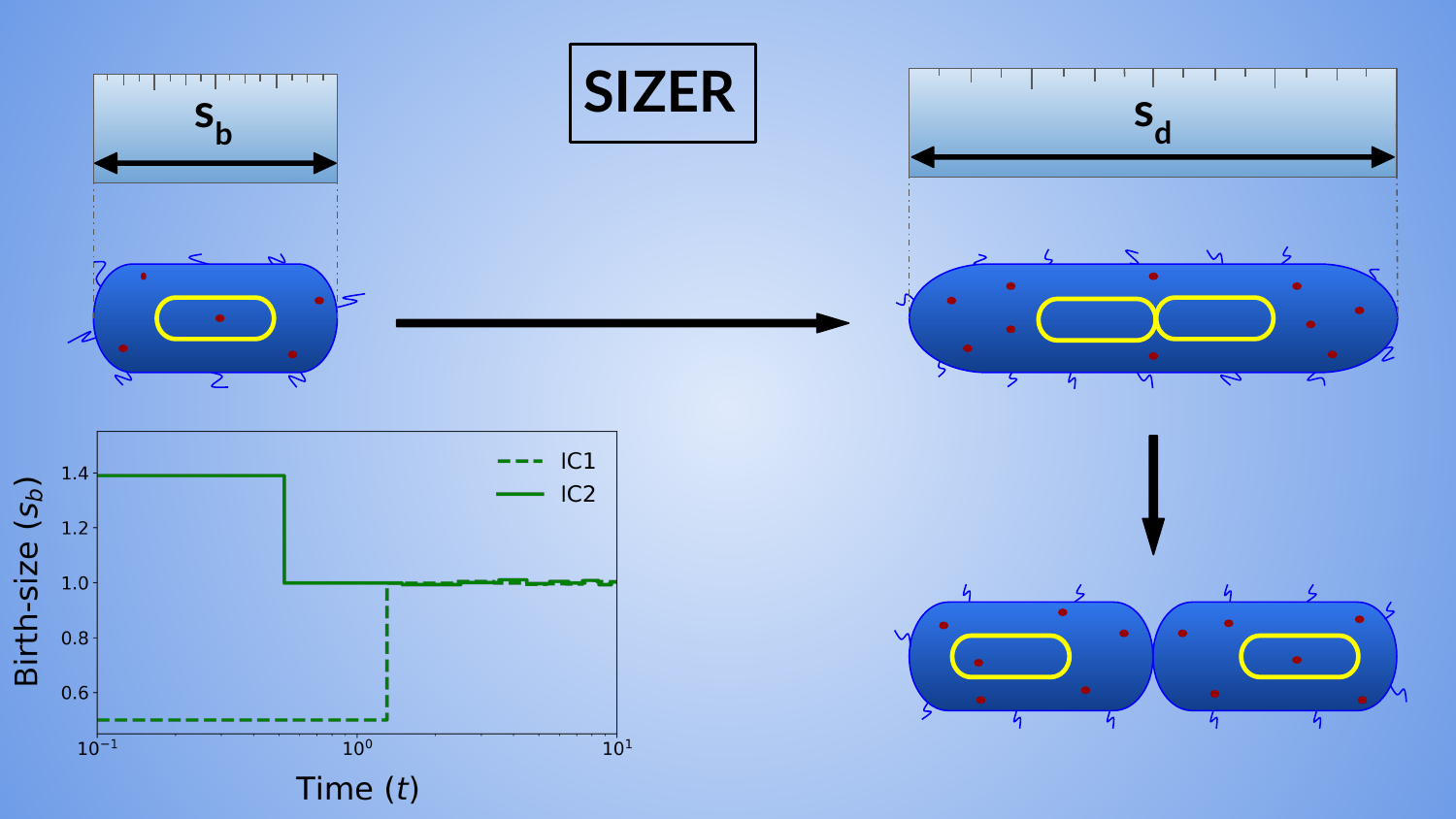}\\
    \end{tabular}
    \includegraphics[width=0.47\textwidth]{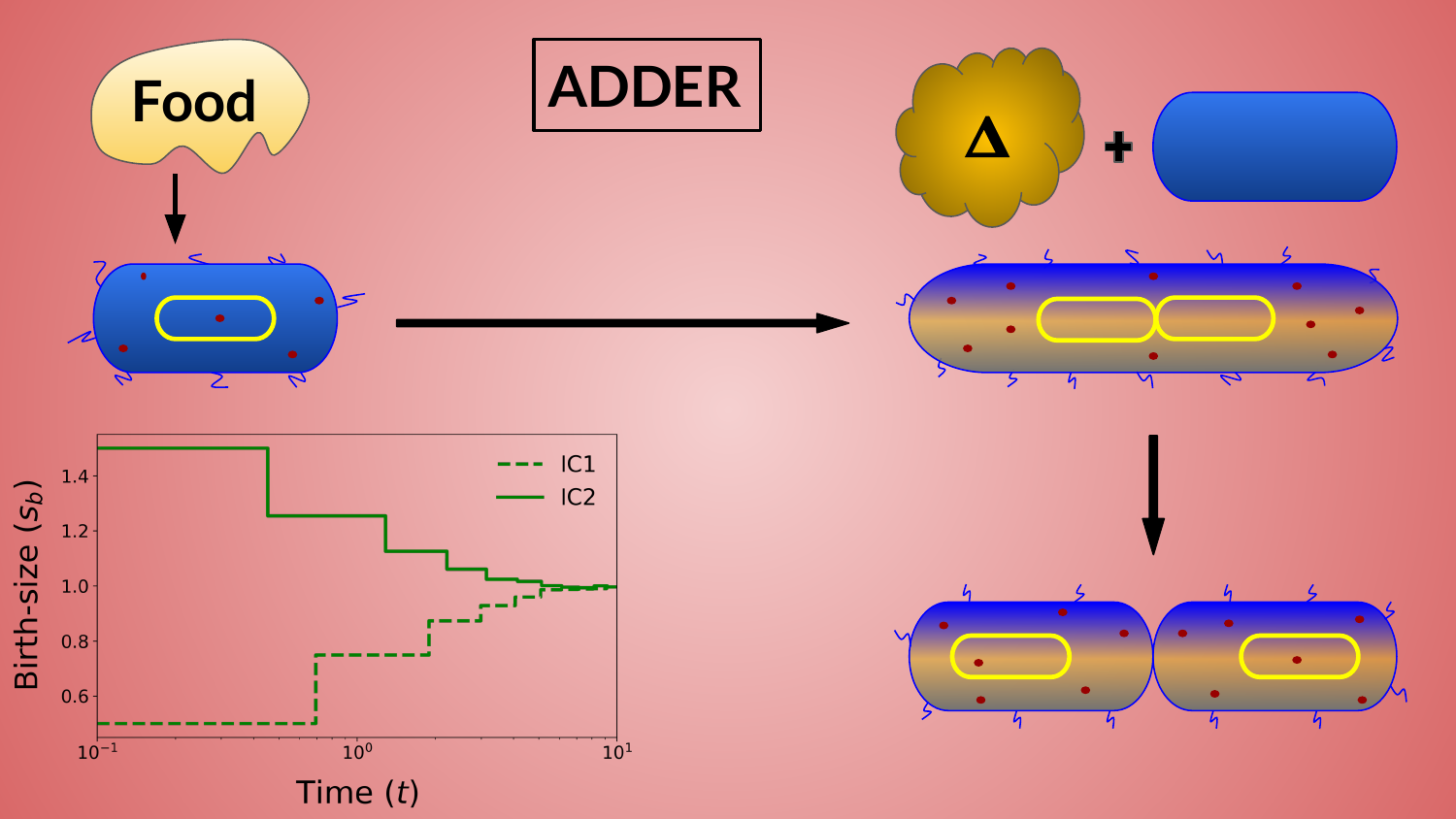}\\
    \caption{Artistic representations of various cell-division models: Timer, Sizer, and Adder. The insets represent the graph of birth-size of a random cell tracked for many generations in a simulated CSTR culture (containing 1 million cells) following the particular cell division model with stochasticity in growth rate and principle of cell division. IC1 and IC2 represent `initial condition 1' (initial birth-size being 0.5) and `initial condition 2' (initial birth-size being 1.5), respectively.}
    \label{fig: cartoon}
\end{figure}

It is interesting to note that while the Sizer and Adder rules enable cell-size homeostasis, Timer does not \cite{taheriJun2015, junReview2015, junReview2016, willisHuang2017, vuaridelDhar2020}. Cell-size homeostasis is fundamental to bacterial survival, ensuring cells maintain a stable birth-size over generations. There is a theoretical limit on the minimum size of a cell set by various factors, such as the minimum volume required to accommodate the essential components of the cell, unusually high concentration of metabolites, and properties of the cell membrane \cite{pirie1973}. Also, bigger cells face other problems, such as a lower surface-to-volume ratio, a larger diffusion time for proteins, enzymes, and metabolites, and a low DNA to cytoplasm ratio resulting in the inability to synthesize more nucleic acids and proteins in proportion to the increase in cell volume \cite{marshallYoung2012, dillGhosh2011, amodeoSkotheim2016, neurohrTerry2019}. Therefore, the cells should have an optimal size to operate appropriately \cite{amodeoSkotheim2016, neurohrTerry2019}, and maintaining cell-size homeostasis ensures that future generations of cells will continue to be in their optimal size.

The Sizer model naturally maintains cell-size homeostasis by enforcing division at a fixed size. Regardless of initial birth-size, subsequent newborn cells are always half the division-size, particularly for symmetrically dividing cells, ensuring a stable birth-size distribution over time. Similarly, the Adder model achieves homeostasis by requiring cells to grow by the addition of a fixed biomass before division. Any size variation at birth diminishes over the successive divisions, leading to convergence toward a stable birth-size. This property of cell-size homeostasis for the Sizer and Adder models can be observed in Fig. \ref{fig: cartoon}, where the average birth-size approaches a constant value with time. However, the Timer model maintains homeostasis only if the growth rate and division-time are fine-tuned, with no stochasticity present in either parameter. The stochasticity in division-time or growth rate causes the cells to attain bigger birth-sizes than their parent cells, which leads to an increase in the average birth-size over time (Appendix \ref{sec: SbirthSize}). Therefore, as shown in Fig. \ref{fig: cartoon}, the average birth-size for the Timer model continues to grow, demonstrating the absence of homeostasis.

Different cell-division strategies, such as the Timer, Sizer, and Adder models, have different underlying biochemical mechanisms to control division. For example, several studies infer the size-sensing mechanisms in S. Cerevisiae, S. Pombe, and E. Coli, which lead to the Sizer-like behavior for cell division \cite{marshallYoung2012, rothfield2005, robert2015, facchettiChang2017}. Additionally, it is shown that the Adder model is a direct consequence of the fact that cells accumulate specific division-related proteins and precursors up to a fixed threshold number before committing to division \cite{siJun2019, pandeyJain2020, nietoGarcia2024}. Knowing which division model is followed by a specific bacterial species can be useful to pinpoint the underlying mechanism for cell-size control and understand its evolutionary relation to other species. Although the different models of cell divisions can be distinguished by the correlations between single-cell-level quantities, such as birth-size, division-time, division-size, and division-added-size \cite{campos2014, taheriJun2015, junReview2015, junReview2016, soiferAmir2016, pandeyJain2020}, in this study, we examine other statistical properties of these single-cell-level quantities and show that, in addition to the correlations, some of the statistical relationships between these quantities can also differentiate between the Sizer and Adder models.

Additionally, population-level distributions, such as age and size distribution for a bacterial population in the culture, are essential in biochemical engineering. For example, they can be useful in optimizing protein production in biochemical reactors \cite{kurthSawodny2023}. Also, they can be useful in analyzing drug efficacy, as the average gene copy number depends on age distribution \cite{junReview2018}. Some of these distributions, such as the age distribution for the Timer model \cite{powell1956} and the size distribution for the Sizer model \cite{kochSchac1962, powell1964, vuaridelDhar2020, genthon2022}, are already known. A few studies also refer to the size and division-time distributions for the Adder model \cite{genthon2022, xiaGreenman2020, kudtarkar2025}. A comprehensive analysis of the three division models, however, is lacking. Also, it is unknown whether the choice of division principle has important implications for the various population-level distributions, such as age and size distributions. In this study, we provide an extensive analysis of bacterial cell division under these three division paradigms and investigate the impact of the different division rules on the distributions of various cellular properties in the culture. We show that even though Timer, Sizer, and Adder models impose different regulatory constraints and have different correlation and statistical patterns at the single-cell level, they ultimately yield statistically indistinguishable distributions at the population level.

Further, while recent studies have established that most of the bacterial species grow in size exponentially \cite{wangJun2010, iyerSrividya2014, taheriJun2015, deforetDitmarsch2015, fievetDucret2015}, some of them indicate linear growth at the level of a single cell \cite{santiDhar2013, chungKarAmir2024, nobsMaerkl2014}. Therefore, we expand our formalism to account for the case of linear biomass growth, derive probability distributions for various cellular properties, and show that while the population-level distributions for linear growth continue to be indistinguishable across different cell-division models, single-cell relationships can be used to differentiate between them. Finally, we compare distributions under exponential and linear growth to summarize the distinguishability between various growth and division models. We compare our analytical predictions with simulations and test them against published experimental results, highlighting the precision level needed to differentiate the underlying growth and division rule. While our results are independent of the exact molecular implementation of the division  and growth rules, the search for molecular implementations remains an exciting endeavor.

\section{Materials and Methods}

\subsection{Various types of distributions for a culture}
In a bacterial culture, various properties can be associated with each cell, whose distributions in the population are of interest. These properties include the age $(a)$ of the cell at a given point in time, the size $(s)$ of the cell at a given point in time (size refers to volume here), and the size that has been added to the cell since its birth, which is indicated by $\Delta$. These quantities evolve for each cell as time progresses. In an asynchronous population, at any given time, different cells will have different ages, sizes, and added-sizes. Therefore, there will be a corresponding probability distribution for each of these quantities of the cells in the culture, at each point in time. They are the age distribution $\Theta(a)$, size distribution $\Phi(s)$, and added-size distribution $\Psi(\Delta)$. The distributions obtained for all the cells in the population at a given time are what we call the `population-level distributions'. The population-level distributions evolve as time progresses; they become stationary for a culture that is in a steady state. Although one can obtain these distributions at each point in time, we are only interested in a bacterial culture that is in steady state, and, therefore, only interested in the stationary forms of the distributions.

In contrast to the quantities that change with time, there are other quantities specific to a cell that do not evolve over its lifetime. These quantities include birth-size $(s_b)$: the size of the cell when it was born; division-time $(\tau_d)$: the time period after which a given cell divides since the time it was born; division-size $(s_d)$: the size of a given cell at which it divides; and division-added-size $(\Delta_d)$: the size a given cell has added starting from the time of birth to the time when it divides. While these quantities do not evolve in time, they do fluctuate from cell to cell, from generation to generation. The corresponding distributions of these quantities, such as birth-size distribution $\zeta(s_b)$,  division-time distribution $\Gamma(\tau_d)$, division-size distribution $\Xi(s_d)$,  and division-added-size distribution $\Omega(\Delta_d)$, are referred to as `single-cell-level distributions'. For each cell division model (Timer, Sizer, and Adder), one of these single-cell-level distributions is the \textit{principal distribution} of that model. This quantity represents the division rule and is a free parameter of the model. For the Timer model, the division-time distribution $\Gamma(\tau_d)$ is the principal distribution. Similarly, division-size distribution $\Xi(s_d)$ is the principal distribution for the Sizer model, and division-added-size distribution $\Omega(\Delta_d)$ is the principal distribution for the Adder model. The principal distribution is assumed to be known beforehand, and it is used to derive all other distributions analytically. The symbols for various quantities of interest and their probability distributions are summarized in Table \ref{table: symbols}.

\begin{table}[h!]
    \centering
    \renewcommand{\arraystretch}{1.5}
    \begin{tabular}{|l||c|c|c|c|c|c|c|}
        \hline
        \textbf{Quantity} & Age & Size & Added-size & Birth-size & Division-time & Division-size & Division-added-size \\ \hline
        \textbf{Symbol} & $a$ & $s$ & $\Delta$ & $s_b$ & $\tau_d$ & $s_d$ & $\Delta_d$ \\ \hline
        \textbf{Probability distribution} & $\Theta(a)$ & $\Phi(s)$ & $\Psi(\Delta)$ & $\zeta(s_b)$ & $\Gamma(\tau_d)$ & $\Xi(s_d)$ & $\Omega(\Delta_d)$ \\ \hline
    \end{tabular}
    \vspace{0.5em}
    \caption{Mathematical symbols for various quantities and their probability distributions. The first three are population-level distributions, while the remaining four are single-cell-level. The population-level properties vary systematically through the life-cycle of the cell, while the single-cell-level properties stay constant throughout the cell-cycle and only update upon division, due to stochasticity.}
    \label{table: symbols}
\end{table}

\subsection{Methods}
Under steady-state growth conditions, we examine various population-level and single-cell-level qualities across different division strategies: the Timer, Sizer, and Adder models. While we also consider the Timer model, we note that the size-related properties in a bacterial culture governed by this division rule, such as $s_b$, $s$, $\Delta$, $s_d$, and $\Delta_d$, do not converge to a steady state under exponential biomass growth due to absence of size-homeostasis. Nevertheless, the age distribution for the Timer model under exponential growth does reach a steady state, and we analyze this aspect in detail. We assume that individual cells grow either exponentially or linearly. For exponential growth, the biomass growth follows the relation $s = s_b \exp{(\alpha a)}$, where $s$, $s_b$, and $a$ are the cell's size, birth-size, and age respectively. For the case of linear growth, the biomass growth is given by the relation $s = s_b + \alpha a$. The growth rate $\alpha$ may be treated as either a fixed constant or it can be a stochastic variable ($\alpha$ can be stochastic in time, or across the population, or both). Note that the stochasticity in the system does not originate solely from variability in the growth rate. A primary source of stochasticity arises from the inherent randomness in the division mechanism specific to each model \cite{powell1955, powell1956, schaechterKoch1962B, miyata1978, taheriJun2015, tanouchi2017}. This division-related variability is captured through model-specific principal distributions, which govern each paradigm's division rules ($\Gamma(\tau_d)$ for Timer, $\Xi(s_d)$ for Sizer, and $\Omega(\Delta_d)$ for Adder). Hence, our model has two free parameters: the principal distribution for the cell division and the growth rate of an individual cell.

To simulate the system in its steady state, we modeled a population of 10 million cells growing in a continuous stirred-tank reactor (CSTR) like environment where, in the steady state, the number of cells is maintained constant. In our simulations, this is achieved by replacing a random cell in the reactor with one of the two daughter cells whenever a cell division takes place \cite{royKlumpp2018}. The cell division is governed by either the Timer, Sizer, or Adder strategy. For each cell, three key properties: age, size, and birth-size, are monitored and updated at each point in time.
At every time step $\Delta t$ (which is taken to be two orders of magnitude smaller than the estimated average division-time), each cell is checked for division, with the criterion for cell division depending on the model: in the Timer model, a cell divides if its age exceeds a value sampled from the division-time distribution $\Gamma(\tau_d)$; in the Sizer model, division occurs when the cell size surpasses a value drawn from the division-size distribution $\Xi(s_d)$; and in the Adder model, division is triggered when the added-size exceeds a value sampled from the division-added-size distribution $\Omega(\Delta_d)$. Whenever a division occurs, the age of the cell is set to zero, the size of the cell is halved, and the birth-size is updated to the new value, which is equal to the size of the cell immediately after its division. These same quantities for the dividing cell (age, size, and birth-size) are then also assigned to a random cell in the culture to keep the number of cells constant, as discussed earlier. However, if a cell-division does not take place, the cell's age is increased by the time step $\Delta t$, and its size is increased by a factor of $\exp{(\alpha a)}$ for exponential growth and by $\alpha a$ for linear growth, while maintaining the birth-size unchanged.

For scenarios involving stochastic growth (stochastic in time), the algorithm does not change, except that at each time step $\Delta t$, $\alpha$ is sampled from a probability distribution as $\alpha_i$, and whenever division does not take place, the age of each cell is incremented by $\Delta t$ and the size is increased by a factor of $\exp(\alpha_i\Delta t)$ for exponential growth and by $\alpha_i \Delta t$ for linear growth. Similarly, for the case of $\alpha$ being stochastic over population, similar procedure is followed, except that each cell is assigned a growth rate $\alpha_j$ (sampled from a probability distribution) that remains constant for the cell throughout its lifetime, and only changes when the cell divides.

As the simulation progresses, the distributions of cellular properties such as age, size, and birth-size evolve. Eventually, these distributions converge to the analytically derived steady-state forms. Moreover, the birth-size $s_b$, division-time $\tau_d$, division-size $s_d$, and division-added-size $\Delta_d$ are recorded for each cell division event. From this data, the single-cell-level distributions, such as $\zeta(s_b)$, $\Gamma(\tau_d)$, $\Xi(s_d)$, and $\Omega(\Delta_d)$ are obtained and analyzed.

To derive the analytical forms of these distributions, we first obtain $\zeta(s_b)$ in terms of the principal distribution by writing the birth-size of a random cell in terms of division-size or division-added-size of cells from previous generations. After that, we use the approach akin to Powell's \cite{powell1956}, viz., survival probability arguments for the cells in the culture, to obtain the analytical form of one of the population-level distributions in terms of the principal distribution --- $\Theta(a)$ is obtained from $\Gamma(\tau_d)$ for the Timer model \cite{euler1977, lotka1922, hoffman1949, powell1956, scherbaumRasch1957}, $\Phi(s)$ is obtained from $\Xi(s_d)$ for the Sizer model \cite{kochSchac1962}, and $\Psi(\Delta)$ is obtained from $\Omega(\Delta_d)$ for the Adder model using the equations given below:
\begin{equation} \label{eq: mainTimer}
    \Theta(a) =  A\; 2^{-a/ \langle \tau_d \rangle} \int_{a}^{\infty} d\tau_d \; \Gamma(\tau_d)
\end{equation}
\begin{equation} \label{eq: mainSizer}
    \Phi(s) = \frac{B}{s^2} \int_{s}^{2s} ds_d \; \Xi(s_d)
\end{equation}
and
\begin{equation} \label{eq: mainAdder}
    \Psi(\Delta) = \frac{C}{(\langle s_b \rangle + \Delta)^2} \int_{\Delta}^{\infty} d\Delta_d \; \Omega(\Delta_d)
\end{equation}
where $\langle \tau_d \rangle$ and $\langle s_b \rangle$ are average division-time and average birth-size, and $A$, $B$, and $C$ are normalization constants. After that, the analytical forms of the other remaining population-level distributions are derived using the method of probability transformation of random variables: For the Sizer model, $\Psi(\Delta)$ and $\Theta(a)$ are obtained from $\Phi(s)$ and $\zeta(s_b)$, and for the Adder model, $\Phi(s)$ and $\Theta(a)$ are obtained from $\Psi(\Delta)$ and $\zeta(s_b)$. For the Timer model, only $\Theta(a)$ is obtained because the other distributions related to size are not stationary in time, as mentioned before.

The details regarding analytical calculations are provided in the Appendices. Appendix \ref{sec: SbirthSize} examines the birth-size distributions for the Timer, Sizer, and Adder models. In Appendix \ref{sec: Stimer}, the derivation of $\Theta(a)$ for the Timer model using Powell's approach is discussed. Appendices \ref{sec: Ssizer} and \ref{sec: Sadder} derive various population and single-cell-level distributions for the Sizer and Adder models respectively. Additionally, Appendix \ref{sec: SalternateMethod} discusses an alternate approach to obtain the analytical forms of population-level distributions. The case of stochastic growth rate is further considered in Appendix \ref{sec: SstochsaticGrowthRate}, and Appendix \ref{sec: SstatRelationships} establishes the statistical relationships among single-cell-level quantities for the Sizer and Adder models. The case of linear growth is addressed in Appendix \ref{sec: SlinearGrowth}.

\section{Results}
\subsection{Population-level distributions are indistinguishable among different cell-division models}
\begin{figure*}[h!] 
    \centering
    \includegraphics[width=\linewidth]{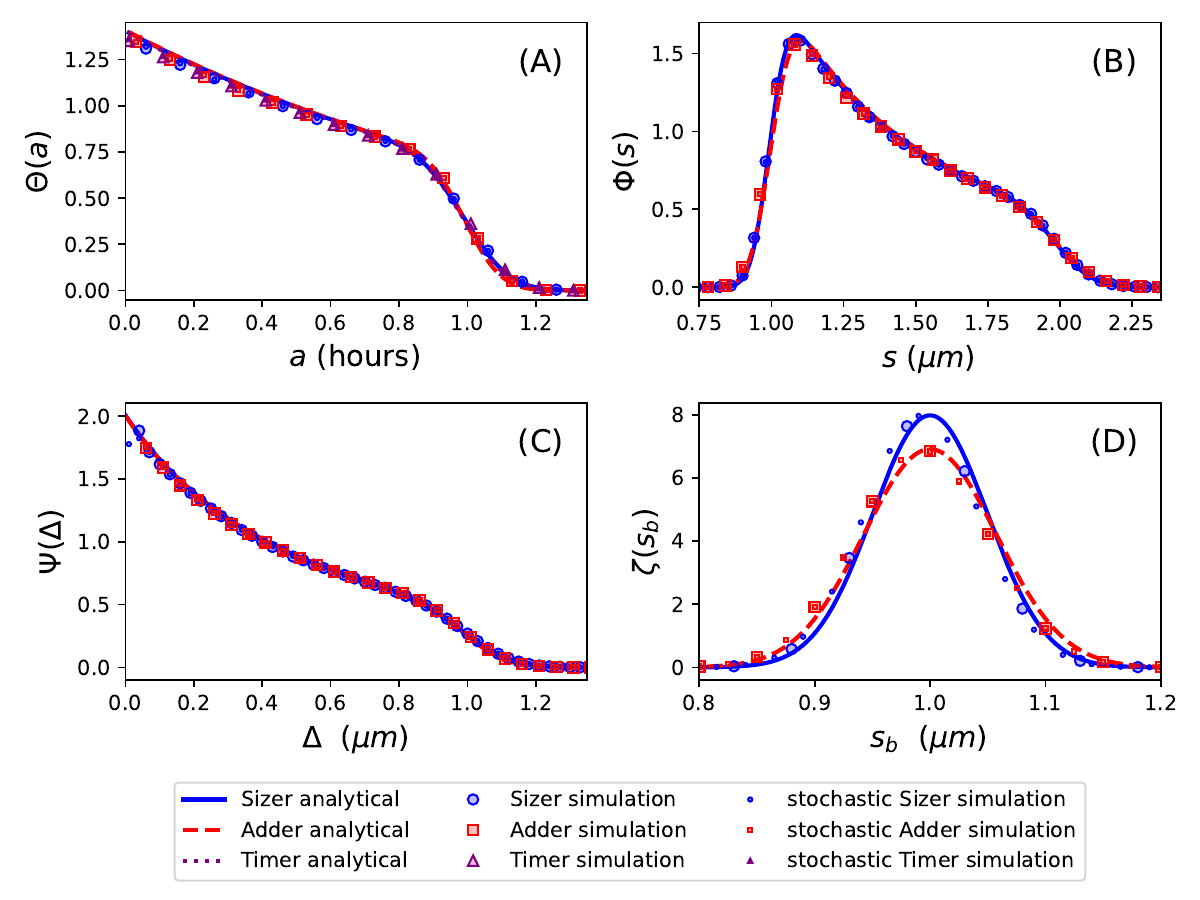}
    \caption{Distribution of various quantities across the cells in CSTR like growth conditions for the three models of cell division, under exponential biomass growth: \textbf{(A)} age distribution, \textbf{(B)} size distribution, \textbf{(C)} added-size distribution, \textbf{(D)} birth-size distribution. The principal distribution for a division model, which is a free parameter based on which the population-level distributions are obtained, are chosen as follows - the division-time distribution $\Gamma(\tau_d)$ for the Timer model is Gaussian with a mean $\langle \tau_d \rangle =1$ hour and a standard deviation $\sigma_{\tau_d} = 0.1$ hour, the Sizer model's division-size distribution $\Xi(s_d)$ is Gaussian with a mean $\langle s_d \rangle =2$ $\mu$m and standard deviation $\sigma_{s_d} = 0.1$ $\mu$m, the division-added-size distribution $\Omega(\Delta_d)$ for the Adder model is taken to be Gaussian with a mean $\langle \Delta_d \rangle =1$ $\mu$m and standard deviation $\sigma_{\Delta_d} = 0.1$ $\mu$m. The lines indicate the analytical curves for the distributions (which are identical regardless of $\alpha$ being stochastic or deterministic), and the markers indicate the distributions obtained from the simulations. Timer, Sizer, and Adder in the legend indicate the results for the corresponding model when the growth rate is fixed. The growth rate for this case is taken to be $\ln(2)$ per hour. In contrast, stochastic Timer, stochastic Sizer, and stochastic Adder in the legend indicate the simulation results for the case when the growth rate is stochastic in time. The mean growth rate for this is taken to be $\ln(2)$ per hour, and the standard deviation in growth rate is taken to be $0.1$ per hour.}
    \label{fig: allModelsExponential}
\end{figure*}
We start with various distributions obtained for the case of exponentially growing cells in a CSTR-like set-up, with the growth rate taken to be constant and the principal distributions sampled from a Gaussian distribution, to simplify the analysis. This is in line with previous studies reporting approximately Gaussian behavior for $\Xi(s_d)$ and $\Omega(\Delta_d)$ \cite{taheriJun2015, fievetDucret2015, soiferAmir2016, priestman2017, siJun2019, nordholtHeerden2020}. 

We find that the population level distributions, such as age, size, and added-size distribution, are indistinguishable among the three cell-division models (Fig. \ref{fig: allModelsExponential}). The close alignment between the simulation data (markers) and analytical predictions (lines) in this figure further underscores the robustness of our findings. This indistinguishability of the population-level distributions is unexpected due to the different division rules for various models. However, this similarity can be easily understood via an alternate method. In this alternate method, one first obtains all the single-cell distributions, such as $\Gamma(\tau_d)$, $\Xi(s_d)$, and $\Omega(\Delta_d)$, through the probability transformations of the principal distributions and $\zeta(s_b)$. After that, all population-level distributions are obtained from the corresponding single-cell distributions using Eq. \ref{eq: mainTimer}, \ref{eq: mainSizer}, and \ref{eq: mainAdder} (Appendix \ref{sec: SalternateMethod}). For the Sizer and the Adder models, we show that if the principal distribution is Gaussian, the other single-cell-level distributions will also be Gaussian up to a very good approximation, in the limit of low stochasticity in principal distribution, as shown in Fig. \ref{fig: SdastauDsizer} and \ref{fig: SdstauDadder} (see Appendices \ref{sec: Ssizer} and \ref{sec: Sadder}). Furthermore, one can always choose the free parameters: the principal distributions and the growth rate for all of the models, such that other single-cell-level distributions have the same mean and similar standard deviations across different models. Since the three formulae to obtain the population-level distributions (Eq. \ref{eq: mainTimer}, \ref{eq: mainSizer}, and \ref{eq: mainAdder}) involve the integration over the single cell distributions, it makes the effect of slightly different standard deviations vanish. Therefore, the population-level distributions for any model are almost indistinguishable from each other.

Note that, in Fig. \ref{fig: allModelsExponential} (D), $\zeta(s_b)$ is different for the Sizer and Adder models. One can indeed choose the principal distributions of the two models such that $\zeta(s_b)$ becomes identical. However, as we will show in a later section, the standard deviations of the single-cell-level quantities are related differently for the two models. Therefore, one cannot make all of the single-cell-level distributions identical simultaneously. If we choose the principal distributions to be such that $\zeta(s_b)$ is identical, other single-cell-level distributions will become slightly different. One may expect this to cause differences in the population-level distributions, owing to the alternate method. However, due to the integration involved, this difference turns out to be not so substantial (Fig. \ref{fig: SallModelsExponentialWithSameSbDisbn}).


\subsection{This indistinguishability is robust against stochasticity in the growth rate}
Biological processes are inherently stochastic, leading to variability in the growth rate of a cell. Temporal fluctuations in the growth rate arise from the intrinsic randomness of biochemical reactions within individual cells, while population-level variability results from differences in extrinsic factors such as nutrient availability. We find that when the growth rate $\alpha$ varies stochastically in time over the cell's life-cycle, the analytical forms of the distributions and the statistical relationships between various quantities are identical to the case when the growth rate is non-stochastic (Fig. \ref{fig: allModelsExponential}). This is so because the stochastic variable $\alpha$ is realized numerous times within the lifetime of a given cell, such that the effect of the stochasticity is diminished, which can be seen as a manifestation of the central limit theorem in this case (Appendix \ref{subsec: SstochsaticGrowthRateTemporal}). Therefore, one can replace $\alpha$ in the mathematical equations of the deterministic case with the time average value $\overline{\alpha}$ to get the results for the case of temporal stochastic growth rate.

However, for the case of growth rate being a stochastic parameter across the population, our analysis reveals that while the analytical forms of all other population and single-cell-level distributions remain the same, the forms of $\Gamma(\tau_d)$ and $\Theta(a)$ for the Sizer and Adder models are different from the ones obtained for the case with fixed growth rate (Appendix \ref{subsec: SstochsaticGrowthRatePopulation}). This is so because, in contrast to $\Gamma(\tau_d)$ and $\Theta(a)$, the size-related distributions have no explicit dependence on $\alpha$. While the $\Gamma(\tau_d)$ distribution for stochastic $\alpha$ becomes skewed, which is also observed in experiments \cite{powell1955, powell1956, schaechterKoch1962B, santiDhar2013, osellaNugent2014, taheriJun2015, chungKarAmir2024}, both of these distributions become flatter, which reflects higher stochasticity in the system relative to the case of non-stochastic growth rate (Fig. \ref{fig: SstochAlphaExpo}). Interestingly, $\Theta(a)$ continues to be indistinguishable between the Sizer and Adder models. 

For the Timer model, $\Gamma(\tau_d)$ and $\Theta(a)$ remain unchanged as they are either the principal distribution or obtained directly from the principal distribution. Note that the size-related distributions for the Timer model are not obtained due to the lack of cell-size homeostasis. 

\subsection{The relationships between various single-cell-level quantities distinguish different cell-division models}
\begin{figure*}[h!]
    \centering
    \includegraphics[width=\textwidth]{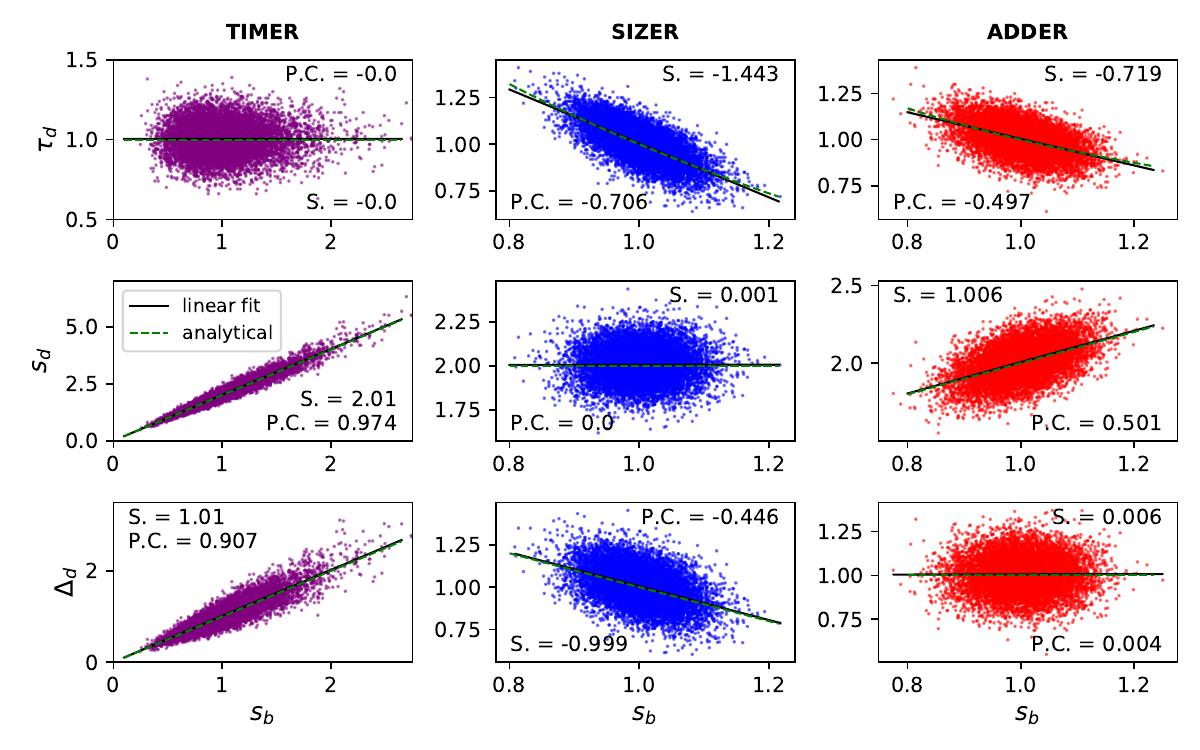}
    \caption{Various correlations for the three models, obtained using simulations under exponential biomass growth: the first column is for the Timer model, the second is for the Sizer model, and the third is for the Adder model; the first row is division-time versus birth-size, the second is division-size versus birth-size, and the third is division-added-size versus birth-size. The slope of the linear fit to the data is shown in each figure, indicated by S, along with the Pearson Correlation Coefficient, indicated by P.C. The analytical curve corresponds to the correlations inferred analytically from the relations $s_d = s_b \exp{(\alpha \tau_d)}$ and $s_d = s_b + \Delta_d$. The parameters of the simulations are the same as mentioned in Figure \ref{fig: allModelsExponential}. }
    \label{fig: corrExp}
\end{figure*}

\begin{figure}[h!]
    \centering
    \includegraphics[width=0.9\textwidth]{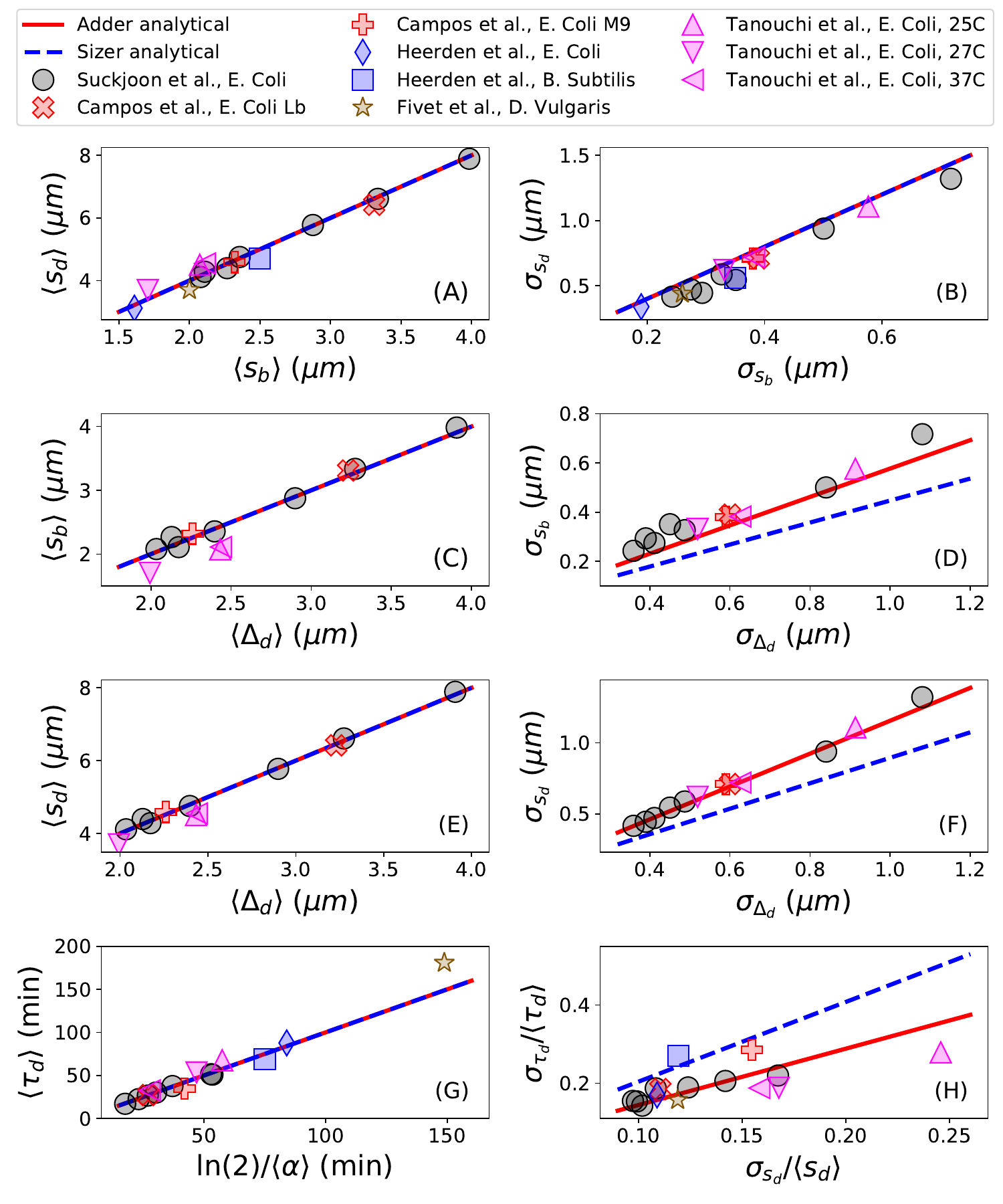}
    \caption{Mean and standard deviations of various single-cell-level quantities plotted against each other. The symbols are for the experimental data on different bacterial species grown in different media, taken from the previously published studies \cite{taheriJun2015, campos2014, heerdenKempe2017, fievetDucret2015, tanouchi2017}. \textbf{(A)} mean division-size versus mean birth-size, \textbf{(B)} standard deviation in division-size versus standard deviation in birth-size,\textbf{(C)} mean birth-size versus mean division-added-size, \textbf{(D)} standard deviation in birth-size versus standard deviation in division-added-size, \textbf{(E)} mean division-size versus mean division-added-size, \textbf{(F)} standard deviation in division-size versus standard deviation in division-added-size, \textbf{(G)} mean division-time versus $\log{(2)}$ times inverse of mean growth rate, \textbf{(H)} coefficient of variation of $\tau_d$ versus coefficient of variation of $s_d$.}
    \label{fig: experimentalData}
\end{figure}

It is well known that the correlations between various single-cell properties vary across different models of bacterial cell division \cite{campos2014, taheriJun2015, junReview2015, junReview2016, soiferAmir2016, pandeyJain2020}. In Fig. \ref{fig: corrExp}, these correlations are plotted for the three models: the Timer, Sizer, and Adder, and it is evident that the correlations can be used to distinguish between  the different models. The relations $s_d= s_b \exp{(\alpha \tau_d)}$ and $s_d = s_b + \Delta_d$ explain these correlations. In the Timer model, there is no correlation between division-time and birth-size, as cells divide after a fixed time interval regardless of their initial size. In contrast, the Sizer and Adder models negatively correlate division-time and birth-size. This is so because larger cells at birth reach the critical threshold for size (Sizer) or added-size (Adder) more quickly due to exponential growth, reducing the time to divide for the cell. Similarly, in the Sizer model, division-size is independent of birth-size, as all cells divide upon reaching a fixed threshold of size, irrespective of their size at birth. However, in the Timer model, division-size positively correlates with birth-size—larger cells accumulate more mass than smaller cells in the same time and thus divide at larger sizes. In the Adder model, division-size also shows a positive correlation with birth-size, since cells born larger, upon adding a fixed amount of biomass, will end up being of larger size at the time of their division. Furthermore, there is no correlation between division-added-size and birth-size for the Adder model, because cells divide after adding a constant biomass, irrespective of the biomass at the time of their birth. This correlation is positive for the Timer model because larger cells add more biomass to them over the fixed time for division. However, this correlation is negative for the Sizer model because larger cells require less additional growth to reach the fixed division-size.

Here, we show that some of the statistical relationships between these quantities are also different for the three models. We find that the mean values of the single-cell-level quantities, such as birth-size, division-size, and division-added-size, are related identically between the Sizer and Adder models, but some of the relationships between the standard deviations of these quantities enable discrimination between these two mechanisms. In a bacterial culture, the following relationship is well known to be true irrespective of cell division strategy (Sizer or Adder) and the type of single-cell growth (exponential or linear) \cite{campos2014, taheriJun2015}:
\begin{equation} \label{eq: mainMeanSingleCell}
    \langle s_d \rangle \;=\; 2\langle s_b \rangle \;=\; 2\langle \Delta_d \rangle
\end{equation}
However, the relationships between the standard deviations of these quantities are model-dependent. For the Adder model, the standard deviations satisfy the following relationship \cite{campos2014, taheriJun2015}:
\begin{equation} \label{eq: mainStdAdder}
    \sigma_{s_d} \;=\; 2\sigma_{s_b} \;=\; 2\sigma_{\Delta_d}/\sqrt{3}
\end{equation}
Whereas for the Sizer model, we show that the following relationship holds (Appendix \ref{sec: SstatRelationships}):
\begin{equation} \label{eq: mainStdSizer}
    \sigma_{s_d} \;=\; 2\sigma_{s_b} \;=\; 2\sigma_{\Delta_d}/\sqrt{5}
\end{equation}
Also, for both Adder and Sizer models, the mean division-time $\langle \tau_d \rangle$ for exponential biomass growth is given by \cite{taheriJun2015}:
\begin{equation} \label{eq: mainMeanTau_d}
    \langle \tau_d \rangle = \ln{2}/\alpha
\end{equation}
But the coefficient of variation of $\tau_d$ differs for the two models. For the Adder model, the following relationship holds \cite{taheriJun2015}:
\begin{equation} \label{eq: mainCOVadderExpo}
    \frac{\sigma_{\tau_d}}{\langle \tau_d \rangle} \;=\; \frac{1}{\ln{2}} \frac{\sigma_{s_d}}{\langle s_d \rangle} \;=\; \frac{1}{\ln{2}} \frac{\sigma_{s_b}}{\langle s_b \rangle} \;=\; \frac{1}{\sqrt{3} \;\ln{2}} \frac{\sigma_{\Delta_d}}{\langle \Delta_d \rangle}
\end{equation}
Whereas, for the Sizer model, we establish that the following relationship holds (Appendix \ref{sec: SstatRelationships}):
\begin{equation} \label{eq: mainCOVsizerExpo}
    \frac{\sigma_{\tau_d}}{\langle \tau_d \rangle} \;=\; \frac{\sqrt{2}}{\ln{2}} \frac{\sigma_{s_d}}{\langle s_d \rangle} \;=\; \frac{\sqrt{2}}{\ln{2}} \frac{\sigma_{s_b}}{\langle s_b \rangle} \;=\; \frac{\sqrt{2}}{\sqrt{5} \;\ln{2}} \frac{\sigma_{\Delta_d}}{\langle \Delta_d \rangle}
\end{equation}

These statistical relationships involving mean, standard deviation, and coefficient of variation, between the single-cell-level quantities, are plotted in Fig. \ref{fig: experimentalData}, where the solid lines indicate the analytical predictions and the markers indicate the single-cell experimental data taken from the published studies \cite{taheriJun2015, campos2014, heerdenKempe2017, fievetDucret2015, tanouchi2017}. As illustrated by Fig. \ref{fig: experimentalData} (D, F, and H), the relationships described by Eq. \ref{eq: mainStdAdder}, \ref{eq: mainStdSizer}, \ref{eq: mainCOVadderExpo}, and \ref{eq: mainCOVsizerExpo} can be used to distinguish between the Sizer and Adder models. Further, this distinguishability based on the single-cell relations does not get spoiled because of stochasticity in growth rate (see Fig. \ref{fig: SstochAlphaExpoCOVtauD}). Although the majority of experimental data points are consistent with the theoretical predictions of the models, certain points exhibit substantial deviations. Additionally, certain data points appear to support the Adder model, whereas the corresponding points in other plots are more consistent with the Sizer model (Campos et al, E. Coli, M9 data point in Fig. \ref{fig: experimentalData}). These discrepancies may be attributed to measurement errors; however, the possibility of alternative theoretical explanations cannot be excluded.

\subsection{Distinguishability in the case of linear growth} \label{subsec: linearGrowth}
\begin{figure*}[t!]
    \centering
    \includegraphics[width=\linewidth]{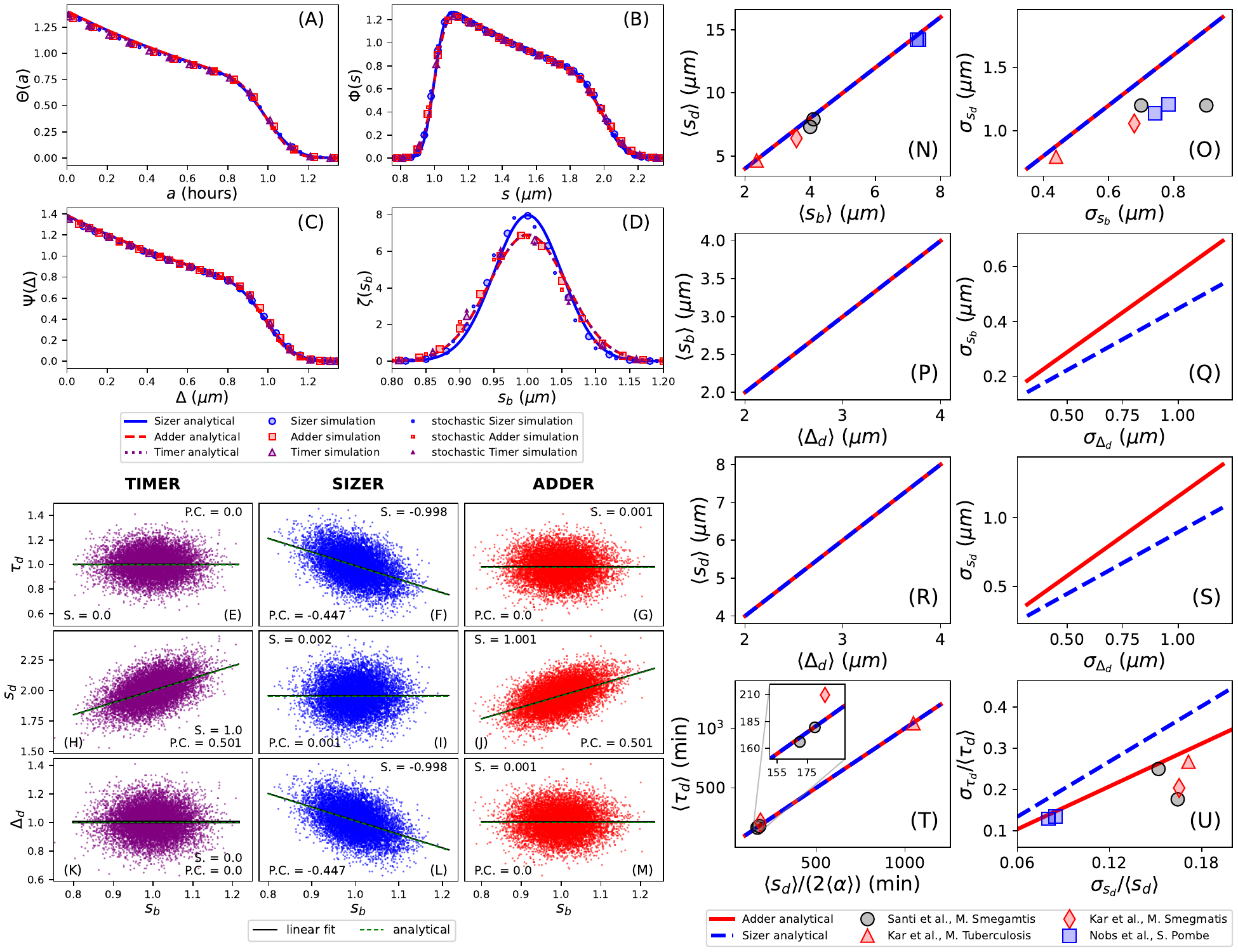}
    \caption{\textbf{(A-D)} Various distributions for the three models of cell division under linear biomass growth: \textbf{(A)} age distribution, \textbf{(B)} size distribution, \textbf{(C)} added-size distribution, \textbf{(D)} birth-size distribution; the division-time distribution $\Gamma(\tau_d)$ for the Timer model is taken to be Gaussian with a mean $\langle \tau_d \rangle =1$ hour and a standard deviation $\sigma_{\tau_d} = 0.1$ hour. The Sizer model's division-size distribution is Gaussian with a mean $\langle s_d \rangle =2$ $\mu m$ and a standard deviation $\sigma_{s_d} = 0.1$ $\mu m$. The Adder model's division-added-size distribution is Gaussian with a mean $\langle \Delta_d \rangle =1$ $\mu m$ and a standard deviation $\sigma_{\Delta_d} = 0.1$ $\mu m$. The growth rate for the case of non-stochastic growth is taken to be $1 \;\mu$m per hour. For the case of stochastic growth rate (in time), the mean growth rate is taken to be $1 \; \mu$m per hour, and the standard deviation in growth rate is taken to be $0.1 \; \mu$m per hour. \textbf{(E-M)} correlations between various single-cell-level quantities: \textbf{(E, F, G)} correlations between division-time and birth-size for Timer, Sizer, and Adder, respectively, \textbf{(H, I, J)} correlations between division-size and birth-size for Timer, Sizer, and Adder, \textbf{(K, L, M)} correlations between division-added-size and birth-size for Timer, Sizer, and Adder. The slope of the best-fit line is indicated by S, along with the Pearson Correlation Coefficient indicated by P.C. The parameters of the simulations are the same as mentioned for the plots (A-D). \textbf{(N-U)} Mean and standard deviations of various single-cell-level quantities plotted against each other for different bacterial species grown in different media:  \textbf{(N)} mean division-size versus mean birth-size, \textbf{(O)} standard deviation in division-size versus standard deviation in birth-size, \textbf{(P)} mean birth-size versus mean division-added-size, \textbf{(Q)} standard deviation in birth-size versus standard deviation in division-added-size, \textbf{(R)} mean division-size versus mean division-added-size, \textbf{(S)} standard deviation in division-size versus standard deviation in division-added-size, \textbf{(T)} mean division-time versus $\sigma_{s_d} / 2 \langle \alpha \rangle $, \textbf{(U)} coefficient of variation of $\tau_d$ versus coefficient of variation of $s_d$. The experimental data have been taken from the previously published studies \cite{santiDhar2013, chungKarAmir2024, nobsMaerkl2014}.}
    \label{fig: linearGrowth}
\end{figure*}
As discussed earlier, although most of the bacterial species grow exponentially at the single-cell level, linear growth is also observed for some species. Therefore, we extend our model for the case of linear growth and obtain various single-cell and population-level distributions (Appendix \ref{sec: SlinearGrowth}). We show that, similar to the case of exponential growth, the population-level distributions are indistinguishable across different cell-division models here as well (Fig. \ref{fig: linearGrowth} (A-D)). This is, again, due to the similarity of the single-cell-level distributions across different models with properly chosen free parameters. Moreover, here too, this indistinguishability is maintained when the linear biomass growth rate $\alpha$ is taken to be stochastic over time. The close alignment between the simulation data (markers) and analytical predictions (lines) in Fig. \ref{fig: linearGrowth} again underscores the robustness of our findings.

One of the interesting aspects of linear growth, also pointed out by \cite{vuaridelDhar2020}, is that the Timer and Adder models become functionally equivalent, which is also visible from the correlation patterns for the Timer and Adder models (Fig. \ref{fig: linearGrowth} (E-M)). Consider a culture where cells divide according to the Timer model, where each cell undergoes division after a fixed division-time $\tau_d$. Under linear growth with a fixed growth rate $\alpha$, the size of a cell grows as $s=s_b + \alpha a$. Therefore, the added-size from birth to division $\Delta_d = s_d - s_b  = \alpha\tau_d$ remains constant for all cells because $\tau_d$ is a constant. This implies that, despite being governed by a time-based division rule, each cell adds a fixed amount of biomass before dividing, which is the definition of the Adder model. This equivalence between Timer and Adder implies that the Timer mechanism will also give cell size homeostasis, and all of the population-level distributions obtained for the Timer model are identical to the ones derived for the Adder model. Additionally, one can write for the added-size at any time under linear biomass growth as $\Delta = s-s_b= \alpha a$. Since $\alpha$ is a fixed number in a given medium, the probability distribution of added-size $\Psi(\Delta)$ is just a scaled version of the probability distribution of age $\Theta(a)$ for all the models (as visible from Fig. \ref{fig: linearGrowth} (A and C)).

The statistical relationships between birth-size $s_b$, division-size $s_d$, and division-added-size $\Delta_d$ under linear biomass growth are identical to the ones obtained for exponential growth (Eq. \ref{eq: mainMeanSingleCell}, \ref{eq: mainStdAdder}, \ref{eq: mainStdSizer}, and Fig. \ref{fig: linearGrowth} (N-S)) because these quantities have no dependence on the type of single-cell growth. However, the statistical relationships involving $\tau_d$ are different from the ones that are derived for the case of exponential growth. Since $\tau_d = \Delta_d / \alpha$ for linear growth, one can write:
\begin{equation} \label{eq: mainMeanTau_dRawLinear}
    \langle \tau_d \rangle = \langle \Delta_d \rangle /  \alpha
\end{equation}
and
\begin{equation} \label{eq: mainStdTau_dRawLinear}
    \sigma_{\tau_d} = \sigma_{\Delta_d}/\alpha
\end{equation}
Therefore, using Eq. \ref{eq: mainMeanSingleCell} and \ref{eq: mainMeanTau_dRawLinear}, one can write:
\begin{equation} \label{eq: mainMeanTau_dLinear}
    \langle \tau_d \rangle = \langle \Delta_d \rangle /  \alpha= \langle s_b \rangle /  \alpha = \langle s_d \rangle /(2 \alpha)
\end{equation}
which is true irrespective of the Sizer or the Adder model. And similarly, using Eq. \ref{eq: mainStdAdder}, \ref{eq: mainMeanTau_dRawLinear}, and \ref{eq: mainStdTau_dRawLinear}, one can write for the Adder model:
\begin{equation} \label{eq: mainCOVadderLinear}
    \frac{\sigma_{\tau_d}}{\langle \tau_d \rangle} \;=\; \sqrt{3} \; \frac{\sigma_{s_d}}{\langle s_d \rangle} \;=\; \sqrt{3} \; \frac{\sigma_{s_b}}{\langle s_b \rangle} \;=\; \frac{\sigma_{\Delta_d}}{\langle \Delta_d \rangle}
\end{equation}
Whereas for the Sizer model, one can show using Eq. \ref{eq: mainStdSizer}, \ref{eq: mainMeanTau_dRawLinear}, and \ref{eq: mainStdTau_dRawLinear}:
\begin{equation} \label{eq: mainCOVsizerLinear}
    \frac{\sigma_{\tau_d}}{\langle \tau_d \rangle} \;=\; \sqrt{5} \; \frac{\sigma_{s_d}}{\langle s_d \rangle} \;=\; \sqrt{5} \; \frac{\sigma_{s_b}}{\langle s_b \rangle} \;=\; \frac{\sigma_{\Delta_d}}{\langle \Delta_d \rangle}
\end{equation}

We see that, analogous to the case of exponential growth, the mean values of the single-cell-level quantities are related to each other in a similar manner regardless of the Sizer or Adder model (Eq. \ref{eq: mainMeanSingleCell} and \ref{eq: mainMeanTau_dLinear}), but most of the relationships between the standard deviations of these quantities are different for the two models (Eq. \ref{eq: mainStdAdder}, \ref{eq: mainStdSizer}, \ref{eq: mainCOVadderLinear}, \ref{eq: mainCOVsizerLinear}), which can be used to differentiate between the Sizer and Adder models for linear growth as well. This is illustrated in Fig. \ref{fig: linearGrowth} (N-U), where the mean, standard deviation, and coefficient of variation for various single-cell-level quantities are plotted against each other (the experimental data is taken from previous studies \cite{santiDhar2013, chungKarAmir2024, nobsMaerkl2014}). As in the case of exponential growth, here too, some of the experimental data points (markers) align closely with the predictions of the model (lines), but a significant deviation is observed in other plots. Although such discrepancies might reflect experimental errors or measurement inaccuracies, they may also indicate the presence of additional factors not incorporated into the current models.

The results for linear growth are robust to temporal stochasticity in the growth rate (Appendix \ref{subsec: SlinearStochGrowth}). Similar to the case of exponential growth, the temporal fluctuations in the growth rate are averaged out and one can just replace $\alpha$ with its time-averaged value $\overline{\alpha}$ in the equations for the case with fixed growth rate. However, for the case of growth rate being a quantity that is stochastic over population, similar to the case of exponential growth, only $\Theta(a)$ and $\Gamma(\tau_d)$ have different analytical forms for the Sizer and Adder models, while other distributions remain invariant (Appendix \ref{subsec: SlinearStochGrowth}). While $\Gamma(\tau_d)$ distribution becomes skewed in comparison to the case of a fixed growth rate, both distributions become flatter, reflecting higher stochasticity in the system (see Fig. \ref{fig: SstochAlphaLinearAgeTauD}). Furthermore, our analysis reveals that, for the Timer model, while the analytical forms of $\Gamma(\tau_d)$ and $\Theta(a)$ remain unchanged, the size-related distributions are different from the case of fixed growth rate (Appendix \ref{subsec: SlinearStochGrowth}). Here too, the population-level distributions continue to be almost indistinguishable across the three models (see Fig. \ref{fig: SstochAlphaLinear}).

\subsection{Comparing linear and exponential growths}
\begin{figure*}[h!]
    \centering
    \includegraphics[width=\linewidth]{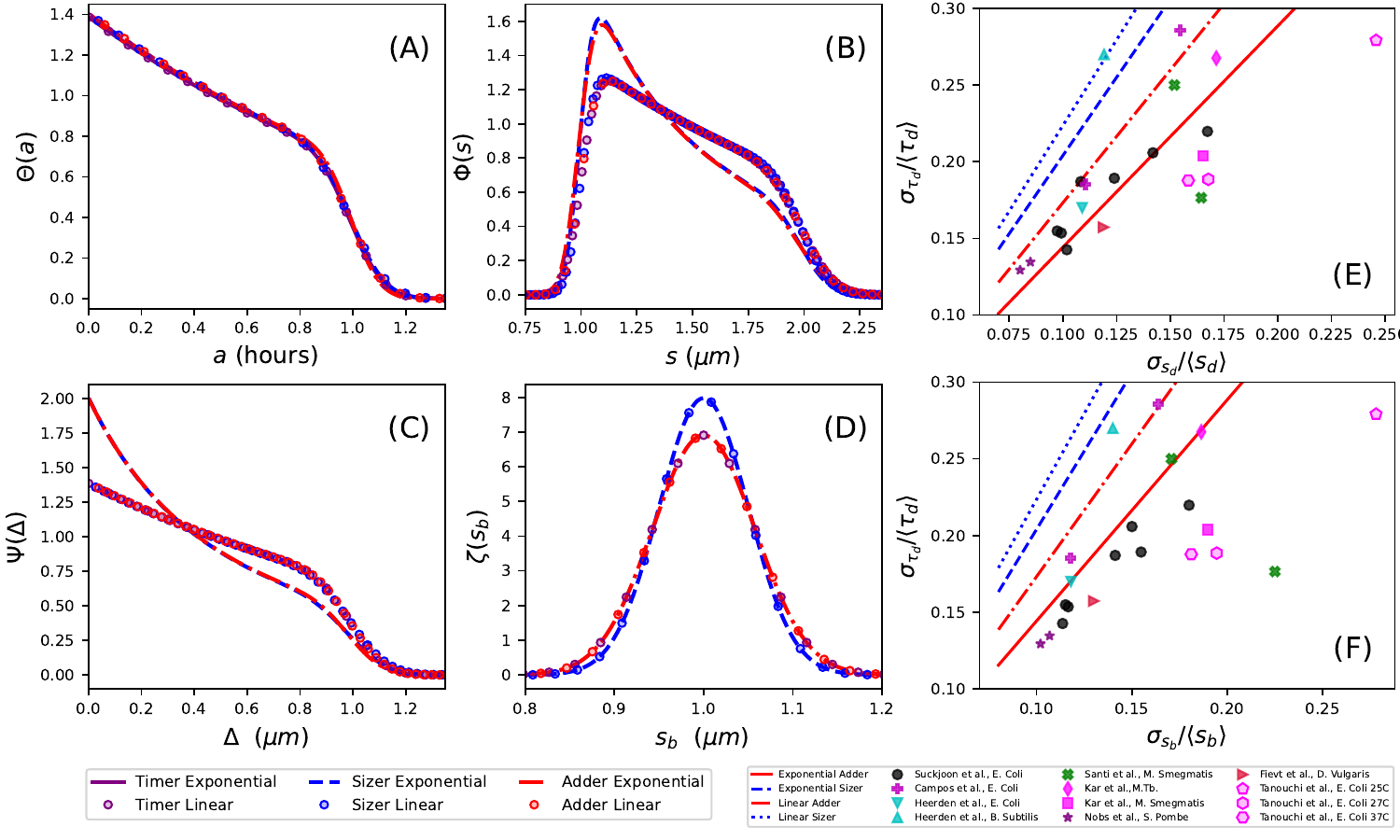}
    \caption{Comparison of various analytical population-level distributions under linear and exponential single cell growth: \textbf{(A)} age distribution, \textbf{(B)} size distribution,  \textbf{(C)} added-size distribution, \textbf{(D)} birth-size distribution, with the parameters of models as mentioned in the captions for the figures \ref{fig: allModelsExponential} and \ref{fig: linearGrowth}; \textbf{(E)} Coefficient of variation of division-time versus that of division-size, \textbf{(F)} Coefficient of variation of division-time versus that of birth-size. The lines in (E, F) indicate the theoretical predictions, and the markers indicate the data obtained from previous experimental studies \cite{taheriJun2015, campos2014, heerdenKempe2017, fievetDucret2015, tanouchi2017, santiDhar2013, chungKarAmir2024, nobsMaerkl2014}.}
    \label{fig: linearVsExponential}
\end{figure*}
In single-cell growth experiments, it remains challenging to determine whether the growth of individual cells is linear or exponential \cite{junReview2018}. While various methods have been traditionally used to differentiate between these two types of growth from single-cell data, such as direct curve fitting, log-transformed plots, growth rate versus size plots, and binning methods, which are discussed and criticized by Kar et al. \cite{karAmir2021}, we observed that one can also use the statistical relationships between single-cell-level quantities to distinguish between linear and exponential growth modes at single-cell-level, along with the correlation patterns. Specifically, the coefficients of variation in $\tau_d$ and $s_d$ can be used to identify the type of single-cell growth using equations \ref{eq: mainCOVadderExpo}, \ref{eq: mainCOVsizerExpo}, \ref{eq: mainCOVadderLinear}, \ref{eq: mainCOVsizerLinear}. Similarly, the relationship between the coefficients of variation of $\tau_d$ and $s_b$ can also be exploited to distinguish between these two types of growth. In Fig. \ref{fig: linearVsExponential} (E and F), the relationships between the coefficient of variation in $\tau_d$ and $s_d$, and $\tau_d$ and $s_b$ are plotted (lines) for the Sizer and Adder models under both linear and exponential growth, along with the single-cell experimental data (markers) \cite{taheriJun2015, campos2014, heerdenKempe2017, fievetDucret2015, tanouchi2017, santiDhar2013, chungKarAmir2024, nobsMaerkl2014}. Here too, we see that some of the experimental data points deviate significantly from the predictions of the models. Again, this variation may arise from experimental uncertainties or imprecision, or it may suggest the influence of factors beyond the scope of the present models.

In addition to the statistical relationships between single-cell-level quantities, correlation patterns can also be used to distinguish between the types of growth themselves. The simultaneous absence of correlation between $\tau_d$ and $s_b$, and between $\Delta_d$ and $s_b$ suggests linear mode of growth, either following the Timer or Adder principle. Note that one can not distinguish between the linear and exponential modes of growth for the Sizer model, based only on the correlations.

Interestingly, analogous to the statistical relationships between the single-cell-level quantities, some population-level distributions also differ between the two growth types. In Fig. \ref{fig: linearVsExponential} (A-D), we show a comparison between the various distributions obtained for the two growth types: exponential and linear, with a non-stochastic growth rate. $\zeta(s_b)$ remains the same for both linear and exponential growth, both under the Sizer and Adder division rule. This is because $\zeta(s_b)$ only depends upon the principal distribution and not the type of growth. $\Phi(s)$ for the Sizer model under linear growth, as obtained by \cite{koch1966}, differs from the one derived for the Sizer model under exponential growth, which can also be deduced from the Collins-Richmond method \cite{collinsRichmond1962}. As $\Phi(s)$ is indistinguishable across the division models, it will also be different between linear and exponential growth regimes for the Adder model. Further, since $\zeta(s_b)$ is identical for linear and exponential growth, and $\Phi(s)$ is different for the two types of growth, $\Psi(\Delta)$ is also different between the two growth modes. However, interestingly, $\Theta(a)$ for linear growth remains indistinguishable from that obtained for exponential growth, for properly chosen parameters. This is so because for the Timer model, where $\Gamma(\tau_d)$ is the principal distribution,  the derivation of $\Theta(a)$ involves only the exponential growth of cell numbers, which is independent of the individual cell-growth mode. In comparison, for the Sizer and Adder models under linear growth, the corresponding single-cell-level distribution, $\Gamma(\tau_d)$, has a similar form to the one for the exponential growth, with similar standard deviations (as given by Eq. \ref{eq: mainCOVadderExpo}, \ref{eq: mainCOVsizerExpo}, \ref{eq: mainCOVadderLinear}, \ref{eq: mainCOVsizerLinear}) which makes $\Theta(a)$ indistinguishable. Note that the same conclusion can be reached via the previous argument as well --- the indistinguishability of $\Theta(a)$ across cell-division models imply that they will be the same across linear and exponential growth for Adder and Sizer models too.

\section{Discussion}
Various cell-division strategies, like the Timer, Sizer, and Adder, have different rules regarding cell division. This study explores the observable differences among various division strategies and helps differentiate between them. It aims to elucidate the implications of various division strategies on the probability distributions of various cellular quantities and the statistical relationships related to single-cell observables. It shows that various division strategies, such as the Timer, Sizer, and Adder, have primary differences in the correlation patterns and the statistical relationships between single-cell-level quantities. However, the differences are not evident in the probability distributions of population-level quantities, which are indistinguishable across different models. Although the result is quite unexpected, it can be attributed to the fact that all single-cell distributions, obtained through the transformation of principal distributions, have similar forms across different division models, with slight differences in standard deviations that do not significantly affect the form of the probability distributions.

Although the model we present here has only two free parameters: growth rate and the principal distribution, some instances are not fully understood. One of these is the observed difference between the two different formulae for size distribution, one obtained through Powell's approach and the other through Shaechter's approach (Appendix \ref{sec: Ssizer}). For the Sizer model, Schaechter's formula \cite{kochSchac1962} gives the correct size distribution. However, when used in probability transformations, it does not give the correct age and added-size distributions. In contrast, while the size distribution obtained using Powell's approach \cite{powell1956} is not of the same form as Schaechter's, it gives the right population-level distributions when used for transformations.

The difference between experimental data and our analytical estimates is also a point of concern. While this may be attributed to experimental imprecision, a possible alternative theoretical explanation can also not be ruled out. Indeed, some studies have suggested mixed models for cell division and growth, such as bilinear growth for Escherichia coli \cite{donachie1976, kubitschek1981}, linear growth until the cell reaches a critical size, followed by an exponential growth for a constant time for Bacillus subtilis \cite{nordholtHeerden2020}, and a mixed size and age-based division strategy for Escherichia coli \cite{osellaNugent2014}. The present analysis could be extended further to handle such scenarios, which remains an open avenue for investigation.

Further, it may also be interesting to include a third kind of stochasticity in our model, called `partitioning stochasticity' \cite{pandeyJain2020}, which reflects the randomness in cell partitioning during cell division, where one of the cells can get slightly more or less biomass and resources than the other one. Moreover, some bacterial species, such as Caulobacter crescentus, divide asymmetrically, where the two distinct types of cells are created after the division: stalked cells (bigger) and swarmer cells (smaller) \cite{campos2014}. Our framework can be adapted to examine these types of organisms as well.

In recent years, many studies have focused on the intrinsic molecular details of cell-division models \cite{siJun2019, pandeyJain2020, serbanescuBanerjee2020, nietoGarcia2024, nietoGarcia2024B}. Our study here adopts a coarse-grained approach and does not explicitly incorporate these molecular details. This makes our results robust to the details of molecular implementations of these growth and division strategies.

\appendix
\renewcommand{\thefigure}{\thesection\arabic{figure}}
\makeatletter
\@addtoreset{figure}{section}
\makeatother

\section{Birth-size distributions} \label{sec: SbirthSize}

Birth-size distributions are critical in determining other distributions using probability transformations of random variables. One can obtain $\zeta(s_b)$ for the Sizer and Adder models from the principal distribution. However, given that the individual cells grow exponentially, it can not be obtained for the Timer model due to the lack of cell-size homeostasis, as mentioned in the main text. In this appendix, we derive the analytical forms of $\zeta(s_b)$ for the Sizer and Adder models. But we first start with why the Timer model does not give cell-size homeostasis.

The birth-size for a daughter cell after a symmetric division is given as $s_{b_d} = s_b \exp{(\alpha \tau_d)}/2$, where $s_b$, $\alpha$, and $\tau_d$ are the birth-size, growth rate, and the division-time of the parent cell. The growth rate $\alpha$ and division-time $\tau_d$ are not correlated for the Timer model, and one can choose the desired values for both of them independently (the same is not true for Sizer and Adder models, where the division-time is directly correlated with growth rate). If $\exp{(\alpha \tau_d)} \neq 2$, the birth-size for the daughter cell will be either smaller or bigger than that of the parent cell. And therefore, the average birth-size will not remain constant in time. But,  with properly chosen values of $\alpha$ and $\tau_d$ as $\tau_d = \ln{(2)/\alpha}$, along with no stochasticity in these parameters (also no stochasticity in halving), two daughter cells are created with the size equal to that of the parent cell after the cell-division, and the average birth-size remains constant for the Timer model. However, it will be shown below that the stochasticity in either $\tau_d$ or $\alpha$ further destroys the cell-size homeostasis for the Timer model.

Let us consider a culture following the Timer model for cell-division with a fixed growth rate $\alpha$ and having a Gaussian form for the division-time distribution $\Gamma(\tau_d)$ (mean $\langle \tau_d \rangle$ and standard deviation $\sigma_{\tau_d}$). The ratio of the birth-size of the daughter cell to that of the parent cell is given by $\beta = \exp{(\alpha \tau_d)}/2$.  Hence, the distribution of $\beta$, $\Upsilon(\beta)$ is given by:
\begin{equation}
    \Upsilon(\beta) = \frac{1}{\alpha \beta\sqrt{2 \pi \sigma_{\tau_d}^2}} \exp{\left(  \frac{-(\frac{\ln{(2\beta)}}{\alpha} - \langle \tau_d \rangle)^2}{2\sigma_{\tau_d}^2}\right)}
\end{equation}
Hence, the average value of $\beta$ will be given as:
\begin{equation}
    \langle \beta \rangle \;=\; \int_{0}^{\infty} d\beta \; \beta \; \Upsilon(\beta) \;=\; \frac{1}{\alpha \sqrt{2 \pi \sigma_{\tau_d}^2}}\int_{0}^{\infty} d\beta \; \exp{\left(  \frac{-(\frac{\ln{(2\beta)}}{\alpha} - \langle \tau_d \rangle)^2}{2\sigma_{\tau_d}^2}\right)}
\end{equation}
After doing the manipulation of $y=\ln{(2 \beta)}$, one can get:
\begin{equation}
    \langle \beta \rangle \;=\; \frac{1}{2\alpha \sqrt{2 \pi \sigma_{\tau_d}^2}}\int_{0}^{\infty} d y \; \exp{\left(y -  \frac{(y-\alpha \langle \tau_d \rangle)^2}{2 \alpha^2 \sigma_{\tau_d}^2}\right)}
\end{equation}
Using the method of completing the square, the integral above can be evaluated, and it can be further shown that
\begin{equation}
    \langle \beta \rangle \;=\; \frac{e^{\alpha \langle \tau_d \rangle}}{2} \exp{\left(\frac{\alpha^2 \sigma_{\tau_d}^2}{2}\right)} .
\end{equation}
First of all, the equation above shows that the values of $\alpha$ and $\langle \tau_d \rangle$ should be fine tuned to make $\langle \beta \rangle =1$. Otherwise, the average value of the ratio of birth-sizes of daughter cell and parent cell will be bigger or smaller than $1$, destroying cell-size homeostasis. Further, since $\Gamma(\tau_d)$ is Gaussian, it is equally likely for a cell to divide in less time than $\langle \tau_d \rangle$ than it is to divide in more time than $\langle \tau_d \rangle$. However, it is implied from the above equation that even for properly chosen value of $\alpha$ and $\langle \tau_d \rangle$, the average value of $\beta$ is greater than $1.0$ for any value of stochasticity in $\tau_d$. This means that although a daughter cell is equally likely to be born earlier or later than its parent cell, its birth size will be greater than that of its parent on average. This is so because the cells with bigger division-times will accumulate disproportionately more biomass in comparison to the cells with smaller division-times (due to the growth factor $\exp(\alpha \tau_d)$). This results in the average birth-size in the population increasing with time, as seen in the inset for the Timer model in Fig. \ref{fig: cartoon}.

For the Sizer and Adder models, we present some arguments that are used to obtain $\zeta(s_b)$ from $\Xi(s_d)$ and $\Omega(\Delta_d)$. The first argument is to consider a random cell in a culture. The birth-size of this cell $s_b$ is exactly half the size of its parent cell at the time of division, which had divided into two after achieving a random size $s_d$ selected from a probability distribution $\Xi(s_d)$. Note that the two variables here, $s_b$ and $s_d$, are not for the same cell but for two different cells from different generations (for the same generation, $s_d$ and $s_b$ are not correlated for the Sizer model). Hence, $\zeta(s_b)$, the distribution for $s_b$, is given by:
\begin{equation} \label{eq: zetaXi}
    \zeta(s_b) \;=\; 2 \; \Xi(2s_b) 
\end{equation}
Moreover, we have $s_b = s_d /2$, from which the mean, standard deviation, and the $n$th central moments of $s_d$ and $s_b$ can be related as follows:
\begin{equation} \label{eq: sbSdMean}
    \langle s_b \rangle \;=\; \frac{1}{2} \langle s_d \rangle
\end{equation}
\begin{equation} \label{eq: sbSdSTD}
    \sigma_{s_b} = \frac{1}{2} \sigma_{s_d} 
\end{equation}
\begin{equation} \label{eq: nthCentralMomentSbSd}
    \mu_n(s_b) = \frac{\mu_n(s_d)}{2^n}
\end{equation}
where $\langle s_d \rangle$ and $\langle s_d \rangle$ represent the mean values of $s_b$ and $s_d$, and $\sigma$ and $\mu_n$ represent the standard deviation and $n$th central moment, respectively. The above equations will be used multiple times throughout the text. Additionally, if $\Xi(s_d)$ is known beforehand, such as in the case of the Sizer model, where it is a principal distribution, $\zeta(s_b)$ can be found from $\Xi(s_d)$ using Eq. \ref{eq: zetaXi}. Otherwise, if $\Xi(s_d)$ is a Gaussian distribution with mean $\langle s_d \rangle$ and standard deviation $\sigma_{s_d}$, then $\zeta(s_b)$ is a Gaussian distribution with mean $\langle s_b \rangle = \langle s_d \rangle /2$ and standard deviation $\sigma_{s_b} = \sigma_{s_d}/2$.

To derive the relationship between $\zeta(s_b)$ and $\Omega(\Delta_d)$, consider a random cell in the culture with birth-size $s_b$ at a given time. The cell was created at some point in time in the past by the cell division of a parent cell with a birth-size $s_1$ after it had added a certain size $\Delta_{d_1}$. So, $\Delta_{d_1}$ is the division-added-size for the parent cell. Hence, one can write $s_b = s_{1}/2 + \Delta_{d_1}/2$. Further, tracking the parent cell in the past, it was created from the grandparent cell with a birth-size $s_2$ and a division-added-size $\Delta_{d_2}$. Thus $s_1 = s_{2}/2 + \Delta_{d_2}/2$, and $s_b = s_2/4 + \Delta_{d_2}/4 + \Delta_{d_1}/2$. Moreover, the grandparent cell can be further tracked in the past, and this process goes on, and one can write:
\begin{equation} \label{eq: sbRaw}
    s_b = \lim_{ n\rightarrow\infty} \left(\frac{s_{n}}{2^{n}} + \sum_{k=1}^{n}\frac{\Delta_{d_k}}{2^k}\right)
\end{equation}
The term $\frac{s_{n}}{2^{n}}$ can be neglected in the large $n$ limit and $s_b$ is just a linear sum of identically distributed random variables $\Delta_{d_k}$ with probability distribution $\Omega(\Delta_d)$. Hence, $s_b$ can be written as:
\begin{equation} \label{eq: sbDeltaDSeries}
    s_b \;=\; \lim_{ n\rightarrow\infty} \sum_{k=1}^{n}\frac{\Delta_{d_k}}{2^{k}} \;=\; \sum_{k=1}^{n} a_k x_k
\end{equation}
where $a_k = 2^{-k}$ and $x_k = \Delta_{d_k}$. Now, taking average of the above equation, the average birth-size $\langle s_b \rangle$ can be written in terms of average division-added-size $\langle x \rangle$ as:
\begin{equation} \label{eq: sbDeltaDmean}
    \langle s_b \rangle \;=\;  \sum_{k=1}^{n} a_k \langle x_k \rangle \;=\; \langle x \rangle \sum_{k=1}^{n} a_k \;=\; \langle x \rangle \;=\;\langle \Delta_d \rangle
\end{equation}
where $\langle x_k \rangle = \langle x \rangle = \langle \Delta_d \rangle$ because $x_k$ are identically distributed random variables. The sum of the series converges to 1 in the large n limit. Additionally, other higher-order moments, such as standard deviation, can be found by similar arguments. For example, using the above two equations, the standard deviation $\sigma_{s_b}$ for birth-size distribution can be written as:
\begin{equation}
    \sigma_{s_b}^2 = \langle (s_b - \langle s_b \rangle )^2 \rangle = \sum_{k=0}^n \sum_{j=0}^n \langle a_j a_k (x_j - \langle x \rangle)(x_k - \langle x \rangle) \rangle
\end{equation}
which can be simplified as:
\begin{equation}\label{eq: sbDeltaDstdRaw}
    \sigma_{s_b}^2 = \sum_{k=1}^n \sum_{j=1}^n a_j a_k ( \langle x_j x_k \rangle - \langle x \rangle^2 ) .
\end{equation}
$\langle x_j x_k \rangle \;-\; \langle x \rangle^2$ represents the correlation between the parent and the $\left| i-j\right|$th generation daughter cells. So far, the two arguments presented here to relate $\zeta(s_b)$ and other single-cell-level distributions are not specific to any division or growth model. Therefore, the results derived so far are valid for both the Sizer and Adder models under exponential or linear biomass growth.

For the case of the Sizer model, to calculate $\sigma_{s_b}^2$, one has to consider the correlations between $\Delta_d$'s of the parent cell and ancestor/descendant cells (see Appendix \ref{sec: Scorrelations} for correlations). After putting Eq. \ref{eq: corrDeltaDdeltaDmotherDaughterSizer} and \ref{eq: corrDeltaDdeltaDSizer} in Eq. \ref{eq: sbDeltaDstdRaw}, we get:
\begin{equation}
\begin{split}
    \sigma_{s_b}^2 &\;=\; \sigma_{\Delta_d}^2 \sum_{k=1}^n a_k^2 \;+\; 2 \sum_{k=1}^n a_k a_{k+1} (-\sigma_{s_d}^2/2)\\
    & = \frac{\sigma_{\Delta_d}^2}{3} - \frac{\sigma_{s_d}^2}{6}
\end{split}
\end{equation}
Using the relation $\sigma_{s_d} = 2 \sigma_{s_b}$ (Eq. \ref{eq: sbSdSTD}), we finally get:
\begin{equation} \label{eq: sbDeltaDstdSizer}
    \sigma_{s_b}^2 = \frac{\sigma_{\Delta_d}^2}{5}
\end{equation}
Now, for the Adder model, the division-added-sizes for the parent and daughter cells are uncorrelated. Therefore, $x_j$ and $x_k$ are independent of each other. Hence, $\langle x_j x_k \rangle - \langle x \rangle^2  =  \delta_{jk} \sigma_{x}^2$ (Eq. \ref{eq: corrDeltaDdeltaDadder}). On further simplification, it gives:
\begin{equation} \label{eq: sbDeltaDstdAdder}
    \sigma_{s_b}^2 = \sum_{k=1}^n a_k^2 (\langle x^2 \rangle-\langle x \rangle^2) = \frac{\sigma_{x}^2}{3} = \frac{\sigma_{\Delta_d}^2}{3} 
\end{equation}
One can further calculate the other higher moments for $\zeta(s_b)$ in terms of those of $\Omega(\Delta_d)$ for the Adder model using the method discussed above. However, the calculations are lengthy using this approach. Therefore, one can use the method of cumulants. By using the properties of cumulants and moment generating functions, one can show that if we are given with the linear combination of identically distributed independent random variables, such as in Eq. \ref{eq: sbDeltaDSeries}
, the cumulants for the distribution of $s_b$ can be written in terms of those of $\Delta_d$ as:
\begin{equation}
    \kappa_r(s_b) \;=\; \kappa_r(\Delta_d) \sum_{n=0}^{\infty} a_n^{r} \;=\; \frac{\kappa_r(\Delta_d)}{2^r -1}
\end{equation}
Note that we have made an assumption of independent random variables here, which is valid only for the Adder model. The central moments for the distribution of $s_b$ can be further found in terms of those of $\Delta_d$ using the standard relationships between the central moments and cumulants. For example, the central moments for $\zeta(s_b)$ up to sixth order are given as:
\begin{equation}
    \mu_{3}(s_b) \;=\; \frac{\mu_{3}(\Delta_d)}{7} 
\end{equation}
\begin{equation} \label{eq: fourthCentralMomentsSbDeltaD}
    \mu_{4}(s_b)  \;=\; \frac{\mu_{4}(\Delta_d)}{15} + \frac{2}{15} \sigma_{\Delta_d}^4
\end{equation}
\begin{equation}
    \mu_5(s_b) \;=\; \frac{\mu_5(\Delta_d)}{31} + \frac{100}{651} \mu_3(\Delta_d)\sigma_{\Delta_d}^2
\end{equation}
\begin{equation}
    \mu_6(s_b) \;=\; \frac{\mu_6(\Delta_d)}{63} + \frac{2}{21} \mu_4(\Delta_d)\sigma_{\Delta_d}^2 + \frac{20}{441} \mu_3(\Delta_d)^2 + \frac{2}{63} \sigma_{\Delta_d}^6
\end{equation}
where $\mu_{n}(s_b)$ and $\mu_{n}(\Delta_d)$ are $n$th central moment for $\zeta(s_b)$ and $\Omega(\Delta_d)$ respectively. Using the above equations, the skewness ($\gamma$) and kurtosis ($K$) are related for the two quantities as:
\begin{equation}
    \gamma_{s_b} = \frac{3\sqrt{3}}{7} \gamma_{\Delta_d}
\end{equation}
\begin{equation}
    K_{s_b} = \frac{3}{5} ( K_{\Delta_d} +2 )
\end{equation}
Other higher order moments can also be calculated, but it is apparent from the equations including central moments and dimensional analysis that $n$th order central moment for $\zeta(s_b)$ can be written as a linear combination of the product of powers of the central moments for $\Omega(\Delta_d)$, such that the dimensionality of each term is size raised to the power $n$. This can be written as:
\begin{equation}
    \mu_n(s_b) \;=\; \sum_{i} c_i \prod_{j} \mu_j(\Delta_d)^{\alpha_j}  
\end{equation}
where $j \geq 2$ and the product contains the terms for which $\sum_{j} j\alpha_j = n$. In the equation above, the summation is over all such possible combinations of these products. The factor $c_i$ can be determined from the method of cumulants as discussed above. Dividing the above equation by $\langle s_b \rangle^n$, we get:
\begin{equation}
    \frac{\mu_n(s_b)}{\langle s_b \rangle^n} \;=\; \sum_{i} c_i \prod_{j=2}^{n} \left( \frac{\mu_j(\Delta_d)}{\langle \Delta_d \rangle^j} \right)^{\alpha_j}
\end{equation}

For a random variable $x$, if $\mu_n(x)/\langle x \rangle^n \approx 0$ for all $n \in \{2,3,4,... \infty \}$, we say that its probability distribution is sharply spiked. Suppose we are given that $\mu_m(\Delta_d)/\langle \Delta_d \rangle^m \approx 0$ for all $m \in \{ 2,3,4,...,\infty \}$, then the equation above implies $\mu_n(s_b)/\langle s_b \rangle^n \approx 0$ for all $n \in \{ 2,3,4,...,\infty \}$. Therefore, we have proved that for the adder model, if $\Omega(\Delta_d)$ is sharply spiked, it implies that $\zeta(s_b)$ will also be sharply spiked. This result will be useful to derive the mean and standard deviation of $\Gamma(\tau_d)$ in Appendix \ref{sec: SstatRelationships}. 

If $\Omega(\Delta_d)$ is known beforehand, such as in the case of the Adder model where it is the principal distribution, the analytical form of $\zeta(s_b)$ can be found in terms of $\Omega(\Delta_d)$ using characteristic functions. But in comparison to the Sizer model, where simple analytical form of $\zeta(s_b)$ is known in terms of the principal distribution (Eq. \ref{eq: zetaXi}), a more complicated form of $\zeta(s_b)$ is expected here. However, since one can always find out the first four central moments for $s_b$, $\zeta(s_b)$ can also be approximated by a Pearson Type distribution. Additionally, if $\Omega(\Delta_d)$ is given to be a Gaussian distribution, then $\zeta(s_b)$ will also be a Gaussian distribution with the previously mentioned mean and standard deviation. This is so because the linear combination of independent Gaussian random variables is also a Gaussian random variable.

\section{Age distribution for the Timer model} \label{sec: Stimer}

As mentioned in the main text, the size-related distributions are not obtained for the Timer model. However, we discuss here the most conclusive derivation for age distribution which obtains $\Theta(a)$ in terms of $\Gamma(\tau_d)$ (Eq. \ref{eq: mainTimer}). This derivation was done by Powell \cite{powell1956} (also discussed by \cite{junReview2018}), whose arguments are continued in the following appendices for the Sizer and Adder models. We have $\Gamma(\tau_d)d\tau_d$ as the probability that a cell divides after reaching an age between $\tau_d$ and $\tau_d + d\tau_d$. $F_>(a) \;=\; \int_{a}^{\infty} \Gamma(\tau_d) d\tau_d$ is the probability that a cell will not divide before it reaches an age $a$. Then the survival probability is defined as:
\begin{equation}
    \eta \;=\; \frac{F_>(a+t)}{F_>(a)}
\end{equation}
which is the conditional probability that given a cell does not divide before it reaches an age $a$, the likelihood of surviving until it reaches an age $a+t$. $\Theta(a)$ is the probability distribution for the age of the cells in the culture, such that $\Theta(a) da$ represents the fraction of the cells that have an age between $a$ and $a+da$. The number of cells that reach age $a$ and do not divide until reaching an age $a+t$ is equal to $N \;\Theta(a) \; da \;\eta $, which is also equal to the number of cells with age $a+t$ after $t$ time, and equating both of them gives:
\begin{equation} \label{eqn:rawTimer}
    N \; \Theta(a) da \; \eta \;=\; N \; e^{\lambda t}\; \Theta(a+t)da 
\end{equation}
where $\lambda$ is the cell number growth rate. Considering the Taylor expansion of the functions in the above equation and neglecting higher-order terms under the assumption that $t$ is very small, we get:
\begin{equation}
    1+t \frac{F_{>}^{'}(s)}{F_{>}(s)} = ( 1+ \lambda t) \left( 1+t \frac{\Theta^{'}(a)}{\Theta(a)} \right)
\end{equation}
Comparing the coefficients of $t$ on both sides, we get:
\begin{equation}
    \frac{F_>^{'}(s)}{F_>(s)} = \lambda + \frac{\Theta^{'}(a)}{\Theta(a)}
\end{equation}
And after integration, we get :
\begin{equation} \label{eq: Timer}
    \Theta(a) = C\exp(-\lambda a) \int_a^{\infty} \Gamma(\tau_d) d \tau_d
\end{equation}
where $C$ is some constant, which can be found from the normalization condition $\int_0^{\infty} \Theta(a) da \;=\; 1$. Using Eq. \ref{eq: lambda}, which describes the relationship between cell number growth rate and mean division-time as $\lambda = \ln{(2)}/\langle \tau_d \rangle$ (see Appendix \ref{sec: SrelationshipGrowthRates}), one finally obtains Eq. $\ref{eq: mainTimer}$.

\section{Various probability distributions for the Sizer model} \label{sec: Ssizer}

The approach we have used here to get various population-level distributions for the Sizer and Adder models is quite different from the general approach of population balance models, which is taken in the recent studies \cite{giomettoAltermatt2013, osellaNugent2014, taheriJun2015, xiaGreenman2020, genthon2022}. In that approach, one solves partial differential equations to find the time evolution of the population-level distributions, such as age, size, and added-size distribution. Although it is very powerful, it is not so intuitive. Further, we are only interested in the steady-state behavior of these distributions, not the time evolution of the distributions. Hence, we use a completely different approach, which is just the extensions of the arguments proposed by \cite{powell1956, kochSchac1962}.

\begin{figure}[hbt]
    \centering
    \includegraphics[width=0.5\textwidth]{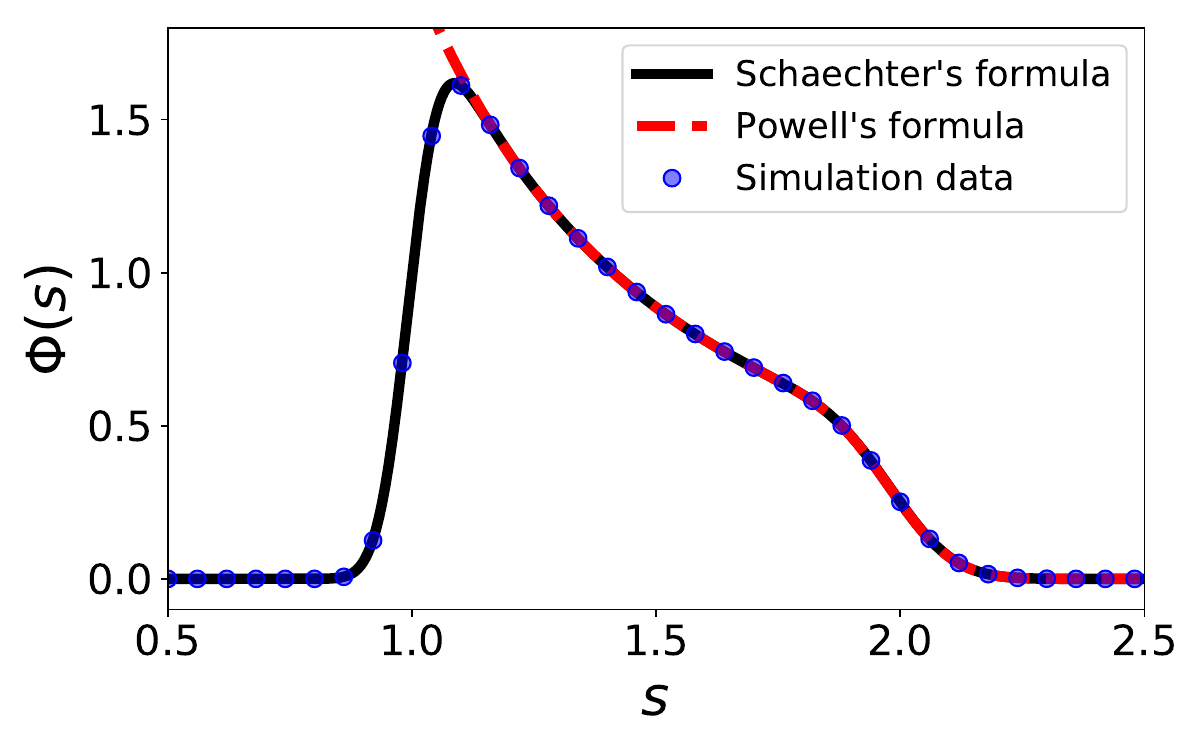}
    \caption{Analytical size distribution for the Sizer model as obtained from the two formulae (Schaechter's formula i.e. Eq. \ref{eq: mainSizer} and Powell's formula i.e. \ref{eq: sizer}) for a Gaussian division-size distribution $\Xi(s_d)$ with mean $\langle s_d \rangle$ = $2 \; \mu$m and standard deviation $\sigma_{s_d}$ = $0.1 \;\mu$m. Both of them are further compared with the simulation results.}
    \label{fig: SsizeAnalyticalSizer}
\end{figure}

For the Sizer model, we are already given the principal distribution, i.e., division-size distribution $\Xi(s_d)$, and we can find $\zeta(s_b)$ from $\Xi(s_d)$ using Eq. \ref{eq: zetaXi}, as discussed in Appendix \ref{sec: SbirthSize}. Moreover, we have two formulae for size distribution in steady-state population in terms of $\Xi(s_d)$ for the Sizer model derived by Koch and Schaechter \cite{kochSchac1962}. One of them is Eq. \ref{eq: mainSizer}, which is rewritten here for easy reference:
\begin{equation}
    \Phi(s) = \frac{A}{s^2} \int_s^{2s} \Xi(s_d) d s_d
\end{equation}
The other formula for the size distribution for the Sizer model (also derived by \cite{kochSchac1962}) can be obtained from the extension of the arguments presented by Powell \cite{powell1956}. We have $\Xi(s_d) d s_d$ as the probability that a cell divides after reaching a size between $s_d$ and $s_d + d s_d$. $F_>(s) \;=\; \int_{s}^{\infty} \Xi(s_d) d s_d$ is the probability that a cell will not divide before it reaches size $s$. Then the survival probability is defined for Sizer as :
\begin{equation}
    \eta \;=\; \frac{F_>(s+\Delta s)}{F_>(s)}
\end{equation}
which is the conditional probability that given a cell does not divide before it reaches a size $s$, the probability of it surviving until it reaches a size $s+\Delta s$. Here $\Delta$ is not added-size, but rather $\Delta s$ is an increment in size. The fraction of cells that reach size s and do not divide until reaching a size $s'=s+\Delta s$ is given as:
\begin{equation} \label{eqn:rawSizer}
    N \; \Phi(s) ds \; \eta \;=\; N \; e^{\lambda t}\; \Phi_S(s')ds' 
\end{equation}
where $\alpha$ is the size growth rate, $\lambda$ is the cell number growth rate, and $t$ is the time taken to add a size of $\Delta$. We have $\lambda = \alpha$ (from Appendix \ref{sec: SrelationshipGrowthRates}). Also, $s' = s e^{\alpha  t}$ and thus $\frac{ds'}{ds} = e^{ \lambda t } = \frac{s'}{s}$. Thus, one can write:
\begin{equation}
    e^{\lambda t} \frac{ds'}{ds} = \left(\frac{s'}{s}\right)^2 = \left ( 1 + \frac{\Delta s}{s}\right )^2
\end{equation}
Using this equation in the Eq. \ref{eqn:rawSizer}, we get:
\begin{equation}
    \frac{F_>(s+\Delta s)}{F_{>}(s)} = \left ( 1+\frac{\Delta s}{s} \right )^2 \frac{\Phi(s+\Delta s)}{\Phi(s)}
\end{equation}
Taking the Taylor expansion of the functions above and neglecting the higher-order terms:
\begin{equation}
    1+\Delta s \frac{F_{>}^{'}(s)}{F_{>}(s)} = \left( 1+2 \frac{\Delta s}{s} \right) \left( 1+\Delta s \frac{\Phi^{'}(s)}{\Phi(s)} \right)
\end{equation}
Comparing terms up to the first order of $\Delta s$:
\begin{equation}
    \frac{F_>^{'}(s)}{F_>(s)} = \frac{2}{s} + \frac{\Phi^{'}(s)}{\Phi(s)}
\end{equation}
After integration, we get:
\begin{equation} \label{eq: sizer}
    \Phi(s) = \frac{A}{s^2} \int_s^{\infty} \Xi(s_d) d s_d
\end{equation}
Let us call Eq. \ref{eq: mainSizer} as `Schaechter's formula' and Eq. \ref{eq: sizer} as `Powell's formula' because the later one is obtained using the extensions of Powell's method. The size distributions according to both of these formulae are shown in Fig. \ref{fig: SsizeAnalyticalSizer}. Out of the two formulae, Schaechter's formula gives the correct size distribution for the Sizer model for all values of size. Whereas Powell's formula gives the correct size distribution only for values of sizes bigger than the average birth-size. For the values of size smaller than or near the average birth-size, it starts gives incorrect results. Although both of these formulae explode at values of size very close to zero, in comparison to Powell's formula, Schaechter's formula does so very sharply at $s=0$ such that it not even visible in Fig. \ref{fig: SsizeAnalyticalSizer}. However, for high values of stochasticity in $s_d$, this expulsion is not so sharp and it is visible when compared with the Powell's formula.

We have obtained $\Phi(s)$ and $\zeta(s_b)$ for the Sizer model. We can also get $\Theta(a)$ and $\Psi(\Delta)$ starting from these two distributions, by the method of probability transformations of random variables (Eq. \ref{eq: probTrans}). Suppose we have to find $\Theta(a)$ from $\Phi(s)$ and $\zeta(s_b)$. We know that $s=s_b e^{\alpha a}$. Using this relation and putting it into Eq. \ref{eq: probTrans}, we get:
\begin{equation} \label{eq: ageSizer}
\begin{split}
    \Theta(a) \;&=\; \alpha e^{\alpha a} \; \int_0^{\infty} d s_b \; s_b \; \zeta(s_b) \; \Phi(s_b e^{\alpha a}) \\
    &=\;\alpha e^{-\alpha a} \; \int_0^{\infty} ds \; s \; \Phi(s) \; \zeta(s e^{-\alpha a}) 
\end{split}
\end{equation}
Both of these are equivalent forms and give the same distribution when plotted. The same treatment can be done for $\Psi(\Delta)$ using the relation $\Delta = s-s_b$. That gives us:
\begin{equation} \label{eq: addedSizeSizer}
\begin{split}
    \Psi(\Delta) \;&=\;  \int_0^{\infty} \; d s_b \; \zeta(s_b) \; \Phi(s_b+\Delta) \\
    &=\; \int_0^{\infty} \; d s \; \Phi(s) \; \zeta(s-\Delta)
\end{split}
\end{equation}
We have two analytical forms (Schaechter's formula and Powell's formula) for $\Phi(s)$ for the Sizer model. We can use both of them in the equations above (while doing probability transformations) to get $\Theta(a)$ and $\Psi(\Delta)$, and they give different results. Although Schaechter's formula gives more accurate size distribution than Powell's formula, surprisingly, when used in probability transformations, it gives incorrect age and added-size distribution. On the other hand, despite being a poorer choice for the size distribution, Powell's formula is a better choice for $\Phi(s)$ in order to get these population-level distributions via probability transformations, because it gives the correct analytical forms of these distributions, as can be verified by comparing with the simulation data. The reason for this is not well understood.

\begin{figure}[hbt]
    \centering
    \includegraphics[width=1\linewidth]{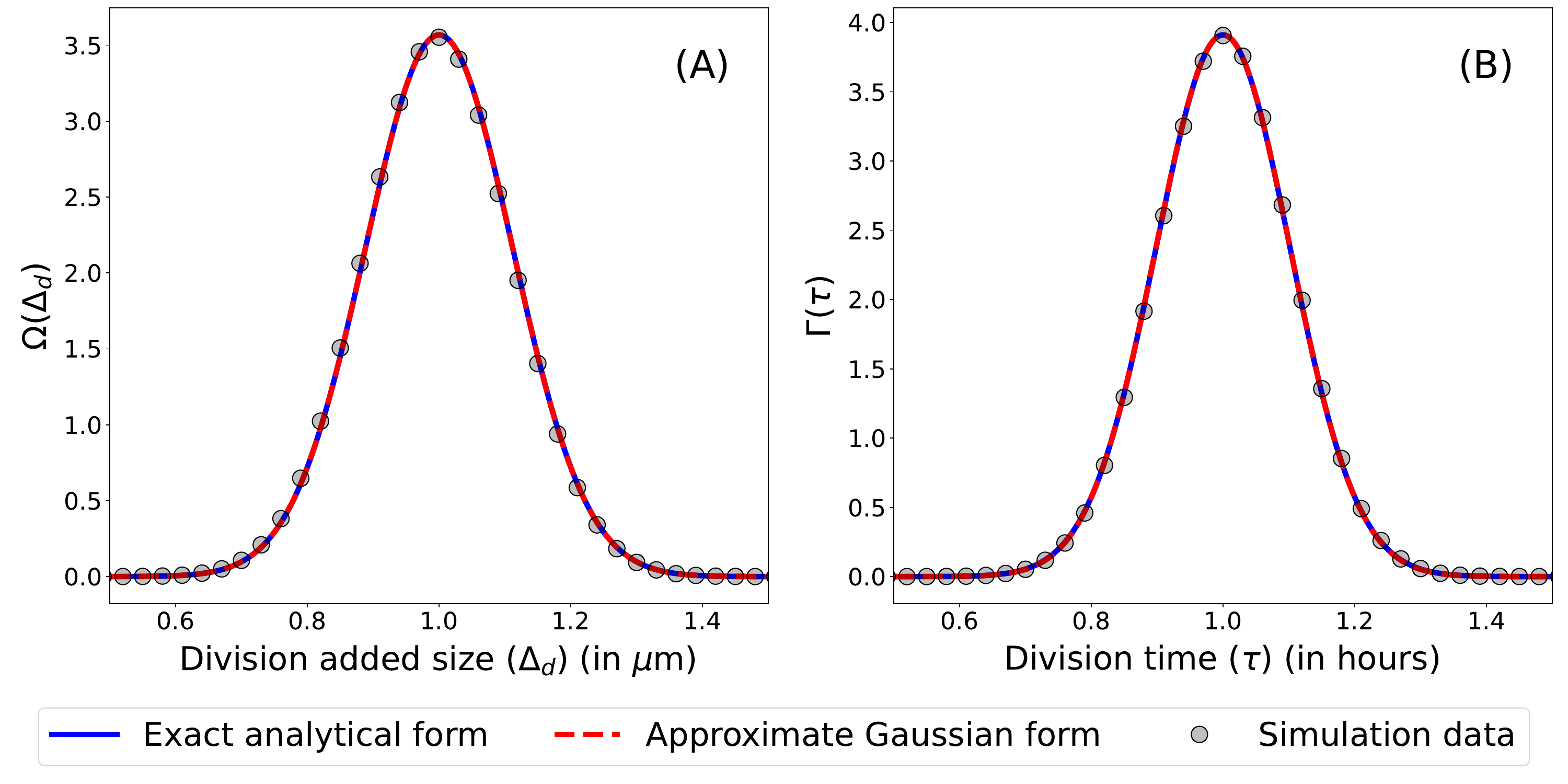}
    \caption{A comparison of the exact analytical forms of the single-cell-level distributions with their approximate Gaussian counterparts and the distributions obtained via simulations for the Sizer model: \textbf{(A)} Division-added-size distribution, \textbf{(B)} Division-time distribution. The parameters are identical to the ones mentioned in Fig. \ref{fig: allModelsExponential}}
    \label{fig: SdastauDsizer}
\end{figure}

The other single-cell-level distributions for the Sizer model, i.e., $\Gamma(\tau_d)$ and $\Omega(\Delta_d)$, can also be obtained from $\Xi(s_d)$ and $\zeta(s_b)$ using probability transformations. We already have $\Xi(s_d)$ and $\zeta(s_b)$ for the Sizer model. From the relation $\Delta_d \;=\; s_d - s_b$, using probability transformation, we have:

\begin{equation} \label{eq: deltaDsizer}
\begin{split}
    \Omega(\Delta_d) \;&=\; \int_{0}^{\infty} d s_b \; \zeta(s_b) \; \Xi(\Delta_d+s_b) \\
    &=\; \int_{0}^{\infty} d s_d \; \Xi(s_d) \; \zeta(s_d-\Delta_d)
\end{split}
\end{equation}
And from the relation $s_d \;=\; s_b e^{\alpha \tau_d}$, we have:
\begin{equation} \label{eq: tau_dSizer}
\begin{split}
    \Gamma(\tau_d) \;&=\; \alpha e^{\alpha \tau_d} \int_{0}^{\infty} d s_b \; \zeta(s_b) \; s_b \; \Xi(s_b e^{\alpha \tau_d}) \\
    &=\; \alpha e^{-\alpha \tau_d} \int_{0}^{\infty} d s_d \; \Xi(s_d) \; s_d \; \zeta(s_d e^{-\alpha \tau_d})
\end{split}
\end{equation}
If the principal distribution i.e. $\Xi(s_d)$ is taken to be a Gaussian distribution, it can be proved that all other single-cell-level distributions will also be Gaussian up to a very good approximation, with the mean and standard deviations identical to the ones predicted by Eq. \ref{eq: mainMeanSingleCell}, \ref{eq: mainStdSizer}, \ref{eq: mainMeanTau_d}, and \ref{eq: mainCOVsizerExpo}. This can be easily confirmed from the comparison between the exact analytical form and the approximate Gaussian form of $\Omega(\Delta_d)$ and $\Gamma(\tau_d)$ in Fig. \ref{fig: SdastauDsizer}. Now, let us explicitly prove that they will be approximately Gaussian.
From the properties of the error function, it can be proved that
\begin{equation}
    \int_{0}^{\infty} \exp{(- \alpha(x-\beta)^2 )} dx \;=\; \frac{\sqrt{\pi}}{2\sqrt{\alpha}} [ 1+ \erf{(\beta \sqrt{\alpha})}]
\end{equation}
And after doing a bit of algebra, using above equation, one can further show that:
\begin{equation} \label{eq: integralGaussians}
    \int_{0}^{\infty} e^{-a(x-b)^2} e^{-c(x-d)^2} dx = \frac{1}{2} \sqrt{\frac{\pi}{a+c}} \left[\erf{\left(\frac{ab+cd}{\sqrt{a+c}}\right)+1} \right] \exp{\left( - \frac{(b-d)^2}{\frac{1}{a} +\frac{1}{c}} \right)}
\end{equation}
and
\begin{equation} \label{eq: integralXgaussians}
\begin{split}
    \int_{0}^{\infty} x\;e^{-a(x-b)^2} e^{-c(x-d)^2} dx \;=\; & \exp{\left( - \frac{(b-d)^2}{\frac{1}{a} +\frac{1}{c}} \right)} \left[ \frac{1}{2(a+c)} \exp{\left( -\frac{(ab+cd)^2}{a+c} \right) } \;+\; \frac{(ab+cd)}{2(a+c)} \sqrt{\frac{\pi}{a+c}} \left(1+ \erf{\left(\frac{ab+cd}{\sqrt{a+c}}\right)} \right) \right]
\end{split}
\end{equation}

We know (from Appendix \ref{sec: SbirthSize}) that for the Sizer model, if $\Xi(s_d)$ is Gaussian, $\zeta(s_b)$ will also be Gaussian. Further, given that $\Xi(s_d)$ and $\zeta(s_b)$ are Gaussian, one can show using Eq. \ref{eq: deltaDsizer} and \ref{eq: integralGaussians}:
\begin{equation}
    \Omega(\Delta_d) = \frac{1}{2 \sqrt{2\pi (\sigma_{s_b}^2 + \sigma_{s_d}^2)}} \exp{\left[ - \frac{(\Delta_d - (\langle s_d \rangle - \langle s_b \rangle))^2}{2(\sigma_{s_b}^2 + \sigma_{s_d}^2)}\right]} \left[1+\erf{\left( \frac{\frac{\langle s_b \rangle}{2\sigma_{s_b}^2}+\frac{\langle s_d \rangle -\Delta_d}{2\sigma_{s_d}^2}}{\sqrt{\frac{1}{2 \sigma_{s_b}^2}+\frac{1}{2\sigma_{s_d}^2}}} \right)} \right]
\end{equation}
Putting in the values of $\langle s_b \rangle$ and $\sigma_{s_b}$ in terms of $\langle s_d \rangle$ and $\sigma_{s_d}$, we get:
\begin{equation}
    \Omega(\Delta_d) = \sqrt{\frac{2}{5 \pi \sigma_{s_d}^2}} \exp{\left[ - \frac{(\Delta_d - \langle s_d \rangle /2)^2}{5 \sigma_{s_d}^2 /2}\right]} \left[\frac{1+\erf{\left(\frac{3 \langle s_d \rangle-\Delta_d}{\sqrt{10}\sigma_{s_d}}\right)}}{2} \right]
\end{equation}

This shows that the probability distribution for $\Delta_d$ is Gaussian with a mean $\langle s_d \rangle/2$ and a standard deviation $\sqrt{5} \sigma_{s_d}/2$. But there is an extra factor multiplied by the Gaussian function. However, one can show that this factor is $1$ up to a very good approximation, within the limits where the Gaussian function is non-zero. Therefore, $\Omega(\Delta_d)$ is also a Gaussian distribution. Moreover, given that $\Xi(s_d)$ and $\zeta(s_b)$ are Gaussian, one can show using Eq. \ref{eq: tau_dSizer} and \ref{eq: integralXgaussians}:

\begin{equation}
\begin{split}
    \Gamma(\tau_d) \;=\;& \frac{\alpha e^{\alpha \tau_d}}{\pi \sigma_{s_d}^2} \exp{\left( -\frac{\langle s_d\rangle^2(1-2e^{-\alpha \tau_d})^2}{2\sigma_{s_d}^2 (1+4e^{-2\alpha \tau_d})} \right)} \left[ \; A+B \; \right]
\end{split}
\end{equation}
where
\begin{equation}
    A \;=\;  \frac{\sigma_{s_d}^2}{4+e^{2\alpha \tau_d}} \exp{\left( -\frac{\langle s_d \rangle^2}{2 \sigma_{s_d}^2} \left( 1+ \frac{4 e^{\alpha \tau_d}}{4+e^{2 \alpha \tau_d}} \right) \right)}
\end{equation}
and 
\begin{equation}
    B \;=\; \frac{\langle s_d \rangle}{2} \left( \frac{2+e^{\alpha \tau_d}}{4+e^{2 \alpha \tau_d}} \right) \sqrt{ \frac{2 \pi \sigma_{s_d}^2}{4+e^{2\alpha \tau_d}}} \left( 1+ \erf{\left( \frac{\langle s_d \rangle(2+e^{\alpha \tau_d})}{\sqrt{2\sigma_{s_d}^2} \sqrt{4+e^{2\alpha \tau_d}}} \right)} \right)
\end{equation}
Now, from the independent arguments presented in Appendix \ref{sec: SstatRelationships}, we know that the mean generation time $\langle \tau_d \rangle = \ln{(2)}/ \alpha$. Therefore, let us find out the functional form of $\Gamma(\tau_d)$ around the mean value of $\tau_d$. Assume, $\tau_d = \langle \tau_d \rangle + \delta$, where $\delta \ll 1/\alpha$. Therefore, $\exp{(\alpha \tau_d)} = \exp(\alpha \langle \tau_d \rangle) \exp(\alpha \delta) = 2 \exp(\alpha \delta)$, where $\exp(\alpha \delta) \approx 1+\alpha \delta$. Putting these values in the three equations above, we get:
\begin{equation}
    \Gamma(\tau_d) \;=\; \frac{2 \alpha}{\pi \sigma_{s_d}^2} \exp{\left( - \frac{\langle s_d \rangle^2 \alpha^2 \delta^2}{4 \sigma^2} \right)} \left[ \frac{\sigma_{s_d}^2}{8} \exp{\left( -\frac{\langle s_d \rangle^2}{\sigma_{s_d}^2} \right)} \;+\; \frac{\langle s_d \rangle \sqrt{2 \pi \sigma_{s_d}^2}}{8 \sqrt{2}} \left( 1 + \erf{\left( \frac{\langle s_d \rangle}{\sigma_{s_d}} \right)} \right)\right] 
\end{equation}
We have used the binomial expansion of $\exp(\alpha \delta)$ in the limit $\alpha \delta \ll 1$  such that only the first-order term is retained, which justifies our assumption that $\delta \ll 1/\alpha = \langle \tau_d \rangle /\ln{2}$. The first term in the square brackets vanishes because $\langle s_d \rangle \gg \sigma_{s_d}$, and therefore $\exp{(-\langle s_d \rangle^2 / \sigma_{s_d}^2)} \approx 0$. Moreover, the error function outputs 1 because $\erf{(x)} \approx 1$ for very large $x$. Therefore, we get:
\begin{equation}
    \Gamma(\tau_d) = \frac{1}{\sqrt{2 \pi \frac{2 \sigma_{s_d}^2}{\langle s_d \rangle^2 \alpha^2}}} \exp{\left( - \frac{(\tau_d - \langle \tau_d \rangle)^2}{2  \frac{2 \sigma_{s_d}^2}{\langle s_d \rangle^2 \alpha^2}} \right)}
\end{equation}
where we have replaced $\delta$ by $\tau_d - \langle \tau_d \rangle$. One can see that it is a Gaussian distribution with variance $\sigma_{\tau_d}^2 = 2 \sigma_{s_d}^2 / \langle s_d \rangle ^2 \alpha^2$.

\section{Various probability distributions for the Adder model} \label{sec: Sadder}

For the Adder model, the division-added-size distribution $\Omega(\Delta_d)$ is the principal distribution and is assumed to be known beforehand. Also, $\zeta(s_b)$ can be found in terms of $\Omega(\Delta_d)$ (see Appendix \ref{sec: SbirthSize}). Additionally, we obtain $\Psi(\Delta)$ in terms of $\Omega(\Delta_d)$ and then further derive other remaining distributions by probability transformations of random variables using $\Omega(\Delta_d)$, $\zeta(s_b)$, and $\Psi(\Delta)$.

\begin{figure}[hbt]
    \centering
    \includegraphics[width=1\linewidth]{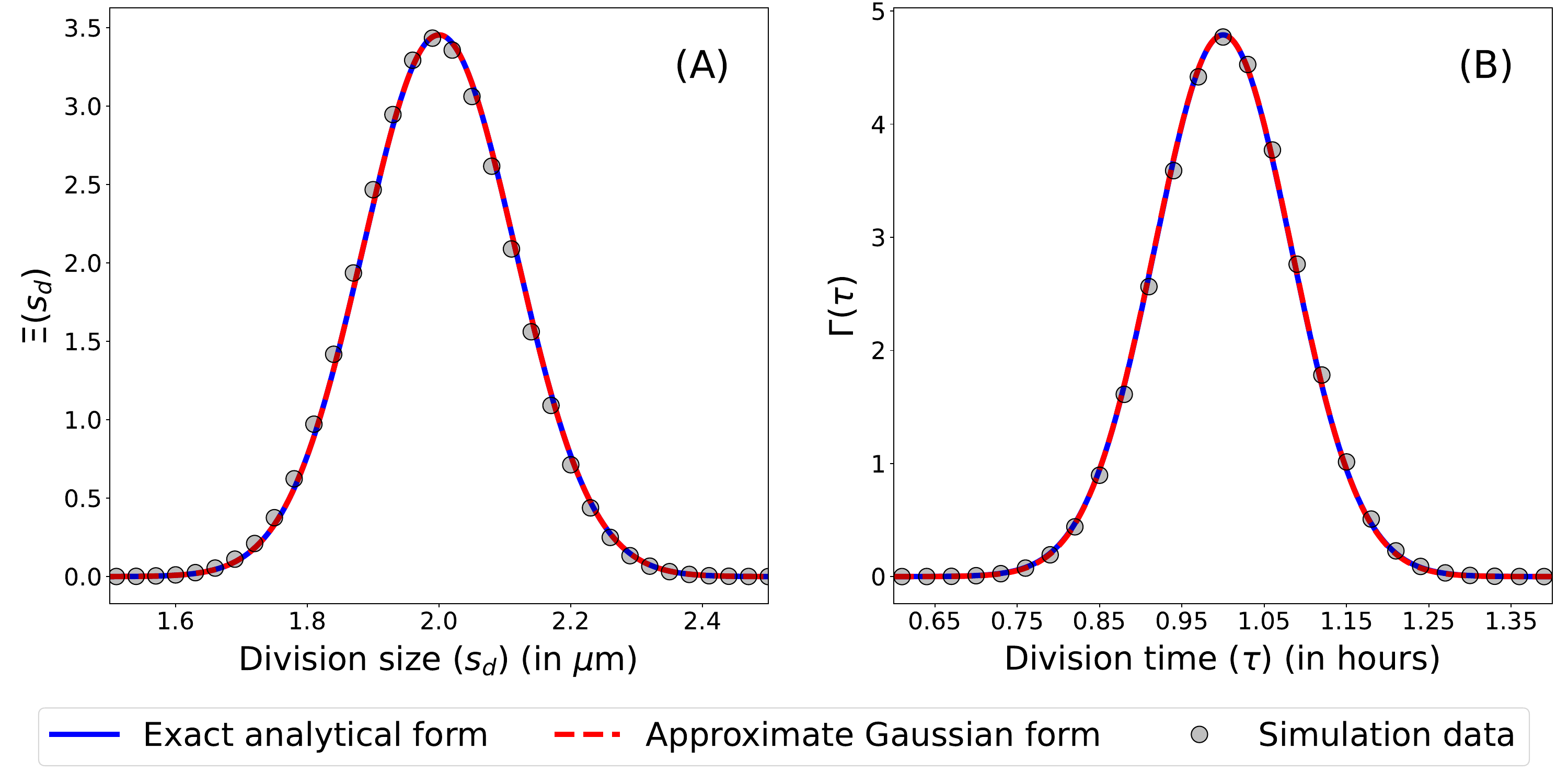}
    \caption{A comparison of the exact analytical forms of the single-cell-level distributions with their approximate Gaussian counterparts and simulation data for the Adder model: \textbf{(A)} Division-size distribution, \textbf{(B)} Division-time distribution. The parameters are identical to the ones mentioned in Fig. \ref{fig: allModelsExponential}}
    \label{fig: SdstauDadder}
\end{figure}

Similar to the arguments presented by Powell \cite{powell1956}, one can follow the survival probability arguments for the Adder model also in order to find out $\Psi(\Delta)$ in terms of $\Omega(\Delta_d)$. Let us say, $\Omega(\Delta_d)  d\Delta_d$ is the probability that a cell divides after it has added a size between $\Delta_d$ and $\Delta_d + d\Delta_d$ since its birth. $F_>(\Delta)$ = $\int_{\Delta}^{\infty} \Omega(\Delta_d) d\Delta_d$ is the probability that a given cell divides after it has added a size $\Delta$ since its birth, and the survival probability is defined as:
\begin{equation}
    \eta \;=\; \frac{F_>(\Delta +x)}{F_>(\Delta)}
\end{equation}
which is the conditional probability that a given cell does not divide before it has added a size $\Delta+x$, given that it has already added a size $\Delta$. Now, the number of cells that reach an added-size $\Delta$ and do not divide until they add another size $x$ is given by:
\begin{equation}
    N \; \Psi(\Delta) d\Delta \; \eta \;\; =\;\; N e^{\lambda t} \; \Psi(\Delta') d\Delta'
\end{equation}
where $\lambda$ is the cell number growth rate, $t$ is the time to add an extra amount of size $x$, and $\Psi(\Delta) d\Delta$ is the probability that a given cell has added-size between $\Delta$ and $d\Delta$. We have $\lambda = \alpha$ (from Appendix \ref{sec: SrelationshipGrowthRates}). Also $\Delta'$ = $\Delta+x$. Further simplifying the equation above :
\begin{equation} \label{Eqn:rawAdder}
    \frac{F_>(\Delta + x)}{F_>(\Delta)} \;\;=\;\; e^{\alpha t} \;\frac{\Psi(\Delta +x)}{\Psi(\Delta)} \; \frac{d\Delta'}{d\Delta}
\end{equation}
Now, we have
\begin{equation}
    s=s_b+\Delta
\end{equation}
Therefore,
\begin{equation}
    se^{\alpha t} = s_b + \Delta + x
\end{equation}
Hence,
\begin{equation}
    e^{\alpha t} \;=\; \frac{s_b + \Delta + x}{s} \;=\; \left( 1 + \frac{x}{s_b + \Delta} \right) 
\end{equation}
Also, we have:
\begin{equation}
    (s_b + \Delta)e^{\alpha t} \;=\; s_b + \Delta '
\end{equation}
which gives us:
\begin{equation}
    \frac{d\Delta'}{d\Delta} \;=\; e^{\alpha t}
\end{equation}
because $s_b$ does not change over the lifetime of the cell, and it only depends on the principal distribution $\Omega(\Delta_d)$.  Hence,
\begin{equation}
    e^{\alpha t} \frac{d\Delta'}{d\Delta} \;=\; \left( 1 + \frac{x}{s_b+\Delta} \right)^2
\end{equation}
Now, putting this value in the Eq. \ref{Eqn:rawAdder} and expanding all the functions using their Taylor expansion up to first order in $x$, we get:
\begin{equation}
    1 \;+\; x\frac{F_{>}^{'}(\Delta)}{F_>(\Delta)} \;\;=\;\; \left( 1+x\frac{2}{s_b + \Delta} \right) \left( 1+x\frac{\Psi^{'}(\Delta)}{\Psi(\Delta)} \right)
\end{equation}
Comparing L.H.S. and R.H.S. up to the first order of $x$, we get
\begin{equation}
    \frac{F_{>}^{'}(\Delta)}{F_{>}(\Delta)} - \frac{2}{s_b + \Delta} \;=\; \frac{\Psi^{'}(\Delta)}{\Psi(\Delta)}
\end{equation}
After integrating, we get the added-size distribution for the adder model as:
\begin{equation} \label{eq: AdderRaw}
    \Psi(\Delta) \;=\; \frac{B}{(s_b + \Delta)^2} \; \int_{\Delta}^{\infty} \Omega_A(\alpha) d\alpha 
\end{equation}
The equation above gives us the probability distribution for added-size. However, $s_b$ is a variable, not a constant. Therefore, to obtain $\Psi(\Delta)$, we integrate over $s_b$ as follows:
\begin{equation}
    \Psi(\Delta) \;=\; \int_{0}^{\infty} ds_b \; \zeta(s_b) \; \frac{B}{(s_b + \Delta)^2} \; \int_{\Delta}^{\infty} \Omega_A(\alpha) d\alpha 
\end{equation}
To further simplify the expression, we can assume that $\zeta(s_b)$ is a very sharply spiked distribution peaked at $\langle s_b \rangle$, and it can be replaced by the Dirac delta function such that one obtains Eq. \ref{eq: mainAdder}, which is rewritten here for easy reference:
\begin{equation} 
    \Psi(\Delta) \;=\; \frac{B}{(\langle s_b \rangle + \Delta)^2} \; \int_{\Delta}^{\infty} \Omega_A(\alpha) d\alpha 
\end{equation}
Now, using probability transformations, i.e., Eq. \ref{eq: probTrans}, we can also get $\Theta(a)$ and $\Phi(s)$ for the Adder model as:
\begin{equation} \label{eq: sizeAdder}
\begin{split}
    \Phi(s) \;&=\; \int_{0}^{\infty} d s_b \; \zeta(s_b) \; \Psi(s-s_b) \\
    &=\;\int_{0}^{\infty} d\Delta \; \Psi(\Delta) \; \zeta(s-\Delta)
\end{split}
\end{equation}
and
\begin{equation} \label{eq: ageAdder}
\begin{split}
    \Theta(a) \;&=\; \alpha e^{\alpha a} \int_{0}^{\infty} d s_b \; s_b \; \zeta(s_b) \; \Psi(s_b(e^{\alpha a}-1)) \\
    &\;= \frac{\alpha e^{\alpha a}}{(e^{\alpha a}-1)^2} \;\; \int_{0}^{\infty} d\Delta \; \Delta \; \Psi_A(\Delta) \; \zeta_A \left( \frac{\Delta}{e^{\alpha a}-1} \right)
\end{split}
\end{equation}
Equation \ref{eq: AdderRaw} should be used in the equations above instead of Eq. \ref{eq: mainAdder} for probability transformations, because the former one is more accurate. The other single-cell-level distributions, i.e., $\Gamma(\tau_d)$ and $\Xi(s_d)$, can also be obtained from $\Omega(\Delta_d)$ and $\zeta(s_b)$ via probability transformations as:
\begin{equation} \label{eq: sdAdder}
\begin{split}
\Xi(s_d) \;&=\; \int_{0}^{\infty} d\Delta_d \; \Omega(\Delta_d) \; \zeta(s_d -\Delta_d) \\
&=\; \int_{0}^{\infty} ds_b \; \zeta(s_b) \; \Omega(s_d -s_b)
\end{split}
\end{equation}
And
\begin{equation} \label{eq: tau_dAdder}
\begin{split}
    \Gamma(\tau_d)\;&=\;  \frac{\alpha e^{\alpha \tau_d}}{(e^{\alpha \tau_d}-1)^2} \int_{0}^{\infty} d\Delta_d \;\Delta_d\; \Omega(\Delta_d) \; \zeta \left(\frac{\Delta_d}{e^{\alpha \tau_d}-1} \right)  \\
    &=\; \alpha e^{\alpha \tau_d} \int_{0}^{\infty} ds_b \; s_b\; \zeta(s_b) \; \Omega(s_b(e^{\alpha \tau_d}-1))
\end{split}
\end{equation}
If the principal distribution i.e. $\Omega(\Delta_d)$ is taken to be a Gaussian distribution, it can be proved that all other single-cell-level distributions will also be Gaussian up to a very good approximation, with the mean and standard deviations identical to the ones predicted by Eq. \ref{eq: mainMeanSingleCell}, \ref{eq: mainStdAdder}, \ref{eq: mainMeanTau_d}, and \ref{eq: mainCOVadderExpo}. This can be easily confirmed from the comparison between the exact analytical forms and the approximate Gaussian forms of the distributions in Fig. \ref{fig: SdstauDadder}. Now, let us explicitly prove that they will be approximately Gaussian.

We know (from Appendix \ref{sec: SbirthSize}) that for the Adder model, if $\Omega(\Delta_d)$ is Gaussian, $\zeta(s_b)$ will also be Gaussian. Further, given that $\Omega(\Delta_d)$ and $\zeta(s_b)$ are Gaussian, one can show using Eq. \ref{eq: integralGaussians} and \ref{eq: sdAdder}:
\begin{equation}
    \Xi(s_d) = \frac{1}{2 \sqrt{2\pi (\sigma_{\Delta_d}^2 + \sigma_{s_b}^2)}} \exp{\left[ - \frac{(s_d - (\langle \Delta_d \rangle + \langle s_b \rangle))^2}{2(\sigma_{\Delta_d}^2 + \sigma_{s_b}^2)}\right]} \left[1+\erf{\left( \frac{\frac{\langle s_b \rangle}{2\sigma_{s_b}^2}+\frac{s_d - \langle \Delta_d \rangle}{2\sigma_{\Delta_d}^2}}{\sqrt{\frac{1}{2 \sigma_{s_b}^2}+\frac{1}{2\sigma_{\Delta_d}^2}}} \right)} \right]
\end{equation}
Putting in the value of $\langle s_b \rangle$ and $\sigma_{s_b}$ in terms of $\langle \Delta_d \rangle$ and $\sigma_{\Delta_d}$, we get:
\begin{equation}
    \Xi(s_d) = \sqrt{\frac{3}{8 \pi \sigma_{\Delta_d}^2}} \exp{\left[ - \frac{(s_d-2\langle \Delta_d \rangle)^2}{8 \sigma_{\Delta_d}^2 /3}\right]} \left[\frac{ 1+ \erf{\left( \frac{s_d + 2 \langle\Delta_d \rangle}{2 \sqrt{2} \sigma_{\Delta_d}}  \right)}}{2} \right]
\end{equation}
This shows that the probability distribution for $s_d$ is Gaussian with a mean $2\langle \Delta_d \rangle$ and a standard deviation. But there is an extra factor multiplied by the Gaussian function. However, if one assumes that the stochasticity in the principal distribution ($\sigma_{\Delta_d}$) is small, it can be shown that this factor is $1$ for a very good approximation within the limits where the Gaussian function is non-zero. Hence, to a very good approximation, the overall shape of $\Xi(s_d)$ is Gaussian. Moreover, given that $\Omega(\Delta_d)$ and $\zeta(s_b)$ are Gaussian, one can show using Eq. \ref{eq: tau_dAdder} and \ref{eq: integralXgaussians}:
\begin{equation}
\begin{split}
    \Gamma(\tau_d) \;=\;& \frac{\sqrt{3} \alpha e^{\alpha \tau_d}}{2 \pi \sigma_{\Delta_d}^2} \exp{\left( -\frac{3 \langle \Delta_d \rangle^2 (e^{\alpha \tau_d}-2)^2}{2\sigma_{\Delta_d}^2 (e^{2\alpha \tau_d} -2 e^{\alpha \tau_d}+4 )} \right)} \left[ \; A+B \; \right]
\end{split}
\end{equation}
where
\begin{equation}
    A \;=\;  \frac{\sigma_{\Delta_d}^2}{(e^{2\alpha \tau_d}-2e^{\alpha \tau_d}+4)} \exp{\left( -\frac{\langle \Delta_d \rangle^2 (2+e^{\alpha \tau_d})^2}{2 \sigma_{\Delta_d}^2 (e^{2 \alpha \tau_d} -2e^{\alpha \tau_d}+4)}\right)}
\end{equation}
and 
\begin{equation}
    B \;=\; \frac{\langle \Delta_d \rangle (2+e^{\alpha \tau_d})}{2 (e^{2 \alpha \tau_d} -2e^{\alpha \tau_d}+4)}\sqrt{ \frac{2 \pi \sigma_{\Delta_d}^2}{e^{2 \alpha \tau_d} -2e^{\alpha \tau_d}+4}} \left( 1+ \erf{\left( \frac{\langle \Delta_d \rangle(2+e^{\alpha \tau_d})}{\sqrt{2\sigma_{\Delta_d}^2 (e^{2 \alpha \tau_d} -2e^{\alpha \tau_d}+4)}} \right)} \right)
\end{equation}
Now, from the independent arguments presented in Appendix \ref{sec: SstatRelationships}, we know that the mean generation time $\langle \tau_d \rangle = \ln{(2)}/ \alpha$. Therefore, let us find out the functional form of $\Gamma(\tau_d)$ around the mean value of $\tau_d$. Assume, $\tau_d = \langle \tau_d \rangle + \delta$, where $\delta \ll 1/\alpha$. Therefore, $\exp{(\alpha \tau_d)} = \exp(\alpha \langle \tau_d \rangle) \exp(\alpha \delta) = 2 \exp(\alpha \delta)$, where $\exp(\alpha \delta) \approx 1 + \alpha \delta$. Putting these values in the three equations above, we get:
\begin{equation}
    \Gamma(\tau_d) \;=\; \frac{\sqrt{3} \alpha}{\pi \sigma_{\Delta_d}^2} \exp{\left( - \frac{3 \langle \Delta_d \rangle^2 \alpha^2 \delta^2}{2 \sigma^2} \right)} \left[ \frac{\sigma_{\Delta_d}^2}{4} \exp{\left( -\frac{2 \langle \Delta_d \rangle^2}{\sigma_{\Delta_d}^2} \right)} \;+\; \frac{\langle \Delta_d \rangle \sqrt{2 \pi \sigma_{\Delta_d}^2}}{4} \left( 1 + \erf{\left( \frac{2 \langle \Delta_d \rangle}{\sigma_{\Delta_d}} \right)} \right)\right] 
\end{equation}
where we have used the binomial expansion of $\exp(\alpha \delta)$ in the limit $\alpha \delta \ll 1$  such that only the first order term is retained, which justifies our assumption that $\delta \ll 1/\alpha = \langle \tau_d \rangle /\ln{2}$. The first term in the square brackets vanishes because $\langle \Delta_d \rangle \gg \sigma_{\Delta_d}$, and therefore $\exp{(-2 \langle \Delta_d \rangle^2 / \sigma_{\Delta_d}^2)} \approx 0$. Moreover, the error function outputs 1 because $\erf{(x)} \approx 1$ for very large $x$. Therefore, we get:
\begin{equation}
    \Gamma(\tau_d) = \frac{1}{\sqrt{2 \pi \frac{\sigma_{\Delta_d}^2}{3 \langle \Delta_d \rangle^2 \alpha^2}}} \exp{\left( - \frac{(\tau_d - \langle \tau_d \rangle)^2}{2 \frac{\sigma_{\Delta_d}^2}{3 \langle \Delta_d \rangle^2 \alpha^2}} \right)}
\end{equation}
where we have replaced $\delta$ by $\tau_d - \langle \tau_d \rangle$. One can see that it is a Gaussian distribution with variance $\sigma_{\tau_d}^2 = \sigma_{\Delta_d}^2 /(3 \langle \Delta_d \rangle^2 \alpha^2)$.

\section{Alternate method to obtain population-level distributions} \label{sec: SalternateMethod}

\begin{figure}[h]
    \centering
    \includegraphics[width=1\linewidth]{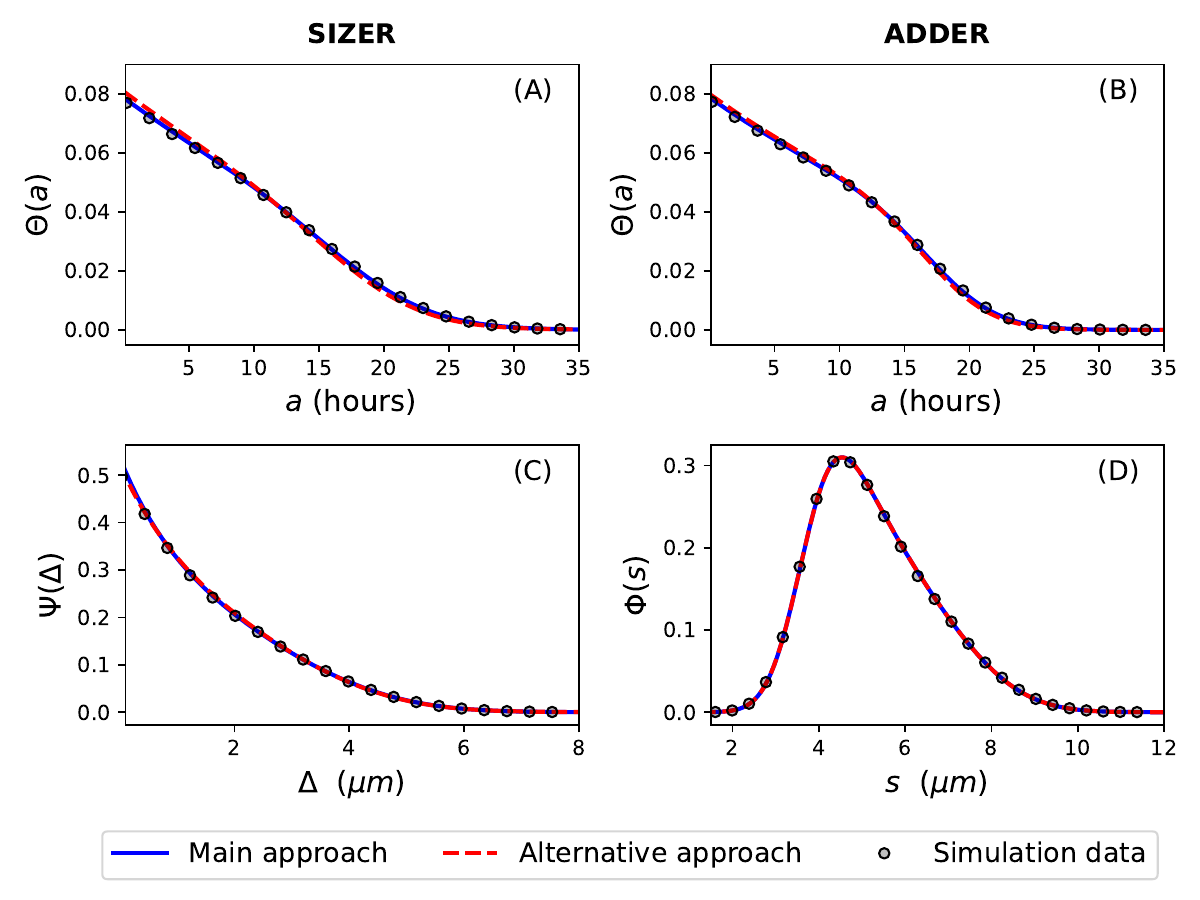}
    \caption{A comparison between the two approaches to get the population-level distributions: \textbf{(A)} $\Theta(a)$ for the Sizer model, \textbf{(B)} $\Theta(a)$ for the Adder model, \textbf{(C)} $\Psi(\Delta)$ for the Sizer model, \textbf{(D)} $\Phi(s)$ for the Adder model. The parameters for the Sizer model are: $\langle s_d \rangle = 7.8912 \; \mu m$ and $\sigma_{s_d} = 1.3201 \; \mu m$ with the exponential growth rate $\alpha = \ln(2)*(0.0569)$ 1/min. The parameters for the Adder model are: $\langle \Delta_d\rangle = 3.9073 \; \mu m$ and $\sigma_{\Delta_d} = 1.0801 \;\mu m$ with the exponential growth rate $\alpha = \ln(2)*(0.0569)$ 1/min. The parameters of the two models are taken from the experimental data for E. Coli strain K12 NCM3722 in TSB medium \cite{taheriJun2015} }
    \label{fig: StwoApproaches}
\end{figure}
In the main method, discussed in Appendices \ref{sec: Ssizer} and \ref{sec: Sadder}, the population-level distributions for the Sizer and Adder models, such as $\Theta(a)$, $\Phi(s)$, and $\Psi(\Delta)$ are obtained using $\zeta(s_b)$ and one of the population-level distributions (which we obtain from the principal distribution using survival probability arguments) via probability transformations. For example, for the Sizer model, $\Theta(a)$ and $\Psi(\Delta)$ are obtained from $\zeta(s_b)$ and $\Phi(s)$ via probability transformations, where $\Phi(s)$ is obtained from the principal distribution $\Xi(s_d)$ using Eq. \ref{eq: mainSizer}. Similarly, for the Adder model, $\Theta(a)$ and $\Phi(s)$ are obtained from $\zeta(s_b)$ and $\Psi(\Delta)$ via probability transformations, where $\Psi(\Delta)$ is obtained from the principal distribution $\Omega(\Delta_d)$ using Eq. \ref{eq: mainAdder}.

However, there is also an alternative approach to get these distributions. In this alternative method, we first obtain the single-cell-level distribution corresponding to the required population-level distribution via probability transformation of $\zeta(s_b)$ and the principal distribution, as shown in Appendices \ref{sec: Ssizer} and \ref{sec: Sadder}. After that, we can use the three equations, Eq. \ref{eq: mainTimer}, \ref{eq: mainSizer}, \ref{eq: mainAdder}, to get the population-level distributions. These three equations are derived using arguments of survival probability, which are independent of the cell-division model.  For example, to find the distributions $\Theta(a)$ and $\Psi(\Delta)$ for the Sizer model, we first find out the single-cell-level distributions $\Gamma(\tau_d)$ and $\Omega(\Delta_d)$ from $\zeta(s_b)$ and $\Xi(s_d)$ using probability transformations, and then $\Theta(a)$ and $\Psi(\Delta)$ are found from $\Gamma(a)$ and $\Omega(\Delta_d)$ using Eq. \ref{eq: mainTimer} and \ref{eq: mainAdder} , respectively. Similarly, to obtain the distributions $\Theta(a)$ and $\Phi(s)$ for the Adder model, we first find out the single-cell-level distributions $\Gamma(\tau_d)$ and $\Xi(s_d)$ from $\zeta(s_b)$ and $\Omega(\Delta_d)$ using probability transformations, and then $\Theta(a)$ and $\Phi(s)$ are found from $\Gamma(a)$ and $\Xi(s_d)$ using Eq. \ref{eq: mainTimer} and \ref{eq: mainSizer} , respectively. The population-level distributions for the Sizer and Adder models obtained using the two approaches are compared with each other and with simulation results in Fig. \ref{fig: StwoApproaches}.

\begin{figure}[h]
    \centering
    \includegraphics[width=1\linewidth]{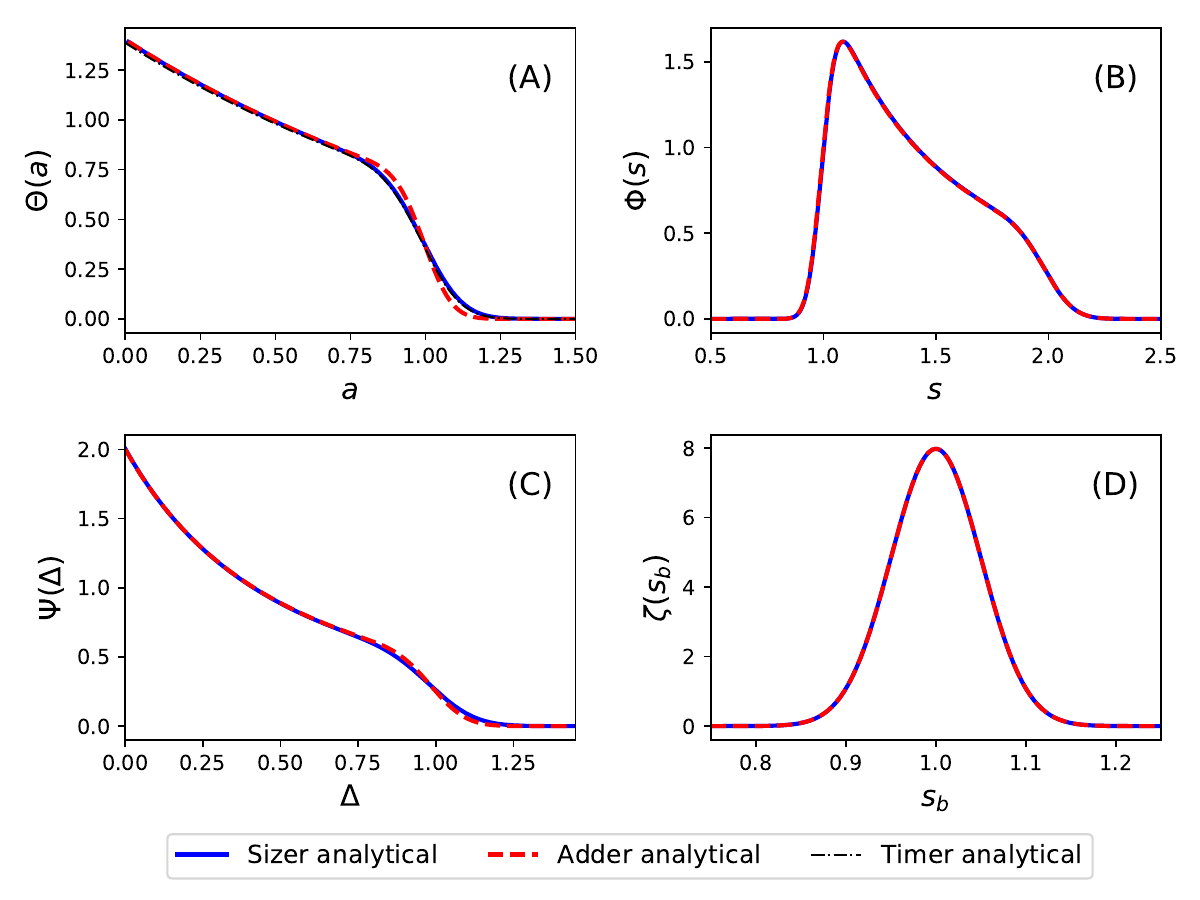}
    \caption{A comparison between the population-level distributions for the three models under exponential growth with parameters chosen such that $\zeta(s_b)$ is identical for the Sizer and Adder models: \textbf{(A)} Age distribution, \textbf{(B)} Size distribution, \textbf{(C)} Added-size distribution, \textbf{(D)} Birth-size distribution. The parameters for the three models are identical to those mentioned in Fig. \ref{fig: allModelsExponential}, except for the standard deviation in the principal distribution for the Adder model, which is taken to be $\sigma_{\Delta_d} = 0.05\sqrt{3} \mu$m.}
    \label{fig: SallModelsExponentialWithSameSbDisbn}
\end{figure}

In Fig. \ref{fig: SallModelsExponentialWithSameSbDisbn}, we have plotted various population-level distributions where the parameters of the model (mean and standard deviation of the principal distribution and the growth rate) are taken such that $\zeta(s_b)$ and $\Xi(s_d)$ have the identical forms. As described in the main text, this leads to different standard deviations in other single-cell-level quantities, such as $\Delta_d$ and $\tau_d$. However, we show that this results in only minor differences in the population-level distributions according to the alternate method.

\section{Stochasticity in growth rate} \label{sec: SstochsaticGrowthRate}

\subsection{Temporal stochasticity} \label{subsec: SstochsaticGrowthRateTemporal}
As mentioned in the main text, when the growth rate is stochastic in time over the lifetime of the cell, the results are identical to the ones derived for the case of non-stochastic growth rate. Let us prove this. We have the size $s(t)$ of an exponentially growing cell at any time $t$ given by $s(t) = s(0) \exp(\alpha t)$, where $s(0)$ is the size of the cell at the initial time and $\alpha$ is the growth rate. But if the growth rate is stochastic over time, then it varies at each point in time. Therefore, we can break the whole time interval from $t'=0$ to $t'=t$ in $n$ infinitesimally small intervals each of length $\Delta t  = t/n$ and write:
\begin{equation}
\begin{split}
    s(t) \;&=\; \; s(0) \; e^{\alpha_1 \Delta t} e^{\alpha_2 \Delta t} ...\;e^{\alpha_n \Delta t} \\
    &=\; \lim_{n  \rightarrow \infty} s(0) \exp{\left( \sum_{i=0}^{n} \alpha_i \Delta t \right)} \\
    &=\; \lim_{n  \rightarrow \infty} s(0) \exp{\left(\Delta t \sum_{i=0}^{n} \alpha_i \right)} \\
\end{split}
\end{equation}
Let $b = \sum_{i=0}^{n} \alpha_i$ and $I = b \Delta t= \Delta t \sum_{i=0}^{n} \alpha_i$. The stochastic variable $b$ is a sum of $n$ identically distributed random variables under the limit $n \rightarrow \infty$. Using the central limit theorem, one can prove that the mean and standard deviation of $b$ are given by $\overline{b} = n \overline{\alpha}$ and $\sigma_{b} = \sigma_{\alpha}\sqrt{n}$ respectively, where $\overline{\alpha}$ is the time averaged exponential growth rate and $\sigma_{\alpha}$ is the standard deviation in $\alpha$ over the lifetime of the cell. Therefore, the mean and standard deviation of $I = b \Delta t$ are given by $\overline{I} =  n \overline{\alpha} \Delta t  = \overline{\alpha} t$ and $\sigma_{I} =  \sigma_{\alpha} \sqrt{n} \Delta t = \sigma_{\alpha} t / \sqrt{n}$ respectively. Hence, in the large $n$ limit, the variable $I$ behaves like a constant with value $\overline{I}  = \overline{\alpha} t$ because the standard deviation falls by $\sqrt{n}$ and becomes zero in the large $n$ limit. Therefore, the expression for $s(t)$ can simply be written as:
\begin{equation}
    s(t) = s(0) \exp{(\overline{\alpha} t)}
\end{equation}
where $\overline{\alpha}$ is the time average value of the growth rate $\alpha$. To derive various expressions and results for the case of exponential growth with fixed growth rate, two relationships are widely used: $s = s_b \exp(\alpha a)$ and $s_d = s_b \exp(\alpha \tau_d)$. If the growth rate is stochastic in time, due to the argument presented above, the validity of the relationships still holds if one replaces $\alpha$ by $\overline{\alpha}$. Hence, all of the results derived in the text for the case of non-stochastic growth rate are valid for the case of stochastic growth rate if one replaces $\alpha$ by $\overline{\alpha}$.

\subsection{Population stochasticity}  \label{subsec: SstochsaticGrowthRatePopulation}
\begin{figure}[hbt]
    \centering
    \includegraphics[width=1\linewidth]{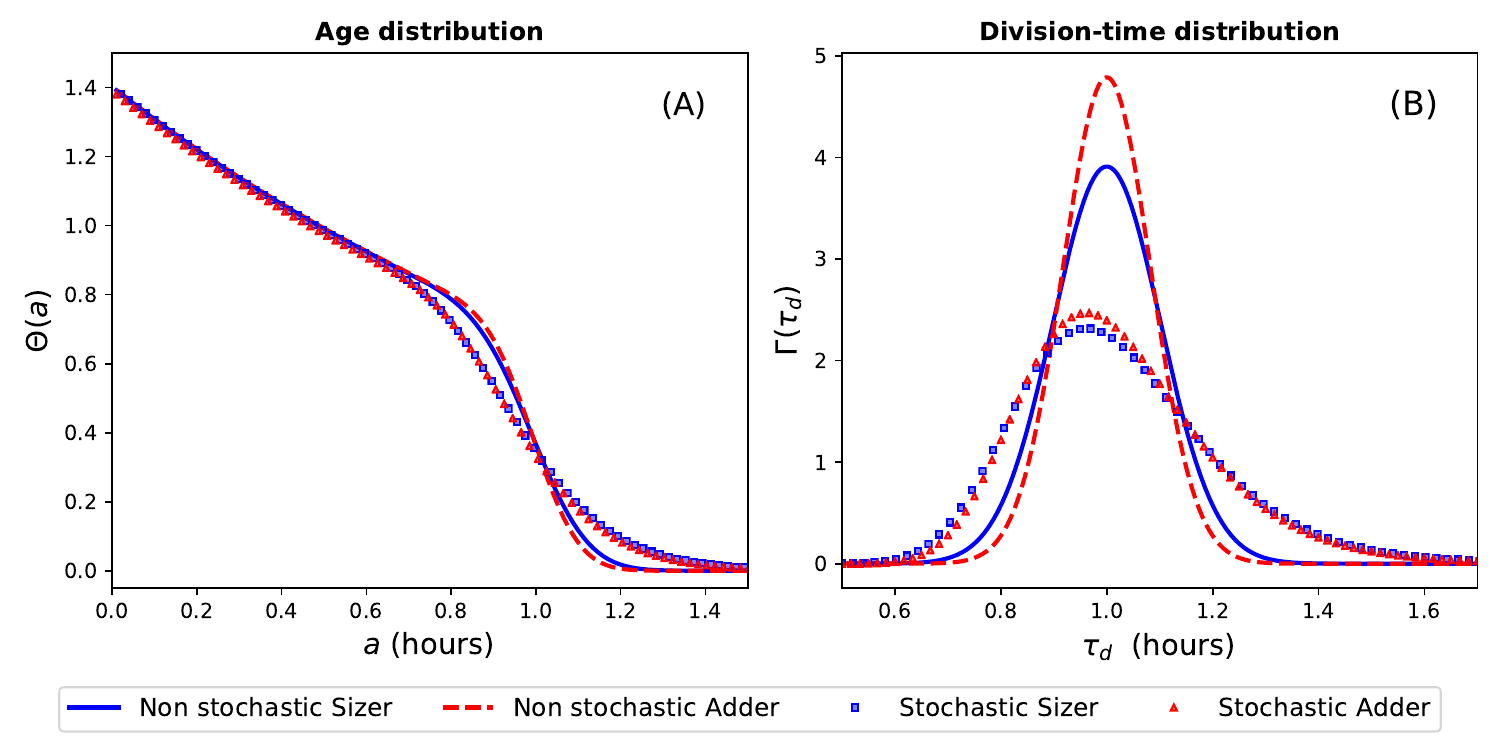}
    \caption{A comparison of the analytical forms of age and division-time distributions obtained for non-stochastic and stochastic (stochastic over the population) exponential growth rate for the Sizer and Adder models: \textbf{(A)} Age distribution, \textbf{(B)} Division-time distribution. The parameters for the models are identical to the ones mentioned in Fig. \ref{fig: allModelsExponential}. For the case of population stochasticity in growth rate, $\alpha$ is sampled randomly from a probability distribution $\Lambda(\alpha)$ with a mean value of $\ln{(2)}$ per hour, and a standard deviation of $0.1$ per hour.}
    \label{fig: SstochAlphaExpo}
\end{figure}
For the second type of stochasticity in the growth rate (growth rate being stochastic over the population), as mentioned in the main text, the results remain the same for the Sizer and Adder model except for $\Gamma(\tau_d)$ and $\Theta(a)$. When the growth rate is taken to be stochastic over the population, the distributions, such as $\zeta(s_b)$, $\Xi(s_d)$, $\Omega(\Delta_d)$, $\Phi(s)$, and $\Psi(\Delta)$, remain identical. The reason is that these distributions do not depend explicitly on the growth rate. They are derived from the principal distribution using survival probability arguments and probability transformations. For the case of survival probability arguments, even when the growth rate is stochastic over population, one can use the population average growth rate $\langle \alpha \rangle$ instead of absolute growth rate $\alpha$  in the arguments (Appendices \ref{sec: Stimer}, \ref{sec: Ssizer}, and \ref{sec: Sadder}) and therefore, the results are valid. Moreover, while doing probability transformations, one has to include an extra stochastic variable $\alpha$ whose probability distribution $\Lambda(\alpha)$ is assumed to be known beforehand, and an extra integration has to be carried over $\alpha$. For example, in order to find out $\Psi(\Delta)$ for the Sizer model, given that the growth rate is non-stochastic, one can simply write Eq. \ref{eq: addedSizeSizer}. However, if the growth rate is stochastic over the population, one has to write:
\begin{equation}
\begin{split}
    \Psi(\Delta) \;&=\; \int_{0}^{\infty} \int_0^{\infty} \; d\alpha \; d s_b\; \Lambda(\alpha)     \; \zeta(s_b) \; \Phi(s_b+\Delta) \\
    &=\; \int_{0}^{\infty}  \int_0^{\infty} \;d\alpha \;d s \; \Lambda(\alpha)  \; \Phi(s) \; \zeta(s-\Delta)
\end{split}
\end{equation}
Since the integrals of $s_b$ and $s$ are independent of $\alpha$, one can take $\alpha$ outside of these integrals, and write:
\begin{equation}
\begin{split}
    \Psi(\Delta) \;&=\; \left(\int_{0}^{\infty} d\alpha \; \Lambda(\alpha)  \right) \int_0^{\infty} \; d s_b \; \zeta(s_b) \; \Phi(s_b+\Delta) \\
    &=\; \left(\int_{0}^{\infty} d\alpha \; \Lambda(\alpha) \right) \int_0^{\infty} \; d s \; \Phi(s) \; \zeta(s-\Delta)
\end{split}
\end{equation}
Since the integration of $\alpha$ gives 1, the older form of $\Psi(\Delta)$ is obtained. Hence, the result remains the same for the added-size distribution. The same can be shown for other size-related distributions as well.

However, the analytical forms of $\Theta(a)$ and $\Gamma(\tau_d)$ depend on $\alpha$. Therefore, the analytical forms of these distributions change for this case. For the case of stochastic growth rate (stochastic over the population), one can write the analytical forms for the division-time and the age distributions for the Sizer model as:
\begin{equation} \label{eq: tau_dSizerStochAlpha}
\begin{split}
    \Gamma(\tau_d) \;&=\; \int_{0}^{\infty} d\alpha \; \Lambda(\alpha)\; \alpha \; e^{\alpha \tau_d} \int_{0}^{\infty} d s_b \; \zeta(s_b) \; s_b \; \Xi(s_b e^{\alpha \tau_d}) \\
    &=\; \int_{0}^{\infty} d \alpha \; \Lambda(\alpha) \; \alpha \;e^{-\alpha \tau_d} \int_{0}^{\infty} d s_d \; \Xi(s_d) \; s_d \; \zeta(s_d e^{-\alpha \tau_d})
\end{split}
\end{equation}
and
\begin{equation} \label{eq: ageSizerStochAlpha}
\begin{split}
    \Theta(a) \;&=\; \int_{0}^{\infty} d\alpha  \; \Lambda(\alpha) \; \alpha \;e^{\alpha a} \; \int_0^{\infty} d s_b \; s_b \; \zeta(s_b) \; \Phi(s_b e^{\alpha a}) \\
    &=\;\int_{0}^{\infty} d\alpha \; \; \Lambda(\alpha) \alpha \;e^{-\alpha a} \; \int_0^{\infty} ds \; s \; \Phi(s) \; \zeta(s e^{-\alpha a}) 
\end{split}
\end{equation}
Similarly, for the Adder model, when $\alpha$ is stochastic over the population, one can write:
\begin{equation} \label{eq: tau_dAdderStochAlpha}
\begin{split}
    \Gamma(\tau_d)\;&=\; \int_{0}^{\infty} d\alpha \; \Lambda(\alpha) \; \frac{\alpha e^{\alpha \tau_d}}{(e^{\alpha \tau_d}-1)^2} \int_{0}^{\infty} d\Delta_d \;\Delta_d\; \Omega(\Delta_d) \; \zeta \left(\frac{\Delta_d}{e^{\alpha \tau_d}-1} \right)  \\
    &=\; \int_{0}^{\infty} d\alpha \; \Lambda(\alpha) \; \alpha \;e^{\alpha \tau_d} \int_{0}^{\infty} ds_b \; s_b\; \zeta(s_b) \; \Omega(s_b(e^{\alpha \tau_d}-1))
\end{split}
\end{equation}
and
\begin{equation} \label{eq: ageAdderStochAlpha}
\begin{split}
    \Theta(a) \;&=\; \int_{0}^{\infty} d\alpha \; \Lambda(\alpha) \; \alpha \;e^{\alpha a} \int_{0}^{\infty} d s_b \; s_b \; \zeta(s_b) \; \Psi(s_b(e^{\alpha a}-1)) \\
    &\;=  \int_{0}^{\infty} d\alpha \; \Lambda(\alpha) \; \frac{\alpha e^{\alpha a}}{(e^{\alpha a}-1)^2} \;\; \int_{0}^{\infty} d\Delta \; \Delta \; \Psi_A(\Delta) \; \zeta_A \left( \frac{\Delta}{e^{\alpha a}-1} \right)
\end{split}
\end{equation}

In Fig. \ref{fig: SstochAlphaExpo}, $\Gamma(\tau_d)$ and $\Theta(a)$ are compared for the case of non-stochastic and stochastic $\alpha$ (stochastic over the population) for the Sizer and Adder models. When there is no stochasticity in $\alpha$, we have proved that the division-time distribution $\Gamma(\tau_d)$ is approximately a Gaussian distribution. However, when we include some stochasticity in the growth rate, the distribution no longer remains Gaussian; rather, it becomes positively skewed, which is the characteristic shape of the division-time distribution observed in the experiments. Additionally, it has a larger standard deviation and excess kurtosis than the one for the case of non-stochastic $\alpha$.

Note that in Fig. \ref{fig: SstochAlphaExpo} (B), $\Gamma(\tau_d)$ is different for the Sizer and Adder models. Here too, just like in the case of $\zeta(s_b)$ discussed in the main text, one can indeed choose the principal distributions such that $\Gamma(\tau_d)$ is made identical for the two models. This will again cause some differences in other single-cell-level distributions, but only minor differences in population-level distributions.

\section{Statistical relationships for single-cell-level quantities} \label{sec: SstatRelationships}

Any statistical relationship between two single-cell-level quantities for the Timer model (under exponential growth) could not be obtained due to the lack of cell size homeostasis, because these quantities' means and standard deviations keep changing with time. However, such relationships can be obtained for the Sizer and Adder models. In this appendix, we derive the statistical relationships (Eq. \ref{eq: mainMeanSingleCell}, \ref{eq: mainStdAdder}, \ref{eq: mainStdSizer}, \ref{eq: mainMeanTau_d}, \ref{eq: mainCOVadderExpo}, and \ref{eq: mainCOVsizerExpo}). One can already obtain Eq. \ref{eq: mainMeanSingleCell} (the relationship between the averages of $s_b$, $s_d$, and $\Delta_d$) from Eq. \ref{eq: sbSdMean} and \ref{eq: sbDeltaDmean} (Appendix \ref{sec: SbirthSize}).

\subsection{Sizer Model} \label{subsec: SstatRelationshipsSizer}
From Eq. \ref{eq: sbSdSTD} and \ref{eq: sbDeltaDstdSizer} (Appendix \ref{sec: SbirthSize}), one obtains Eq. \ref{eq: mainStdSizer}, which represents the relationship between the standard deviations in $s_b$, $s_d$, and $\Delta_d$ for the Sizer model. For the division-time, it is a function of $s_d$ and $s_b$, which can be written as:
\begin{equation}
    \tau_d \;=\; \frac{1}{\alpha}\ln{\left(\frac{s_d}{s_b}\right)} 
\end{equation}
Using Taylor series expansion about $(\langle s_d \rangle,\langle s_b \rangle )$, we get:
\begin{equation} \label{eq: tau_dSizerRaw}
    \tau_d \;=\; \frac{1}{\alpha} \left[ \ln{\left( \frac{\langle s_d \rangle}{\langle s_b \rangle} \right)} \;+\; \sum_{k=1}^{\infty} \frac{(-1)^{n+1}}{k} \left( \frac{(s_d -\langle s_d \rangle)^k}{\langle s_d \rangle^k} \right) \;+\; \sum_{k=1}^{\infty} \frac{(-1)^{n}}{k} \left( \frac{(s_b-\langle s_b \rangle)^k}{\langle s_b \rangle^k} \right)  \right]
\end{equation}
Taking average on both sides, we get:
\begin{equation}
    \langle \tau_d \rangle \;=\; \frac{1}{\alpha} \left[ \ln{\left( \frac{\langle s_d \rangle}{\langle s_b \rangle} \right)} \;+\; \sum_{k=2}^{\infty} \frac{(-1)^{n+1}}{k} \left( \frac{\mu_{k}(s_d)}{\langle s_d \rangle^k} \right) \;+\; \sum_{k=2}^{\infty} \frac{(-1)^{n}}{k} \left( \frac{\mu_{k}(s_b)}{\langle s_b \rangle^k} \right)  \right]
\end{equation}
where $\mu_{k}(s_d) = \langle (s_d-\langle s_d \rangle)^n\rangle$ and $\mu_{k}(s_b) = \langle (s_b-\langle s_b \rangle)^n\rangle$ are the $k$th central moments for distribution of $s_d$ and $s_b$ respectively. Let us assume that the distribution for $s_d$ is very sharply spiked i.e. $\mu_{k}(s_d) /\langle s_d \rangle^k \approx 0$. Then from Eq. \ref{eq: sbSdMean} and \ref{eq: nthCentralMomentSbSd}, it would imply that the distribution for $s_b$ will also be sharply spiked, i.e., $\mu_{k}(s_b) /\langle s_b \rangle^k \approx 0$. Therefore, the terms inside the summations vanish, and the average value of the division-time is written as:
\begin{equation}
    \langle \tau_d \rangle \;=\;  \ln(\langle s_d \rangle/\langle s_b \rangle)/ \alpha = \ln{(2)} / \alpha
\end{equation}
which is just Eq. \ref{eq: mainMeanTau_d}. Further, if one looks in the proximity of the point $(\langle s_d \rangle,\langle s_b \rangle)$ (about where the function is expanded using Taylor series expansion), we have $s_d - \langle s_d \rangle \ll \langle s_d \rangle$ and $s_b - \langle s_b \rangle \ll \langle s_b \rangle$, such that, in the neighborhood of the point, only taking the first order terms in the series is enough . Therefore, up to first order, we have:
\begin{equation}
    \tau_d - \langle \tau_d \rangle \;=\; A (s_d - \langle s_d \rangle) \;+\; B ( s_b - \langle s_b \rangle)  
\end{equation}
where $A=1/(\alpha \langle s_d \rangle)$ and  $B=-1/(\alpha \langle s_b \rangle)$. Therefore, one can write for $\langle(\tau_d - \langle \tau_d \rangle)^2\rangle$ as follows:
\begin{equation} \label{eq: tau_dSTDrawSizer}
\begin{split}
    \langle(\tau_d - \langle \tau_d \rangle)^2\rangle &= A^2 \langle (s_d -\langle s_d \rangle)^2 \rangle + B^2 \langle (s_b -\langle s_b \rangle)^2 \rangle \;+\; 2AB(\langle s_d s_b \rangle -2\langle s_b \rangle \langle s_d \rangle +  \langle s_d \rangle \langle s_b \rangle ) \\
    \sigma_{\tau_d}^2 &= A^2 \sigma_{s_d}^2 + B^2 \sigma_{s_b}^2 
\end{split}
\end{equation}
because $\langle s_d s_b \rangle = \langle s_d \rangle \langle s_b \rangle$ for the Sizer model (Eq. \ref{eq: corrSdSbSizer}). Now, putting in the values of $A$ and $B$, we get: 
\begin{equation}
    \sigma_{\tau_d}^2  =  \frac{\sigma_{s_d}^2}{\alpha^2 \langle s_d \rangle^2} +  \frac{\sigma_{s_b}^2}{\alpha^2 \langle s_b \rangle^2}
\end{equation}
Further, putting the previously known values, such as $\langle s_d \rangle$, $\langle s_b \rangle$, $\sigma_{s_d}^2$, and $\sigma_{s_b}^2$, one obtains Eq. \ref{eq: mainCOVsizerExpo}, which is true for the Sizer model.

\subsection{Adder Model} \label{subsec: SstatRelationshipsAdder}
From Eq. \ref{eq: sbSdSTD} and \ref{eq: sbDeltaDstdAdder} (Appendix \ref{sec: SbirthSize}), one obtains Eq. \ref{eq: mainStdAdder}, which represents the relationship between the standard deviations in $s_b$, $s_d$, and $\Delta_d$ for the Adder model. For division-time, we have $\tau_d = f(s_d,s_b)$ which can be written as:
\begin{equation}
    \tau_d \;=\; \frac{1}{\alpha}\ln{\left(\frac{s_d}{s_b}\right)} 
\end{equation}
From the discussion in Appendix \ref{sec: SbirthSize}, we know that if the principal distribution for the Adder model, i.e., $\Omega(\Delta_d)$, is sharply spiked, then $\zeta(s_b)$ will also be sharply spiked. And from Eq. \ref{eq: sbSdMean} and \ref{eq: nthCentralMomentSbSd}, we know that if $\zeta(s_b)$ is a sharply spiked distribution, $\Xi(s_d)$ will also be sharply spiked. Therefore, a sharply spiked principal distribution for the Adder model guarantees the validity of the arguments presented in the last subsection to calculate the mean and standard deviation of division-time. Therefore, the mean division-time for the Adder model is also given by Eq. \ref{eq: mainMeanTau_d}. However, the correlations for the Adder model are different than those for the Sizer model. Therefore, Eq. \ref{eq: tau_dSTDrawSizer} can be written for the Adder model as:
\begin{equation} \label{eq: tau_dSTDrawAdder}
\begin{split}
    \langle(\tau_d - \langle \tau_d \rangle)^2\rangle &= A^2 \langle (s_d -\langle s_d \rangle)^2 \rangle + B^2 \langle (s_b -\langle s_b \rangle)^2 \rangle \;+\; 2AB(\langle s_d s_b \rangle -2\langle s_b \rangle \langle s_d \rangle +  \langle s_d \rangle \langle s_b \rangle ) \\
    \sigma_{\tau_d}^2 &= A^2 \sigma_{s_d}^2 + B^2 \sigma_{s_b}^2 + 2AB \sigma_{s_b}^2
\end{split}
\end{equation}
where we have used the correlation $C(s_d,s_b)$ for the Adder model (Eq. \ref{eq: corrSdsbAdder}). Further, putting the values of $A$, $B$ and the previously known values, such as $\langle s_d \rangle$, $\langle s_b \rangle$, $\sigma_{s_d}^2$, and $\sigma_{s_b}^2$, in the equation above, one obtains Eq. \ref{eq: mainCOVadderExpo}, which is true for the Adder model.

\subsection{Stochastic growth rate} \label{subsec: SstatRelationshipsStochGrowth}
\begin{figure}[h]
    \centering
    \includegraphics[width=1\linewidth]{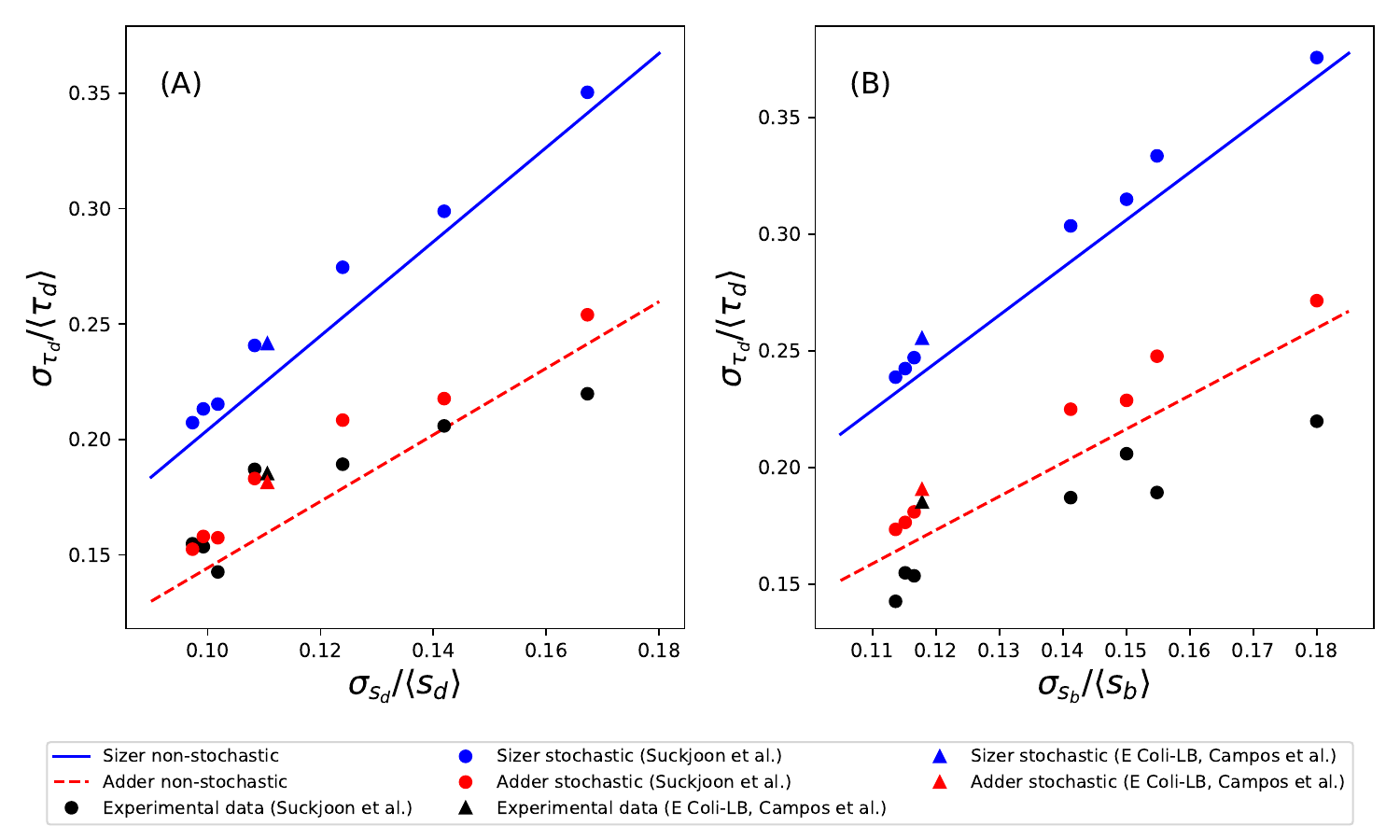}
    \caption{A comparison of \textbf{(A)} Coefficient of variation of $\tau_d$ versus that of $s_d$, \textbf{(B)} Coefficient of variation of $\tau_d$ versus that of $s_b$. The experimental data (black dots and triangles) have been taken from \cite{taheriJun2015, campos2014}. Solid lines represent the analytical prediction for the case of exponential growth with non-stochastic growth according to Eq. \ref{eq: mainCOVadderExpo} and \ref{eq: mainCOVsizerExpo}. Whereas the blue and red dots and triangles represent the analytical prediction for the case of population stochastic growth rate according to Eq. \ref{eq: COVtauStochExpoSizer} and \ref{eq: COVtauStochExpoAdder}.}
    \label{fig: SstochAlphaExpoCOVtauD}
\end{figure}

When the growth rate is stochastic in time, these statistical relationships (Eq. \ref{eq: mainMeanSingleCell}, \ref{eq: mainStdAdder}, \ref{eq: mainStdSizer}, \ref{eq: mainMeanTau_d}, \ref{eq: mainCOVadderExpo}, and \ref{eq: mainCOVsizerExpo}) do not change because one can simply replace $\alpha$ with its time-averaged value $\overline{\alpha}$ due to the arguments presented in Appendix \ref{sec: SstochsaticGrowthRate}. Similarly, when the growth rate is stochastic over population, the relationships for the size-related quantities, such as $s_b$, $s_d$, and $\Delta_d$ (Eq. \ref{eq: mainMeanSingleCell}, \ref{eq: mainStdAdder}, and \ref{eq: mainStdSizer}), remain the same, because they do not depend explicitly on the growth rate $\alpha$ or the type of growth. However, the relationships regarding the standard deviation in $\tau_d$ do change. When $\alpha$ is stochastic, using Taylor series expansion in three variables, one can expand the function $\tau_d = \ln{(s_d/s_b)}/\alpha$ in the vicinity of the point $(\langle s_d \rangle,\langle s_b \rangle, \langle \alpha \rangle)$ as:
\begin{equation}
\begin{split}
    \tau_d \;=\;& \frac{1}{\langle \alpha \rangle} \ln{\left(\frac{\langle s_d \rangle}{\langle s_b \rangle} \right)} \;+\; \frac{1}{\langle \alpha \rangle \langle s_d \rangle} (s_d - \langle s_d \rangle) \;-\; \frac{1}{\langle \alpha \rangle \langle s_b \rangle}(s_b - \langle s_b \rangle) \;-\; \frac{\ln{(\langle s_d \rangle /\langle s_b \rangle)}}{\langle \alpha \rangle^2} (\alpha - \langle \alpha \rangle) \\
    &  +\; \frac{\ln{(\langle s_d \rangle /\langle s_b \rangle)}}{\langle \alpha \rangle} \sum_{m=2}^{\infty} \frac{(-1)^m (\alpha -\langle \alpha \rangle)^m)}{\langle \alpha \rangle^m} + \frac{1}{\langle \alpha \rangle} \sum_{n=2}^{\infty} \frac{(-1)^{n+1}}{n} \left\{  \frac{(s_d -\langle s_d \rangle)^n}{\langle s_d \rangle^n} - \frac{(s_b -\langle s_b\rangle)^n}{\langle s_b \rangle^n} \right\} \\
    & +\; \sum_{n=1}^{\infty} \sum_{m=1}^{\infty} \frac{(-1)^{m+n+1}}{n \langle \alpha \rangle} \left\{  \frac{(s_d -\langle s_d \rangle)^n}{\langle s_d \rangle^n} - \frac{(s_b -\langle s_b\rangle)^n}{\langle s_b \rangle^n} \right\} \left( \frac{(\alpha - \langle \alpha \rangle)^n}{\langle \alpha \rangle^n} \right)
\end{split}
\end{equation}
Taking average of the equation above, we have,
\begin{equation}
\begin{split}
   \langle  \tau_d  \rangle\;=\;& \frac{1}{\langle \alpha \rangle} \ln{\left(\frac{\langle s_d \rangle}{\langle s_b \rangle} \right)} \; +\; \frac{\ln{(\langle s_d \rangle /\langle s_b \rangle)}}{\langle \alpha \rangle} \sum_{m=2}^{\infty} \frac{(-1)^m \mu_{m}(\alpha)}{\langle \alpha \rangle^m} + \frac{1}{\langle \alpha \rangle} \sum_{n=2}^{\infty} \frac{(-1)^{n+1}}{n} \left\{  \frac{\mu_n(s_d)}{\langle s_d \rangle^n} - \frac{\mu_(s_b)}{\langle s_b \rangle^n} \right\} \\
    & +\; \sum_{n=1}^{\infty} \sum_{m=1}^{\infty} \frac{(-1)^{m+n+1}}{n \langle \alpha \rangle} \left\{  \frac{\mu_n(s_d)}{\langle s_d \rangle^n} - \frac{\mu_n(s_b)}{\langle s_b \rangle^n} \right\} \left( \frac{\mu_n(\alpha)}{\langle \alpha \rangle^n} \right)
\end{split}
\end{equation}
where we have used the fact that  $\alpha$ is not correlated with either of $s_d$ and $s_b$ for either the Sizer or Adder model. We further assume that the distribution for $\alpha$ and the principal distributions are sharply spiked for the Adder and Sizer models. As discussed in the previous subsections, for the Sizer and Adder models, a sharply spiked principal distribution implies a sharply spiked $s_d$ and $s_b$ distribution. Therefore, taking all of the three distributions to be sharply spiked will make the terms containing $\mu_n(x)/\langle x \rangle^n$ for $n \geq 1$ vanish. And hence, we obtain the following relation:
\begin{equation}
    \langle \tau_d \rangle \;=\; \frac{\ln(\langle s_d \rangle / \langle s_b \rangle)}{\langle \alpha \rangle} \;=\; \frac{\ln(2)}{\langle \alpha \rangle}
\end{equation}
Additionally, if one looks in the vicinity of the point $(\langle s_d \rangle,\langle s_b \rangle, \langle \alpha \rangle)$ such that $s_d - \langle s_d \rangle \ll \langle s_d \rangle$, $s_b - \langle s_b \rangle \ll \langle s_b \rangle$, and $\alpha - \langle \alpha \rangle \ll \langle \alpha \rangle$, considering only the first-order terms is sufficient. Therefore, one can write: 
\begin{equation}
    \tau_d - \langle \tau_d \rangle = \frac{1}{\langle \alpha \rangle \langle s_d \rangle} (s_d - \langle s_d \rangle) \;-\; \frac{1}{\langle \alpha \rangle \langle s_b \rangle}(s_b - \langle s_b \rangle) \;-\; \frac{\langle \tau_d \rangle}{\langle \alpha \rangle} (\alpha - \langle \alpha \rangle)
\end{equation}
And therefore,
\begin{equation} \label{eq: COVtauStochExpoRaw}
\begin{split}
    \langle (\tau_d - \langle \tau_d \rangle)^2 \rangle \;=\; \sigma_{\tau_d}^2 \;&=\; \frac{1}{\langle \alpha \rangle^2 \langle s_d \rangle^2} \sigma_{s_d}^2 \;+\; \frac{1}{\langle \alpha \rangle^2 \langle s_b \rangle^2}\sigma_{s_b}^2 \;+\; \frac{\langle \tau_d \rangle^2}{\langle \alpha \rangle^2} \sigma_{\alpha}^2  \\ 
    & - \frac{2}{\langle \alpha \rangle^2 \langle s_d \rangle \langle s_b \rangle} \langle (s_d -\langle s_d\rangle)(s_b - \langle s_b \rangle) \rangle - \frac{2 \langle \tau_d \rangle}{\langle \alpha \rangle^2 \langle s_d \rangle} \langle (\alpha -\langle \alpha \rangle)(s_d - \langle s_d \rangle) \rangle  \\
    & + \frac{2 \langle \tau_d \rangle}{\langle \alpha \rangle^2 \langle s_b \rangle} \langle (\alpha -\langle \alpha \rangle)(s_b - \langle s_b \rangle) \rangle
\end{split}
\end{equation}
Since there is no correlation between $\alpha$ and $s_d$, and $\alpha$ and $s_b$, the cross terms containing $\langle (\alpha -\langle \alpha \rangle)(s_d - \langle s_d \rangle) \rangle $ and $\langle (\alpha -\langle \alpha \rangle)(s_b - \langle s_b \rangle) \rangle $ will vanish. Further, $s_d$ and $s_b$ are also uncorrelated for the Sizer model. Therefore, this term will also vanish. Hence, one can write for the coefficient of variation of $\tau_d$ for the Sizer model as:
\begin{equation} \label{eq: COVtauStochExpoSizer}
\begin{split}
     \frac{\sigma_{\tau_d}}{\langle \tau_d \rangle} \;=\; \sqrt{\left( \frac{\sqrt{2} \sigma_{s_b}}{\ln{(2)} \langle s_b \rangle} \right)^2 + \left( \frac{\sigma_{\alpha}}{\langle \alpha \rangle} \right)^2} \;=\; \sqrt{\left( \frac{\sqrt{2} \sigma_{s_d}}{\ln{(2)} \langle s_d \rangle} \right)^2 + \left( \frac{\sigma_{\alpha}}{\langle \alpha \rangle} \right)^2} \;=\; \sqrt{\left( \frac{\sqrt{2} \sigma_{\Delta_d}}{\sqrt{5} \ln{(2)} \langle \Delta_d \rangle} \right)^2 + \left( \frac{\sigma_{\alpha}}{\langle \alpha \rangle} \right)^2} 
\end{split}
\end{equation}
Since $s_d$ and $s_b$ are correlated for the Adder model, putting the value of the correlation between $s_d$ and $s_b$ (Eq. \ref{eq: corrSdsbAdder}) in Eq. \ref{eq: COVtauStochExpoRaw}, one can obtain the following relation for the Adder model:
\begin{equation} \label{eq: COVtauStochExpoAdder}
\begin{split}
     \frac{\sigma_{\tau_d}}{\langle \tau_d \rangle} \;=\; \sqrt{\left( \frac{ \sigma_{s_b}}{\ln{(2)} \langle s_b \rangle} \right)^2 + \left( \frac{\sigma_{\alpha}}{\langle \alpha \rangle} \right)^2} \;=\; \sqrt{\left( \frac{\sigma_{s_d}}{\ln{(2)} \langle s_d \rangle} \right)^2 + \left( \frac{\sigma_{\alpha}}{\langle \alpha \rangle} \right)^2} \;=\; \sqrt{\left( \frac{\sigma_{\Delta_d}}{\sqrt{3} \ln{(2)} \langle \Delta_d \rangle} \right)^2 + \left( \frac{\sigma_{\alpha}}{\langle \alpha \rangle} \right)^2} 
\end{split}
\end{equation}
Although the population stochasticity in growth rate changes the standard deviations for the $\tau_d$, the statistical relationships can still distinguish between the Sizer and Adder models, as shown in Fig. \ref{fig: SstochAlphaExpoCOVtauD}.

\section{Linear Growth} \label{sec: SlinearGrowth}

Some results do not change for the case of linear growth in comparison to the exponential growth (the growth refers to the growth of an individual cell in size/volume here). $\zeta(s_b)$ is identical for linear and exponential growth under both division rules, the Sizer and Adder, because the arguments presented in Appendix \ref{sec: SbirthSize} are independent of the type of growth. Similarly, the analytical forms of $\Xi(s_d)$ and $\Omega(\Delta_d)$ for the Sizer and Adder models are identical to the ones for the exponential growth, due to the fact that the type of growth or growth rate does not play any role in order to derive the analytical forms of these distributions, and the forms of these distributions are solely derived from the principal distribution. Apart from that, the analytical forms of some distributions do change. For example, for the case of linear growth, one can write $s_d = s_b + \alpha \tau_d$. Therefore $\Delta_d = \alpha \tau_d$. Hence, their probability distributions are related as follows:
\begin{equation} \label{eq: OmegaGammaLinear}
    \Gamma(\tau_d) \;=\; \alpha \; \Omega(\alpha  \tau_d)
\end{equation}
Additionally, for linear growth, one can also write $s = s_b + \alpha a$. Therefore $\Delta = \alpha a$. Hence, their probability distributions are also related as follows:
\begin{equation} \label{eq: PsiThetaLinear}
    \Theta(a) \;=\; \alpha \; \Psi(\alpha  a)
\end{equation}
The two equations above are valid for the three models of cell division, namely the Timer, Sizer, and Adder, under linear biomass growth, showing that $\Omega(\Delta_d)$ is a scaled version of $\Gamma(\tau_d)$, and $\Psi(\Delta)$ is a scaled version of $\Theta(a)$, or vice versa.

\subsection{Timer Model}
For the Timer model, $\Gamma(\tau_d)$ is the principal distribution, which is assumed to be known beforehand. As discussed in the main text, the Timer model is equivalent to the Adder model. Therefore, the Timer model gives cell size homeostasis for linear growth, and one can find the distributions for other quantities. Since we already know $\Gamma(\tau_d)$ for the Timer model, $\Omega(\Delta_d)$ is obtained by using Eq. \ref{eq: OmegaGammaLinear}. Once $\Omega(\Delta_d)$ is known, one can also find out $\zeta(s_b)$ using the arguments presented in Appendix \ref{sec: SbirthSize} for the Adder model because the Timer and Adder models are equivalent. Also, Eq. \ref{eq: mainTimer} is valid for the linear growth as well, because the whole derivation (Appendix \ref{sec: Stimer}) does not include the growth rate or the mode of growth for the individual cells. Therefore, $\Theta(a)$ is also known in terms of $\Gamma(\tau_d)$ by Eq. \ref{eq: mainTimer}. Additionally, if one knows $\Theta(a)$, $\Psi(\Delta)$ can also be found using Eq. \ref{eq: PsiThetaLinear}. To find $\Phi(s)$ and $\Xi(s_d)$, one can use the probability transformation of random variables as follows:
\begin{equation}
\begin{split}
    \Phi(s) \;&=\; \frac{1}{\alpha} \int_{0}^{\infty} ds_b \; \zeta(s_b) \; \Theta \left( \frac{s-s_b}{\alpha} \right)\\
    &=\; \alpha \int_{0}^{\infty} da \; \Theta(a) \; \zeta (s-\alpha a)
\end{split}
\end{equation}
and
\begin{equation} \label{eq: XiTimerLinear}
\begin{split}
    \Xi(s_d) \;=&\; \frac{1}{\alpha} \int_{0}^{\infty} ds_b \; \zeta(s_b) \; \Gamma \left( \frac{s_d-s_b}{\alpha} \right) \\
    &=\; \alpha \int_{0}^{\infty} d\tau_d \; \Gamma(\tau_d) \; \zeta (s_d-\alpha \tau_d)
\end{split}
\end{equation}
Further, if $\Gamma(\tau_d)$ is given to be a Gaussian distribution, $\Omega(\Delta_d)$ will also be a Gaussian distribution from Eq. \ref{eq: OmegaGammaLinear}. Additionally, from Eq. \ref{eq: XiTimerLinear}, it is apparent that $\Xi(s_d)$ is a convolution of two Gaussian functions. Therefore, $\Xi(s_d)$ will also be a Gaussian function.

\subsection{Sizer Model}
For the Sizer model, $\Xi(s_d)$ is the principal distribution, which is assumed to be known beforehand. $\zeta(s_b)$ and $\Omega(\Delta_d)$ can be found from $\Xi(s_d)$ using Eq. \ref{eq: zetaXi} and  \ref{eq: deltaDsizer}, having the analytical forms identical to the ones derived for the case of exponential growth, as discussed earlier. Also, one can find $\Gamma(\tau_d)$ using probability transformations as follows:
\begin{equation}
\begin{split}
    \Gamma(\tau_d) \;=&\; \alpha \int_{0}^{\infty} ds_b \; \zeta(s_b) \; \Xi(s_b + \alpha \tau_d) \\
    &=\; \alpha \int_{0}^{\infty} ds_d \; \Xi(s_d) \; \zeta(s_d - \alpha \tau_d)
\end{split}
\end{equation}
This $\Gamma(\tau_d)$ is different from the one obtained for the case of exponential growth. But using the properties of the convolution of two functions, it can be easily shown that the form of $\Gamma(\tau_d)$ will be Gaussian if the principal distribution is Gaussian itself. Furthermore, we have already shown that, for the Sizer model, $\Omega(\Delta_d)$ is identical under both linear and exponential growth, and it is approximately a Gaussian distribution under exponential growth if $\Xi(s_d)$ is Gaussian. Therefore, $\Omega(\Delta_d)$ will also be approximately a Gaussian distribution for linear growth as well. 

To obtain the population-level distributions, one starts from $\Phi(s)$. Using the statistical arguments, one can find $\Phi(s)$ in terms of $\Xi(s_d)$. We can write the probability that a cell divides after reaching a size between $s_d$ and $s_d + d s_d$ as $\Xi(s_d) d s_d$. Therefore, $F_>(s) \;=\; \int_{s}^{\infty} \Xi(s_d) d s_d$ is the probability that a cell will not divide before it reaches size $s$. Then the survival probability is defined for Sizer as:
\begin{equation}
    \eta \;=\; \frac{F_>(s+\Delta s)}{F_>(s)}
\end{equation}
which is the conditional probability that given a cell does not divide before it reaches a size $s$, the probability of it surviving until it reaches a size $s+\Delta s$. Here $\Delta$ is not added-size, but rather $\Delta s$ is an increment in size. The fraction of cells that reach size s and do not divide until reaching a size $s'=s+\Delta s$ is given as:
\begin{equation}
    N \; \Phi(s) ds \; \eta \;=\; N \; e^{\lambda t}\; \Phi(s')ds' 
\end{equation}
where $\lambda$ is the cell number growth rate, and t is the time taken to add a size of $\Delta s$. We have, $\lambda \neq \alpha$, rather they are related by Eq. \ref{eq: lambdaLinear}. Also, $s' = s + \alpha  t$ and thus $ds' = ds$ and $\Delta s /\alpha= t$. Thus, we get:
\begin{equation}
    \frac{F_>(s+\Delta s)}{F_{>}(s)} =\exp{\left( \frac{\lambda \Delta s}{\alpha} \right)}\frac{\Phi(s+\Delta s)}{\Phi(s)}
\end{equation}
Taking the Taylor expansion of the above functions and neglecting higher-order terms :
\begin{equation}
    1+\Delta s \frac{F_{>}^{'}(s)}{F_{>}(s)} = \left( 1+\frac{\lambda\Delta s}{\alpha} \right) \left( 1+\Delta s \frac{\Phi^{'}(s)}{\Phi(s)} \right)
\end{equation}
Comparing up to the first order of $\Delta s$ :
\begin{equation}
    \frac{F_>^{'}(s)}{F_>(s)} = \frac{\lambda}{\alpha} + \frac{\Phi^{'}(s)}{\Phi(s)}
\end{equation}
And after integration, we get :
\begin{equation}
    \Phi(s) = B \exp{\left( - \frac{\lambda}{\alpha} s \right)} \int_s^{\infty} \Xi(s_d) d s_d
\end{equation}
where B is the normalization constant. Further putting Eq. \ref{eq: lambdaLinear} into the equation above, one obtains:
\begin{equation} \label{eq: SizerLinear}
    \Phi(s) = B \;2^{-2s/\langle s_d\rangle} \int_s^{\infty} \Xi(s_d) d s_d
\end{equation}
$\Phi(s)$ for the Sizer model under linear growth is different from that for exponential growth. Furthermore, one can find out $\Psi(\Delta)$ and $\Theta(a)$ from $\Phi(s)$ and $\zeta(s_b)$ using probability transformations as follows:
\begin{equation}
\begin{split}
    \Psi(\Delta) \;&=\; \int_{0}^{\infty} ds_b \; \zeta(s_b) \; \Phi(s_b + \Delta) \\
    &=\; \int_{0}^{\infty} ds \; \Phi(s) \; \zeta(s-\Delta)
\end{split}
\end{equation}
and
\begin{equation}
\begin{split}
    \Theta(a) \;&=\; \alpha \;\int_{0}^{\infty} ds_b \; \zeta(s_b) \; \Phi(s_b + \alpha a) \\
    &=\; \alpha \; \int_{0}^{\infty} ds \; \Phi(s) \; \zeta(s-\alpha a)  
\end{split}
\end{equation}
Analogous to the case of exponential growth, Eq. \ref{eq: SizerLinear} does not correctly represent the real size distribution for the Sizer model under linear growth. But if the limits of the integration are from $s$ to $2s$, instead of $s$ to $\infty$, then it correctly represents the size distribution. Hence, the correct size distribution is given by:
\begin{equation} \label{eq: SizerLinear1}
    \Phi(s) = B \exp{\left( - \frac{\lambda}{\alpha} s \right)} \int_s^{2s} \Xi(s_d) d s_d
\end{equation}
The formula above can be derived using the arguments similar to those presented by \cite{kochSchac1962} to derive Eq. \ref{eq: mainSizer}. However, analogous to the case of the Sizer model for exponential growth, while doing the probability transformations, one has to use Eq. \ref{eq: SizerLinear} without any change of limits, because Eq. \ref{eq: SizerLinear1} gives incorrect results for distributions other than $\Phi(s)$ when used for the probability transformations.

\subsection{Adder Model}
For the Adder model, the division-added-size distribution $\Omega(\Delta_d)$ is the principal distribution, which is assumed to be known beforehand. $\zeta(s_b)$ and $\Xi(s_d)$ have the analytical forms identical to the ones derived for the Adder model under exponential growth, as discussed earlier. Also, since $\Omega(\Delta_d)$ is known, $\Gamma(\tau_d)$ can also be obtained using Eq. \ref{eq: OmegaGammaLinear}. Additionally, if we assume that $\Omega(\Delta_d)$ is a Gaussian distribution, $\Gamma(\tau_d)$ will also be Gaussian from Eq. \ref{eq: OmegaGammaLinear}. Furthermore, for the Adder model, we have shown that the $\Xi(s_d)$ is identical under both linear and exponential growth, and $\Xi(s_d)$ is approximately Gaussian for the Adder model under exponential growth if $\Omega(\Delta_d)$ is assumed to be a Gaussian distribution. Therefore, $\Xi(s_d)$ will also be approximately a Gaussian distribution.

To obtain the population-level distributions, we start from $\Psi(\Delta)$, which can be known in terms of $\Omega(\Delta_d)$ by the arguments of survival probability. Let us say, $\Omega(\Delta_d)  d\Delta_d$ is the probability that a cell divides after it has added a size between $\Delta_d$ and $\Delta_d + d\Delta_d$ since its birth. $F_>(\Delta)$ = $\int_{\Delta}^{\infty} \Omega_A(\Delta_d) d\Delta_d$ is the probability that a given cell divides after it has added a size $\Delta$ since its birth, and the survival probability is defined as:
\begin{equation}
    \eta \;=\; \frac{F_>(\Delta +x)}{F_>(\Delta)}
\end{equation}
which is the conditional probability that a given cell does not divide before it has added a size $\Delta+x$, given that it has already added a size $\Delta$. Now, the number of cells that reach an added-size $\Delta$ and do not divide until they add another size $x$ is given by:
\begin{equation}
    N \; \Psi(\Delta) d\Delta \; \eta \;\; =\;\; N e^{\lambda t} \; \Psi(\Delta') d\Delta'
\end{equation}
where $\lambda$ is the cell number growth rate, $t$ is the time to add an extra amount of size $x$, and $\Psi(\Delta) d\Delta$ is the probability that a given cell has added-size between $\Delta$ and $d\Delta$. We have $\lambda \neq \alpha$, rather they are related by Eq. \ref{eq: lambdaLinear}. Furthermore, $x=\alpha t$, and $d\Delta' = d\Delta$. Hence, simplifying the equation above:
\begin{equation}
    \frac{F_>(\Delta + x)}{F_>(\Delta)} \;\;=\;\; \exp{\left(\frac{\lambda x}{\alpha}\right)} \;\frac{\Psi(\Delta +x)}{\Psi(\Delta)}
\end{equation}
Expanding all the functions using their Taylor expansion up to first order in $x$, we get:
\begin{equation}
    1 \;+\; x\frac{F_{>}^{'}(\Delta)}{F_>(\Delta)} \;\;=\;\; \left( 1+x\frac{\lambda}{\alpha} \right) \left( 1+x\frac{\Psi^{'}(\Delta)}{\Psi(\Delta)} \right)
\end{equation}
Comparing L.H.S. and R.H.S. up to the first order of $x$, we get
\begin{equation}
    \frac{F_{>}^{'}(\Delta)}{F_{>}(\Delta)} - \frac{\lambda}{\alpha} \;=\; \frac{\Psi^{'}(\Delta)}{\Psi(\Delta)}
\end{equation}
After integrating, we get the added-size distribution for the adder model as:
\begin{equation} \label{eq: AdderLinear}
    \Psi(\Delta) \;=\; A \exp{\left( -\frac{\lambda}{\alpha} \Delta\right)} \; \int_{\Delta}^{\infty} \Omega_A(\alpha) d\alpha 
\end{equation}
where $A$ is the normalization constant. Using Eq. \ref{eq: lambdaLinear}, one can further write:
\begin{equation}
    \Psi(\Delta) = B \;2^{-\Delta/\langle \Delta_d\rangle} \int_{\Delta}^{\infty} \Omega(\Delta_d) d \Delta_d
\end{equation}
$\Psi(\Delta)$ for the Adder model under linear growth is different from the one for the case of exponential growth. Since we know the functional form of $\Psi(\Delta)$, we can also find out $\Theta(a)$ using Eq. \ref{eq: PsiThetaLinear}. Additionally, one can find out $\Phi(s)$ from $\Psi(\Delta)$ and $\zeta(s_b)$ using probability transformation as follows:
\begin{equation}
\begin{split}
    \Phi(s) \;&=\; \int_{0}^{\infty} ds_b \; \zeta(s_b) \; \Psi(s-s_b) \\
    &=\; \int_{0}^{\infty} d\Delta \; \Psi(\Delta) \; \zeta(s-\Delta)
\end{split}
\end{equation}

\subsection{Stochasticity in growth rate} \label{subsec: SlinearStochGrowth}
The arguments presented for the case of temporal stochastic exponential growth (Appendix \ref{sec: SstochsaticGrowthRate}) are valid for the case of linear growth as well. For linear growth, one can write for size for any time t as $s(t) = s(0)+ \alpha t$, where $s(0)$ is the initial size at time $t=0$ and $\alpha$ is the linear growth rate. But if the growth rate is stochastic in time, we have to write:
\begin{equation}
\begin{split}
    s(t) \;&=\; \; s(0) \;+\; \alpha_1 \Delta t \;+\; \alpha_2 \Delta t \;+...+\; \alpha_n \Delta t  \\
    &=\; \lim_{n  \rightarrow \infty} s(0) + \sum_{i=0}^{n} \alpha_i \Delta t \\
    &=\; \lim_{n  \rightarrow \infty} s(0) + \Delta t \sum_{i=0}^{n} \alpha_i  \\
\end{split}
\end{equation}
where the whole time interval from $t'=0$ to $t'=t$ is broken into $n$ infinitesimally small intervals of length $\Delta t = t/n $ and the central limit theorem gives $\Delta t \sum_{i=0}^n \alpha_i = \overline{\alpha} t$ in large $n$ limit. To derive various results for the fixed linear growth rate, we have used equations like $s=s_b + \alpha a$ and $s_d = s_b + \alpha \tau_d$. When there is temporal randomness in the growth rate, equations will be updated as $s=s_b + \overline{\alpha} a$ and $s_d = s_b + \overline{\alpha} \tau_d$. Therefore, the results derived for the case of non-stochastic linear growth are valid even when the growth rate is stochastic in time, except one just has to replace $\alpha$ by its time-averaged value $\overline{\alpha}$.

\begin{figure}[hbt]
    \centering
    \includegraphics[width=1\linewidth]{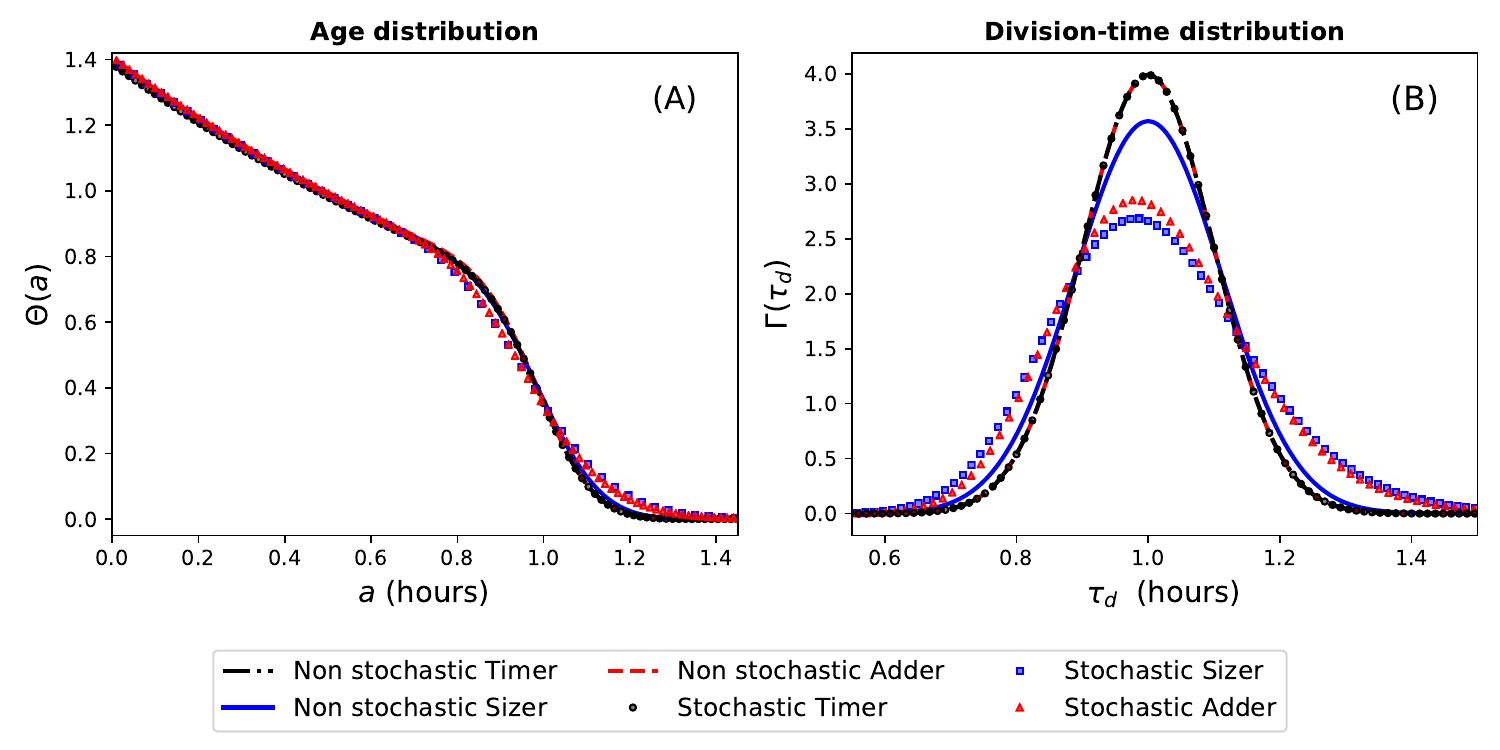}
    \caption{A comparison of the analytical forms of $\Theta(a)$ and $\Gamma(\tau_d)$ obtained for non-stochastic and stochastic (stochastic over the population) linear growth for the Timer, Sizer, and Adder models: \textbf{(A)} Age distribution, \textbf{(B)} Division-time distribution. The parameters for the models are identical to the ones mentioned in Fig. \ref{fig: linearGrowth}. For the case of population stochasticity in growth rate, $\alpha$ is sampled randomly from a probability distribution $\Lambda(\alpha)$ with a mean value $1.0 \; \mu$m per hour, and a standard deviation of $0.1 \; \mu$m per hour.}
    \label{fig: SstochAlphaLinearAgeTauD}
\end{figure}

However, similar to the case of exponential growth, here too, only $\Theta(a)$ and $\Gamma(\tau)$ distributions for the Sizer and Adder models do change their analytical forms when there is population stochasticity in the growth rate. The reason is also the same. The other distributions have no explicit dependence on the growth rate $\alpha$, and the survival probability arguments are still valid after replacing the absolute growth rate $\alpha$ by its population-averaged value $\langle \alpha \rangle$. Hence, their analytical forms do not change.  But, the analytical forms of $\Gamma(\tau_d)$ and $\Theta(a)$  for the Sizer model (when the growth rate is stochastic in time) are as follows:
\begin{equation}
    \begin{split}
    \Gamma(\tau_d) \;=&\;\int_{0}^{\infty} d\alpha \; \Lambda(\alpha) \; \alpha \int_{0}^{\infty} ds_b \; \zeta(s_b) \; \Xi(s_b + \alpha \tau_d) \\
    &=\; \int_{0}^{\infty} d\alpha \; \Lambda(\alpha) \; \alpha \int_{0}^{\infty} ds_d \; \Xi(s_d) \; \zeta(s_d - \alpha \tau_d)
\end{split}
\end{equation}
and
\begin{equation}
\begin{split}
    \Theta(a) \;&=\; \int_{0}^{\infty} d\alpha \; \Lambda(\alpha) \; \alpha \;\int_{0}^{\infty} ds_b \; \zeta(s_b) \; \Phi(s_b + \alpha a) \\
    &=\; \int_{0}^{\infty} d\alpha \; \Lambda(\alpha) \; \alpha \; \int_{0}^{\infty} ds \; \Phi(s) \; \zeta(s-\alpha a)  
\end{split}
\end{equation}
where $\Lambda(\alpha)$ is the probability distribution of $\alpha$. Similarly, for the Adder model, these distributions have the analytical forms as follows:
\begin{equation}
    \Gamma(\tau_d) = \int_{0}^{\infty} d\alpha \; \Lambda(\alpha) \; \alpha \; \Omega(\alpha \tau_d)
\end{equation}
and
\begin{equation}
    \Theta(a) = A \int_{0}^{\infty} d\alpha \; \Lambda(\alpha) \; \alpha \; \exp{\left(-\frac{\ln{(2)} \alpha a}{ \langle \Delta_d \rangle}\right)}  \int_{\alpha a}^{\infty} d\Delta_d \; \Omega(\Delta_d)
\end{equation}
where $A$ is the normalization constant. Fig. \ref{fig: SstochAlphaLinearAgeTauD} shows the comparison of the analytical forms of these distributions between stochastic (over population) and non-stochastic growth rate for the three models under linear growth.

\begin{figure}[hbt]
    \centering
    \includegraphics[width=1\linewidth]{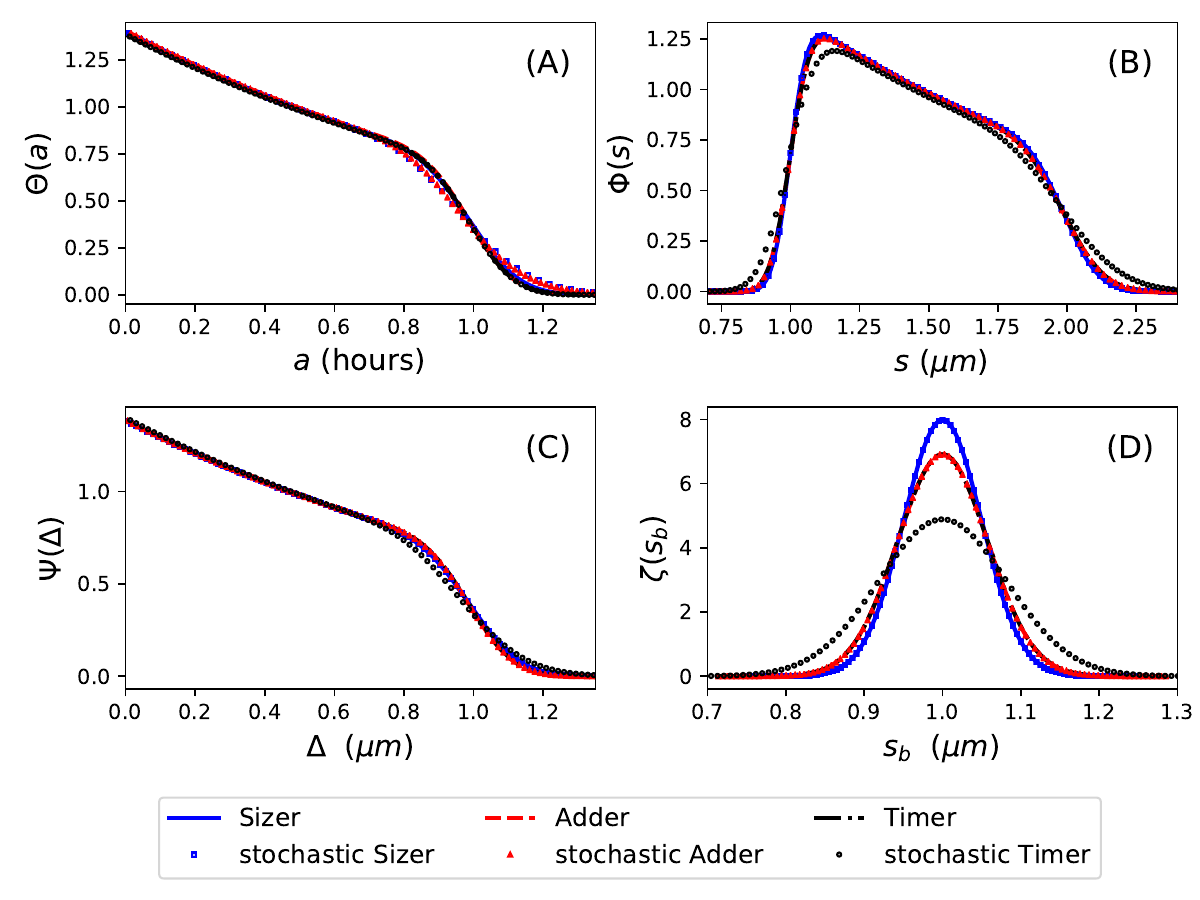}
    \caption{A comparison of the analytical forms of various distributions obtained for non-stochastic and stochastic (stochastic over the population) linear growth for the Timer, Sizer, and Adder models: \textbf{(A)} Age distribution, \textbf{(B)} Size distribution, \textbf{(C)} Added-size distribution, and \textbf{(D)} Birth-size distribution. The parameters for the models are identical to the ones mentioned in Fig. \ref{fig: linearGrowth}. For the case of population stochasticity in growth rate, $\alpha$ is sampled randomly from a probability distribution $\Lambda(\alpha)$ with a mean value $1.0 \; \mu$m per hour, and a standard deviation of $0.1 \; \mu$m per hour.}
    \label{fig: SstochAlphaLinear}
\end{figure}

For the case of the Timer model under population stochastic linear growth, $\Theta(a)$ and $\Gamma(\tau_d)$ are unchanged. This is so because the $\Gamma(\tau_d)$ is the principal distribution, which is a free parameter of the model, and $\Theta(a)$ is obtained using survival probability arguments, which are independent of $\alpha$, as discussed before. On the other hand, $\Omega(\Delta_d)$ is not a scaled version of $\Gamma(\tau_d)$ because $\alpha$ is a stochastic variable now. Rather, $\Omega(\Delta_d)$ can be found by using the probability transformation of $\alpha$ and $\tau_d$ as follows:
\begin{equation}
\begin{split}
    \Omega(\Delta_d) &\;=\; \int_{0}^{\infty} d\alpha \; \Lambda(\alpha) \; \frac{\Gamma(\Delta_d / \alpha)}{\alpha} \\
    &\;=\; \int_{0}^{\infty} d\tau_d \; \Gamma(\tau_d) \; \frac{\Lambda(\Delta_d / \tau_d)}{\tau_d}
\end{split}
\end{equation}
In general, $\Omega(\Delta_d)$ will have a complicated analytical form. However, let us assume that $\Lambda(\alpha)$ and $\Gamma(\tau_d)$ are Gaussian distributions with small standard deviations. Thus, one can write $\alpha = \langle \alpha \rangle + \delta_{\alpha}$ and $\tau_d = \langle \tau_d \rangle + \delta_{\tau_d}$. Further, $\delta_\alpha = \alpha - \langle \alpha \rangle$ is a Gaussian random variable with zero mean and a standard deviation of $\sigma_{\alpha}$. Similarly, $\delta_{\tau_d} = \tau_d - \langle \tau_d \rangle$ is a Gaussian random variable with zero mean and a standard deviation of $\sigma_{\tau_d}$. Therefore,
\begin{equation}
    \Delta_d \;=\; \alpha \tau \;=\; \langle \alpha \rangle \langle \tau_d \rangle + \langle \alpha \rangle \delta_{\tau_d} + \langle \tau_d \rangle \delta_{\alpha} + \delta_{\alpha} \delta_{\tau_d}
\end{equation}
Neglecting the product of small quantities, i.e. $\delta_{\alpha} \delta_{\tau_d}$, we can say that $\Delta_d$ is a random variable with the linear combination of two Gaussian random variables with some additive constant. Therefore, $\Omega(\Delta_d)$ will also be a Gaussian distribution with mean $\langle \Delta_d \rangle = \langle \alpha \rangle \langle \tau_d \rangle$ and variance $\sigma_{\Delta_d}^2 = \langle \alpha \rangle^2 \sigma_{\tau_d}^2 + \langle \tau_d \rangle^2 \sigma_{\alpha}^2$. Hence, $\Omega(\Delta_d)$will have more variance for the case of population stochastic growth in comparison to the case of a fixed growth rate. Additionally, since the Timer model is equivalent to the Adder model under linear growth, $\zeta(s_b)$ can be found because one knows the principal distribution for the Adder model, i.e., $\Omega(\Delta_d)$. The distribution of $s_b$ will be broader in comparison to the case of fixed growth rate (as visible from Fig. \ref{fig: SstochAlphaLinear} (D)) because $\sigma_{s_b} = \sigma_{\Delta_d}/\sqrt{3}$ and $\sigma_{\Delta_d}$ is bigger in comparison to the one obtained for non-stochastic growth. Furthermore, other distributions such as, $\Phi(s)$ and $\Psi(\Delta)$ can be found from the probability transformations of $\zeta(s_b)$ and $\Theta(a)$. A comparison of the population-level distributions among the three division models (for both stochastic and non-stochastic linear growth) is shown in Fig. \ref{fig: SstochAlphaLinear}.
\subsection{Statistical relationships for stochastic growth rate}
Analogous to the exponential growth, the statistical relationships for linear growth (Eq. \ref{eq: mainMeanSingleCell}, \ref{eq: mainStdAdder}, \ref{eq: mainStdSizer}, \ref{eq: mainMeanTau_dLinear}, \ref{eq: mainCOVadderLinear}, and \ref{eq: mainCOVsizerLinear}), are invariant to the temporal stochasticity in the growth rate, except that one has to replace $\alpha$ by its time-averaged value $\overline{\alpha}$ in the statistical relationships for the case of non-stochastic growth rate. Similarly, for the case of population stochastic growth rate, the statistical relationships regarding the size-related quantities (Eq. \ref{eq: mainMeanSingleCell}, \ref{eq: mainStdAdder}, and \ref{eq: mainStdSizer}) are not dependent on the type of growth and growth rate $\alpha$. Therefore, they do not change when there is population stochasticity in the growth rate. However, the relationships regarding standard deviation in $\tau_d$ do change for this case. For linear growth, we have $\tau_d = \Delta_d /\alpha$. Since, both $\Delta_d$ and $\alpha$ are stochastic quantities, we expand the function $\tau_d$ about the point$(\langle \Delta_d \rangle ,\langle \alpha \rangle)$ using Taylor series expansion in two variables:
\begin{equation}
    \tau_d \;=\; \frac{\langle \Delta_d \rangle}{\langle \alpha \rangle} \;+\; \sum_{n=1}^{\infty} (-1)^n \left( \frac{\langle \Delta_d \rangle}{\langle \alpha \rangle} \right) \frac{(\alpha - \langle \alpha \rangle)^n}{\langle \alpha \rangle^n}  \;+\; \sum_{n=0}^{\infty} (-1)^n \left( \frac{\langle \Delta_d \rangle}{\langle \alpha \rangle} \right)  \left( \frac{\Delta_d - \langle \Delta_d \rangle}{\langle \Delta_d \rangle} \right) \frac{(\alpha - \langle \alpha \rangle)^n}{\langle \alpha \rangle^n}
\end{equation}
Taking average on both sides of the above equation, we get:
\begin{equation}
    \langle \tau_d \rangle \;=\; \frac{\langle \Delta_d \rangle}{\langle \alpha \rangle} \;+\; \sum_{n=1}^{\infty} (-1)^n \left( \frac{\langle \Delta_d \rangle}{\langle \alpha \rangle} \right) \frac{\mu_n(\alpha)}{\langle \alpha \rangle^n}  \;+\; \sum_{n=0}^{\infty} (-1)^n \left( \frac{\langle \Delta_d \rangle}{\langle \alpha \rangle} \right)  \left( \frac{\langle \Delta_d - \langle \Delta_d \rangle \rangle}{\langle \Delta_d \rangle} \right) \frac{\mu_n(\alpha)}{\langle \alpha \rangle^n}
\end{equation}
where we have used the fact that $\Delta_d$ and $\alpha$ are not correlated. The term corresponding to the second summation evaluates to zero because $\langle \Delta_d - \langle \Delta_d \rangle \rangle = 0$. Further, if we assume that the distribution of $\alpha$ is sharply spiked, i.e.  $\mu_n(\alpha)/\langle \alpha \rangle^n \approx 0$ for $n \geq 2$, the term corresponding to the first summation also evaluates to zero. Therefore, we obtain:
\begin{equation}
    \langle \tau_d \rangle \;=\; \frac{\langle \Delta_d \rangle}{\langle \alpha \rangle}
\end{equation}
Additionally, while we look in the close neighborhood of the point $(\langle \Delta_d \rangle ,\langle \alpha \rangle)$ such that $\Delta_d - \langle \Delta_d \rangle \ll \langle \Delta_d \rangle $ and $\alpha - \langle \alpha \rangle \ll \langle \alpha \rangle $, higher order terms, being very small, do not contribute much. Therefore, we consider the first-order terms only and write for $\tau_d$ as:
\begin{equation}
    \tau_d - \langle \tau_d \rangle =  \frac{1}{\langle \alpha \rangle}(\Delta_d - \langle \Delta_d \rangle) - \frac{\langle \Delta_d \rangle}{\langle \alpha \rangle^2}(\alpha -\langle \alpha \rangle) 
\end{equation}
Therefore,
\begin{equation}
    \langle (\tau_d - \langle \tau_d \rangle)^2 \rangle =  \frac{1}{\langle \alpha \rangle^2} \langle (\Delta_d - \langle \Delta_d \rangle)^2 \rangle - \frac{\langle \Delta_d \rangle^2}{\langle \alpha \rangle^4} \langle (\alpha -\langle \alpha \rangle)^2 \rangle - \frac{2 \langle \Delta_d \rangle}{\langle \alpha \rangle^3} \langle (\Delta_d - \langle \Delta_d \rangle)(\alpha - \langle \alpha \rangle) \rangle
\end{equation}
Since $\Delta_d$ and $\alpha$ are not correlated. Therefore, the third term on the R.H.S. of the equation above vanishes; therefore, one can write:
\begin{equation}
    \frac{\sigma_{\tau_d}}{\langle \tau_d \rangle} \;=\; \sqrt{\left( \frac{\sigma_{\Delta_d}}{\langle \Delta_d \rangle} \right)^2 + \left( \frac{\sigma_{\alpha}}{\langle \alpha \rangle} \right)^2}
\end{equation}
Therefore, using the relationships between the standard deviation of $s_b$, $s_d$, and $\Delta_d$ for the Sizer model  (Eq. \ref{eq: mainStdSizer}), one can write:
\begin{equation} \label{eq: COVtauStochLinearSizer}
    \frac{\sigma_{\tau_d}}{\langle \tau_d \rangle} \;=\; \sqrt{\left( \frac{\sigma_{\Delta_d}}{\langle \Delta_d \rangle} \right)^2 + \left( \frac{\sigma_{\alpha}}{\langle \alpha \rangle} \right)^2} \;=\; \sqrt{\left( \frac{\sqrt{5} \sigma_{s_d}}{\langle s_d \rangle} \right)^2 + \left( \frac{\sigma_{\alpha}}{\langle \alpha \rangle} \right)^2} \;=\; \sqrt{\left( \frac{\sqrt{5} \sigma_{s_b}}{\langle s_b \rangle} \right)^2 + \left( \frac{\sigma_{\alpha}}{\langle \alpha \rangle} \right)^2}
\end{equation}
Further, using the relationships between the standard deviation of $s_b$, $s_d$, and $\Delta_d$ for the Adder model  (Eq. \ref{eq: mainStdAdder}), one can write:
\begin{equation} \label{eq: COVtauStochLinearAdder}
    \frac{\sigma_{\tau_d}}{\langle \tau_d \rangle} \;=\; \sqrt{\left( \frac{\sigma_{\Delta_d}}{\langle \Delta_d \rangle} \right)^2 + \left( \frac{\sigma_{\alpha}}{\langle \alpha \rangle} \right)^2} \;=\; \sqrt{\left( \frac{\sqrt{3} \sigma_{s_d}}{\langle s_d \rangle} \right)^2 + \left( \frac{\sigma_{\alpha}}{\langle \alpha \rangle} \right)^2} \;=\; \sqrt{\left( \frac{\sqrt{3} \sigma_{s_b}}{\langle s_b \rangle} \right)^2 + \left( \frac{\sigma_{\alpha}}{\langle \alpha \rangle} \right)^2}
\end{equation}

\section{Probability transformation of random variables} \label{sec: SprobabTrans}

Suppose we are given a joint probability distribution of two random variables $x$ and $y$ as $P(x,y)$ and we want to find out the probability distribution of $z=Z(x,y)$. Now, assume that this function is invertible and $x$ can be written as a function of $y$ and $z$ as $x=X(y,z)$,  and the same goes for $y$ as $y=Y(x,z)$. The probability distribution for z can be given as \cite{rileyHobsonBence} :
\begin{equation} \label{eq: probTransRaw}
    Q(z) \;=\; \int_{0}^{\infty} \int_{0}^{\infty} P(x,y) \delta(z-Z(x,y)) dx\;dy
\end{equation}
The lower limit of integration is $0$, not $-\infty$, because the variables we have considered are related to the age and size of the cells. Therefore, they can not have negative values. Now using the property of Dirac delta function:
\begin{equation}
    \delta(f(x)) \;=\; \sum_{x_i} \frac{\delta(x-x_i)}{|f'(x_i)|} \;\;\; \forall f(x_i) = 0 
\end{equation}
For our purposes of transformations, there is only one zero of such functions. This is so because in the probability transformations, we use functions like $s_d = s_b \exp{(\alpha \tau_d)}$ that are invertible and have only one zero for each variable. Hence, the equation can be written as :
\begin{equation}
    \delta(f(x)) \;=\;  \frac{\delta(x-x_0)}{|f'(x_0)|} 
\end{equation}
where $f(x_0)=0$. Using the equation above, Eq. \ref{eq: probTransRaw} can be modified as:
\begin{equation}
    Q(z) \;=\; \int_{0}^{\infty} \int_{0}^{\infty} P(x,y) \; \frac{\delta(x-X(z,y))}{ \left |\frac{\partial Z(x,y)}{\partial x} \right |_{x=X(y,z)}} \;dx\;dy
\end{equation}
Now from $x=X(y,z)$ and $y=Y(x,z)$ , one can write:
\begin{equation}
    dx \;=\; \frac{\partial X(y,z)}{\partial y} dy + \frac{\partial X(y,z)}{\partial z} dz
\end{equation}
and
\begin{equation}
    dz \;=\; \frac{\partial Z(x,y)}{\partial x} dx + \frac{\partial Z(x,y)}{\partial y} dy
\end{equation}
From the two equations above, one can conclude that:
\begin{equation}
    \left |\frac{\partial Z(x,y)}{\partial x} \right |^{-1} \;=\; \left |\frac{\partial X(y,z)}{\partial z} \right |
\end{equation}
which gives us:
\begin{equation} 
    Q(z) \;=\; \int_{0}^{\infty} \int_{0}^{\infty} \; f(x,y,z) \delta(x-X(y,z)) \; dx\;dy
\end{equation}
where $f(x,y,z) = P(x,y) \left |\frac{\partial X(y,z)}{\partial z} \right |$, which finally gives us:
\begin{equation} 
    Q(z) \;=\; \int_{0}^{\infty} P(X(z,y),y) \;\; \left |\frac{\partial X(y,z)}{\partial z} \right | \;\; dy 
\end{equation}
The same thing can be done for $y$ also and one can show that the distribution for $z$ is geiven by:
\begin{equation} \label{eq: probTrans}
\begin{split}
    Q(z) &\;=\; \int_{0}^{\infty} P(X(z,y),y) \;\; \left |\frac{\partial X(y,z)}{\partial z} \right | \;\; dy \\
    &\;=\; \int_{0}^{\infty} P(x,Y(x,z)) \;\; \left |\frac{\partial Y(z,x)}{\partial z} \right | \;\; dx
\end{split}
\end{equation}
Let us call the equation \ref{eq: probTrans} as the `probability transformation equation'. This equation is used to derive the probability distribution of a variable given as a function of two other variables whose joint probability distribution is known. Using similar arguments, one can further show that if the joint probability distribution $P(x,y,w)$ of three variables, $x$, $y$, and $w$ is known, the probability distribution of another variable $z=Z(x,y,w)$ (where the function $Z(x,y,w)$ is invertible in x, y, and w, and has only one zero for each variable) can also be found using the equation below:
\begin{equation} \label{eq: probTransMultiplevariables}
    \begin{split}
    Q(z) &\;=\; \int_{0}^{\infty} \int_{0}^{\infty} dx \; dy \; P(x,y,W(x,y,z)) \;\; \left |\frac{\partial W(x,y,z)}{\partial z} \right | \\
    &\;=\;  \int_{0}^{\infty} \int_{0}^{\infty} dy \; dw \; P(X(z,y,w),y,w) \;\; \left |\frac{\partial X(y,z,w)}{\partial z} \right | \\
    &\;=\;  \int_{0}^{\infty} \int_{0}^{\infty} dw \; dx \; P(x,Y(x,w,z),w) \;\; \left |\frac{\partial Y(x,w,z)}{\partial z} \right | 
    \end{split}
\end{equation}
 
\section{Various types of growth rates in a culture} \label{sec: SrelationshipGrowthRates}

A bacterial culture has two types of growth in steady-state growth conditions: an individual cell's biomass growth and the growth in the number of cells over time. Also, it is well known that the number of cells in a bacterial culture always increases exponentially, regardless of the type of individual cell's biomass growth, i.e., linear or exponential. Therefore, the number of cells in a culture at any time $t$ is given as:
\begin{equation}
    N(t) =  N(0) \; 2^{D t} =  N(0) \; e^{\lambda t}
\end{equation}
where $N(0)$ is the number of cells at time $t=0$. We call $D$ the doubling rate, which is the rate at which the number of cells doubles in the population, and $\lambda$ the cell number growth rate, which is the rate of exponential growth in the number of cells in the culture. Both of them are related as $\lambda = D\ln (2)$. If on average, the number of cells in a culture doubles every $\langle \tau_d \rangle$ minute (because individual cells divide after $\langle \tau_d \rangle$ minutes on average, where $\langle \tau_d \rangle$ is the mean division-time), the doubling rate for the culture is $1/ \langle \tau_d \rangle$ per minute. Therefore, the cell number growth rate $\lambda$ is given as:
\begin{equation} \label{eq: lambda}
    \lambda = \ln{(2)}/\langle \tau_d \rangle
\end{equation}
For the Sizer and Adder models, it can be shown that the average division-time $\langle \tau_d \rangle$ is given by Eq. \ref{eq: mainMeanTau_d}. Therefore, using Eq. \ref{eq: mainMeanTau_d} and \ref{eq: lambda}, one can obtain a relationship between the individual cell's growth rate and the cell number growth rate as:
\begin{equation}
    \lambda =  \alpha
\end{equation}

Additionally, if the biomass growth of an individual cell is stochastic, one can replace the absolute value of growth rate with its average value for the first order approximation ($\overline{\alpha}$ for growth stochastic in time and $\langle \alpha \rangle$ for the growth stochastic over the population). This means the cell number growth rate equals the average exponential biomass growth rate for the Sizer and Adder models. However, for the Timer model, the division-time distribution $\Gamma(\tau_d)$ is the principal distribution and is assumed to be known beforehand. Therefore, the average value of the mean division-time is decided independently of the biomass growth rate $\alpha$, and the cell number growth rate is given by Eq. \ref{eq: lambda}. Similarly, for linear biomass growth, using Eq. \ref{eq: mainMeanTau_dLinear} and \ref{eq: lambda}, we get:
\begin{equation} \label{eq: lambdaLinear}
    \lambda = \alpha \ln{(2)}/\langle \Delta_d \rangle = \alpha \ln{(2)} /\langle s_b \rangle = 2 \alpha \ln{(2)} /\langle s_d \rangle
\end{equation}
This relationship is true for the three models of cell division under linear biomass growth.

\section{Correlations}\label{sec: Scorrelations}

Although Jun et al. \cite{taheriJun2015} have derived important correlations for the three models, we also derive the ones that will be useful for us to prove some other results in this text. In the derivations below, $C(a,b)$ represents the correlation between the quantities $a$ and $b$, which is defined as $C(a,b) = \langle a b \rangle - \langle a \rangle \langle b\rangle$.  Also, we have extensively made use of Eq. \ref{eq: mainMeanSingleCell}, \ref{eq: sbSdSTD}, and \ref{eq: sbDeltaDstdAdder} to simplify the expressions for the correlations.
\subsection{Sizer}
For the Sizer model, the division-size of a cell is postulated to be not correlated with the division-size of its ancestor and descendant cells. Therefore, one can write:
\begin{equation} \label{eq: corrSdSdSizer}
    C(s_{d_i} s_{d_j}) \;=\; \langle s_{d_i} s_{d_j} \rangle - \langle s_d \rangle ^2 = \delta_{ij} \sigma_{s_d}^2
\end{equation}
where $s_{d_i}$, and $s_{d_j}$ are the division-sizes for the $i$th and $j$th generation cells. Using the equation above, one can also show that the division-size of a cell is not correlated with its birth-size:
\begin{equation}\label{eq: corrSdSbSizer}
\begin{split}
    C(s_d s_b) \;&=\; \langle s_d s_b\rangle - \langle s_d \rangle \langle s_b \rangle \\
    &=\; \langle s_d s_{d_m}/2 \rangle - \langle s_d \rangle \langle s_b \rangle \\
    &=\; \langle s_d \rangle  \langle s_{d_m} \rangle /2 - \langle s_d \rangle \langle s_b \rangle  \\
    &=\; 0
\end{split}
\end{equation}
where $s_{d_m}$ is the division-size of the mother cell. This correlation is used in obtaining the standard deviation for $\Gamma(\tau_d)$ for the Sizer model in Appendix \ref{subsec: SstatRelationshipsSizer}.  For the correlation between the division-added-sizes of mother and daughter cells, one can write $C(\Delta_{d_{0}} \Delta_{d_{1}}) = \langle \Delta_{d_{0}} \Delta_{d_{1}} \rangle - \langle \Delta_d \rangle^2$, where the subscript of the subscript denotes the generation of the cell, i.e., 0 for the present generation cell, 1 for the daughter cell, -1 for the mother cell, and so on. One can further simplify the expression using Eq. \ref{eq: corrSdSdSizer} as follows:
\begin{equation}\label{eq: corrDeltaDdeltaDmotherDaughterSizer}
\begin{split}
    C(\Delta_{d_0} \Delta_{d_1}) &= \langle (s_{d_0}-s_{d_{-1}}/2) (s_{d_1}-s_{d_{0}}/2)\rangle - \langle s_d \rangle^2 /4 \\
    &= \langle s_{d_0} s_{d_1}\rangle - \langle s_{d_0}^2 \rangle /2 - \langle s_{d_{-1}} s_{d_1}\rangle /2 + \langle s_{d_{-1}} s_{d_0}\rangle /4 - \langle s_d \rangle^2 /4 \\
    &= - \sigma_{s_d}^2 /2
\end{split}
\end{equation}
Moreover, the correlations between the division-added-sizes for distant generations can be written as $C(\Delta_{d_{i}} \Delta_{d_{i+j}}) = \langle \Delta_{d_{i}} \Delta_{d_{i+j}} \rangle - \langle \Delta_d \rangle^2$, where $j \geq 2$. This can be further simplified using Eq. \ref{eq: corrSdSdSizer} as follows:
\begin{equation}\label{eq: corrDeltaDdeltaDSizer}
\begin{split}
    C(\Delta_{d_i} \Delta_{d_{i+j}}) &= \langle (s_{d_i}-s_{d_{i-1}}/2) (s_{d_{i+j}}-s_{d_{i+j-1}}/2)\rangle - \langle s_d \rangle^2 /4 \\
    &= \langle s_{d_i} s_{d_{i+j}}\rangle - \langle s_{d_i} s_{d_{i+j-1}} \rangle /2 - \langle s_{d_{i-1}} s_{d_{i+j}}\rangle /2 + \langle s_{d_{i-1}} s_{d_{i+j-1}}\rangle /4 - \langle s_d \rangle^2 /4 \\
    &= 0
\end{split}
\end{equation}
The correlation between $\Delta_d$'s of different generations, as derived above, are used to get the relationship between the standard deviations in $s_b$ and $\Delta_d$ for the Sizer model in Appendix \ref{sec: SbirthSize}.

\subsection{Adder}
For the Adder model, the division-added-size of a cell is postulated to be not correlated with the division-added-size of its ancestor and descendant cells. Therefore, one can write:
\begin{equation} \label{eq: corrDeltaDdeltaDadder}
    C(\Delta_{d_i} \Delta_{d_j}) \;=\; \langle \Delta_{d_i} \Delta_{d_j} \rangle - \langle \Delta_d \rangle ^2 = \delta_{ij} \sigma_{\Delta_d}^2
\end{equation}
where $\Delta_{d_i}$ and $\Delta_{d_j}$ are the division-added-sizes for the $i$th and $j$th generation cells. This correlation is used to obtain the relationship between the standard deviations in $s_b$ and $\Delta_d$ for the Adder model in Appendix \ref{sec: SbirthSize}. Using the equation above, one can also show that $s_b$ and $\Delta_d$ are also not correlated with each other for the Adder model:
\begin{equation}\label{eq: corrDeltaDsbAdder}
\begin{split}
    C(\Delta_d s_b) \;&=\; \langle \Delta_d s_b\rangle - \langle \Delta_d \rangle \langle s_b \rangle \\
    &=\; \langle \Delta_d (s_{b_{-1}}+\Delta_{d_{-1}})/2 \rangle - \langle \Delta_d \rangle \langle \Delta_d \rangle \\
    &=\; \langle \Delta_d \lim_{ n\rightarrow\infty} \left(\frac{s_{b_{-n}}}{2^{n}} + \sum_{k=1}^{n}\frac{\Delta_{d_{-k}}}{2^k}\right)\rangle - \langle \Delta_d \rangle \langle \Delta_d \rangle   \\
    &=\; \lim_{ n\rightarrow\infty} \left[ \frac{ \langle \Delta_d s_{b_{-n}} \rangle}{2^{n}} + \frac{\sum_{k=1}^{n} \langle \Delta_d \; \Delta_{d_{-k}} \rangle}{2^k} \right] -\langle \Delta_d \rangle^2 \\
    &=\; \langle \Delta_d \rangle^2 \left[ \left( \sum_{k=1}^{n}\frac{1}{2^k} \right) -1 \right] \\
    &=\;  0
\end{split}
\end{equation}
Additionally, for the correlation between $s_d$ and $s_b$ of the same cell, one can write:
\begin{equation}\label{eq: corrSdsbAdder}
\begin{split}
    C(s_d s_b) &= \langle s_d s_b \rangle - \langle s_d s_b \rangle \\
    &= \langle (\Delta_d + s_b) s_b\rangle - \langle s_d s_b\rangle\\
    &= \langle \Delta_d s_b \rangle + \langle s_b^2 \rangle - 2\langle s_b \rangle^2\\
    &= \langle \Delta_d \rangle \langle s_b \rangle + \langle s_b^2 \rangle - 2\langle s_b \rangle^2\\
    &= \sigma_{s_b}^2
\end{split}
\end{equation}
 This correlation is used in obtaining the standard deviation for $\Gamma(\tau_d)$ for the Adder model in Appendix \ref{subsec: SstatRelationshipsAdder}.

\begin{acknowledgments}
Vikas gratefully acknowledges the University Grants Commission (UGC), India, for financial support through the Junior Research Fellowship Number 221610131832. AR would like to thank Krishnajyoti Mukherjee and Amba Anegundi for discussions during the early part of the work.
\end{acknowledgments}

\bibliography{sources}

\end{document}